\documentstyle[aps,preprint,eqsecnum]{revtex}
\tighten
\begin{document}
\preprint{\parbox{1.25in}{\begin{flushright}
        UB-HET-93-1\\ August 1993\end{flushright}}}
\draft


\title{
Chromo-electroweak interference and parity-violating \\
asymmetries in the production of an \\
electroweak boson $+$ two jets in hadron collisions
}
\author{Richard J.\ Gonsalves}
\address{
Department of Physics \\
State University of New York at Buffalo \\
Buffalo, New York 14260, USA
}
\author{C.F.\ Wai}
\address{
Institute of Physics \\
Academia Sinica \\
Taipei, Taiwan 11529, Republic of China
}
\author{[{\em Published in } 
Phys.\ Rev.\ D {\bf49} 190-218 (1994); 
Erratum: {\it ibid.}, D {\bf 51} 1428 (1995)
]}

\maketitle

\begin{abstract}
In the Standard Model the process $q\bar{q}\rightarrow q\bar{q}V$
(where $V=W^\pm,Z^0$ or $\gamma$) can occur
via gluon exchange and also via $W^\pm$ or $Z^0$
exchange.  The corresponding chromodynamic and electroweak amplitudes
can interfere with one another.  These interference cross sections are
largest when the exchanged $W^\pm$ or $Z^0$ is on-shell when they are
also odd under parity.  Interference cross sections computed using
helicity-amplitude techniques are presented for all interesting
subprocesses as well as for the processes $q\bar{q}\rightarrow
q\bar{q} l\bar{l}$ in which the lepton pair $l\bar{l}$ comes from the decay of
$V$ on-shell.  Parity-violating asymmetries are defined and presented
at the parton level and for the hadronic processes $pp$ or
$p\bar{p}\rightarrow V$ + 2 jets or $l\bar{l}$ + 2 jets.  These
asymmetries are independent of the polarizations of all particles
involved, and do not require that the flavors of the jet partons be
measured.  They are generally of order 0.01 pb at energies $\sqrt{s}
\gtrsim 1$ TeV.
\end{abstract}

\pacs{12.10.Dm, 13.85.Qk, 14.80.Er}
\narrowtext
\newpage
\section{Introduction}

The Standard ${\rm SU(3)_c\otimes SU(2)_L\otimes U(1)}$ Model
\cite{qcd,gsw} of the strong and electroweak interactions has several
features for which a simple and satisfactory theoretical explanation
is lacking.  These include the origin of the masses of the $W$ and $Z$
bosons, and the pattern of the couplings of these bosons to quarks and
leptons---a pattern which involves several unexplained parameters and
violation of symmetries such as parity and possibly CP conjugation.
Experiments which study the production and decay of one or more
electroweak gauge bosons might provide clues that will enable us to
better understand these puzzling features of the Standard Model.  In
this paper we will present some intriguing predictions of the Standard
Model for correlations between the final state particles in the
production of pairs of electroweak bosons in hadron-hadron collisions.
These correlations produce asymmetries in the predicted cross sections
that are very sensitive to the details of the couplings of the gauge
bosons to quarks and in particular to the parity-violating nature of
these couplings.  While these asymmetries are predicted to be rather
small, their measurement would provide an elegant test of the Standard
Model, and deviations from these predictions might provide clues to a
more fundamental structure which might underlie the Standard Model.

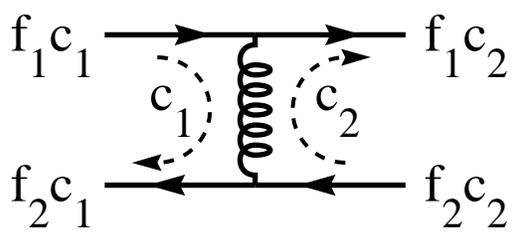
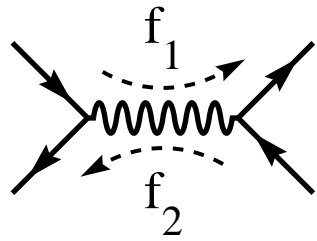
\begin{figure}
\caption{
Chromo-electroweak interference.  A solid line represents a quark or
antiquark with color $c$ and flavor $f$.  The wavy line represents a
color-neutral electroweak boson and the curly line a flavor-neutral
gluon.  The two amplitudes can interfere because their initial states
(on the left) can have identical quantum numbers as can their final
states (on the right).
}
\label{twototwofig}
\end{figure}

The parity-violating asymmetries presented in this paper owe their
existence to a quantum mechanical interference between amplitudes
involving the exchange of gluons and electroweak bosons in $q\bar{q}$
annihilation.  This phenomenon of ``chromo-electroweak interference''
is exemplified by the diagrams in Fig.\ \ref{twototwofig}.  It is
similar in nature to the interference \cite{zeldovich} between a
virtual photon and a virtual $Z^0$ which gives rise for example to the
well known asymmetry \cite{muon} in $e^+e^-\rightarrow\mu^+\mu^-$, to
the parity-violating asymmetry in the scattering of
polarized electrons from nuclei \cite{electron}, 
or to the parity-violation observed in
atomic systems \cite{atomic}.  Unlike electroweak interference between
a photon and a $Z^0$, which have the same quantum numbers,
chromo-electroweak interference between a gluon and an electroweak
boson, which have different color and flavor quantum numbers, can
occur only if the gluon and the electroweak boson are exchanged in
different channels as shown in Fig.\ \ref{twototwofig}.  An
interference between the two diagrams in this figure is interesting
because the magnitude of its contribution to the cross section might
be expected to lie in between that of the strong and the weak exchange
diagrams taken separately: thus if the weak contribution were too
small to be observed, the interference might provide the only
observable signature of the weak exchange.  An extremely interesting
possibility arises if at least one of the quarks is polarized and the
electroweak boson is a $W$ or a $Z^0$, both of which couple
asymmetrically to left- and right-handed fermions: the interference
contributions can then contain terms that are odd under parity, i.e.,
that are proportional to invariants such as $\vec{s}\cdot\vec{p}$
where $\vec{s}$ is a spin and $\vec{p}$ a momentum.  This possibility
has been exploited \cite{hadron} to make predictions for
parity-violating signatures in inclusive hadron production at large
$p_T$ in polarized hadron-hadron scattering.  Unfortunately,
experiments with polarized beams are difficult to perform except at
low energies at which the weak couplings are very small, and the
prospects for observing these asymmetries do not appear to be very
promising.

The prospects for observing chromo-electroweak interference effects
might be considerably better at energies above the threshold for
producing $W$ and $Z$ bosons.  At these energies the weak couplings
are in fact only an order of magnitude smaller than the strong
interaction coupling.  In addition, the cross sections for observing
multiparticle final states including jets of hadrons at large $p_T$
become appreciable at these energies, and this makes it possible to
avoid the need for observing polarizations in order to have parity
violation.  It was pointed out in \cite{rjg86} that if for example a
$W$ boson were radiated off one of the quarks lines in Fig.\
\ref{twototwofig}, the interference cross section would be
parity-violating even if all particles involved were unpolarized.
With two incoming particles and three particles in the final state, it
becomes possible to construct a non-vanishing parity-odd invariant
$\epsilon_{\mu\nu\lambda\sigma} p_1^\mu p_2^\nu p_3^\lambda
p_4^\sigma$ from the momenta $p_i$ of the five particles involved,
four of which can be chosen to be linearly independent. In
\cite{rjg86}, it was shown that this interference gave rise to
parity-violating asymmetries of order 0.01 pb in the process
$p\bar{p}\rightarrow W^- + 2$ jets at energies $\gtrsim$ 1 TeV.  The
asymmetry is largest when the electroweak boson in Fig.\
\ref{twototwofig} and the additional radiated $W$ are on-shell.
Observation of this asymmetry does not require that the flavors of the
jets be determined---only that their momenta be determined with
sufficient accuracy to select pairs of jets with the invariant mass of
the pair equal to that of a $W$ or a $Z$ within a few GeV. Asymmetries
with a similar origin have been studied \cite{clavelli} in $e^+e^-$
annihilation, and an application to the process $e^+e^-\rightarrow
e^+e^-\gamma$ has recently been discussed in \cite{tungwai}. 
Similar parity-violating asymmetries occur in the
process hadron + hadron $\rightarrow$ 3 jets \cite{cwai92}: they originate
from diagrams in which the electroweak boson $V$ is replaced by a
gluon; they are considerably larger than the asymmetries in $V$ +
2 jet production discussed in this paper, 
but may not be easy to detect above the QCD 3-Jet
background.

In \cite{rjg86}, the radiated boson was taken to be a $W^-$ and the
boson which decayed to 2 jets a $Z^0$.  The purpose of this paper is
to present a detailed derivation and discussion of the results which
were summarized in \cite{rjg86}, and to extend these results to all
possible combinations of pairs of electroweak bosons, including real
photons, that appear to be of phenomenological interest.  We also
consider the more general process in which the final state contains a
lepton pair which comes from the decay of an on-shell $W^\pm$ or
$Z^0$, and a $q\bar{q}$ pair which comes from the decay of a second
on-shell $W$ or $Z$.  In section II, we single out the dominant
amplitudes that are expected to contribute to chromo-electroweak
interference, and present compact expressions for amplitudes and
interference cross sections \cite{cwai88} 
derived using the powerful and elegant
helicity-amplitude techniques \cite{helamps} 
that have recently gained popularity.
In section III we first define parity-violating
asymmetries in a way that depends only on the momenta of the jet
partons involved and not on their internal quantum numbers.  We then
present numerical predictions for these asymmetries, both at the
parton and at the hadron levels, first for $V$ + 2 jet final states
where $V=W^\pm$, $Z^0$ or a real photon, and then for $l\bar{l}$ + 2
jet final states in which the lepton pair comes from the decay of an
on-shell $W^\pm$ or $Z^0$.  We also discuss the problem of observing
parity-violating asymmetries above a background \cite{background} that
is expected to be as much as two orders of magnitude larger than the
signal.  Our conclusions are presented in section IV, and some details
pertaining to the calculation of the helicity amplitudes and cross
sections are presented in the Appendixes.

\section{Chromo-electroweak Interference}

\subsection{Subprocesses contributing to $h_1+h_2\rightarrow 
j_1+j_2+V$}

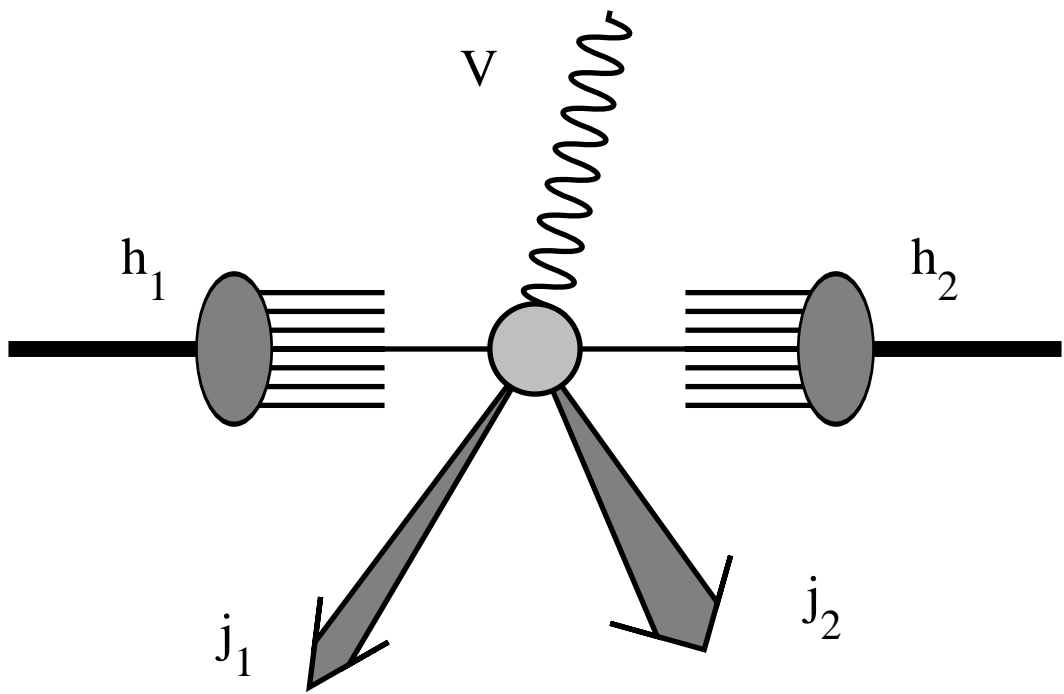
\begin{figure}
\caption{
Hadrons $h_1$ and $h_2$ collide to produce an on-shell electroweak
boson $V$ and two hadron jets $j_1$ and $j_2$ at large transverse
momentum.
}
\label{hadronfig}
\end{figure}

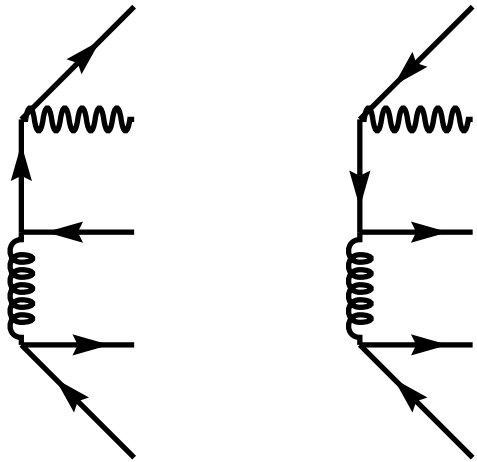
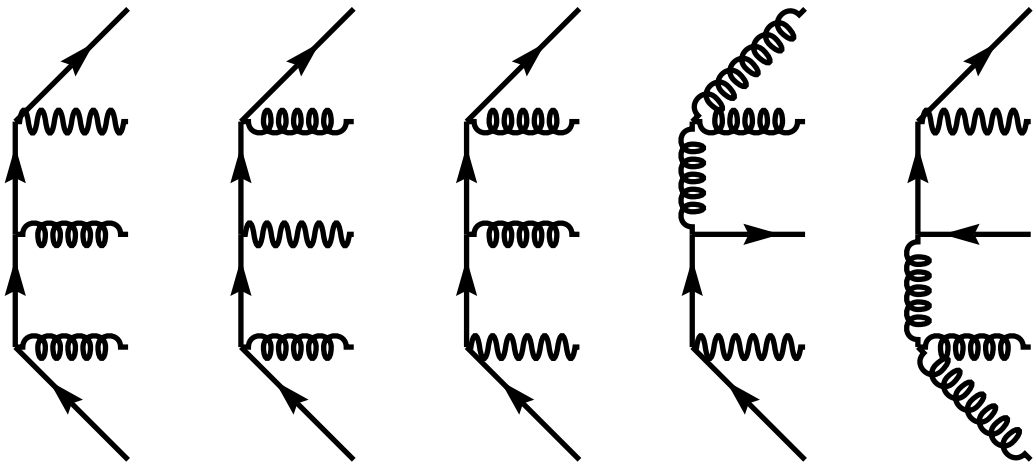
\begin{figure}
\caption{
Generic diagrams with one electroweak boson (wavy line) and four
partons in the final state.  Choosing two partons in all possible
ways, and crossing them to the initial state generates all tree-level
contributions to the processes in Eq.\ (\protect\ref{processeq}).
}
\label{Vjjdiags}
\end{figure}

Consider a collision (Fig.\ \ref{hadronfig}) of two hadrons $h_1$ and
$h_2$ which produces an electroweak boson $V$ and two well-defined
hadron jets $j_1$ and $j_2$.  $V$ stands for one of $W^+,W^-,Z^0$ or
$\gamma$.  $V$, $j_1$ and $j_2$ have transverse momenta, with respect
to the beam axis, of several GeV.  Since $M_V/\Gamma_V \simeq 30$ for
$W^\pm,Z^0$, the cross section is enhanced by a factor of $10^3$ for
on-shell $W$'s or $Z$'s, relative to off-shell production: we will
restrict our attention to this experimentally interesting case.  We
will also consider production of energetic real photons.  The dominant
parton hard-scattering processes which lead to this final state are of
the form
\begin{equation}
a_1+a_2\rightarrow a_3 + a_4 + V\;,
\label{processeq}
\end{equation}
where the $a$'s stand for various allowed combinations of quarks,
antiquarks and gluons.  The ${\cal O}(eg^2)$ contributions to the
hard scattering can be obtained from the generic diagrams of 
Fig.\ \ref{Vjjdiags}.  Here, $e^2=4\pi\alpha$ and $g^2=4\pi\alpha_s$.
Following Refs.\ \cite{helamps}, we use the convention that all
external particles in a generic diagram are taken to be outgoing.
Physical amplitudes are obtained from generic amplitudes by choosing
two of the outgoing partons and crossing them to the initial state.
These amplitudes yield a cross section of ${\cal
O}(\alpha\alpha_s^2)$.

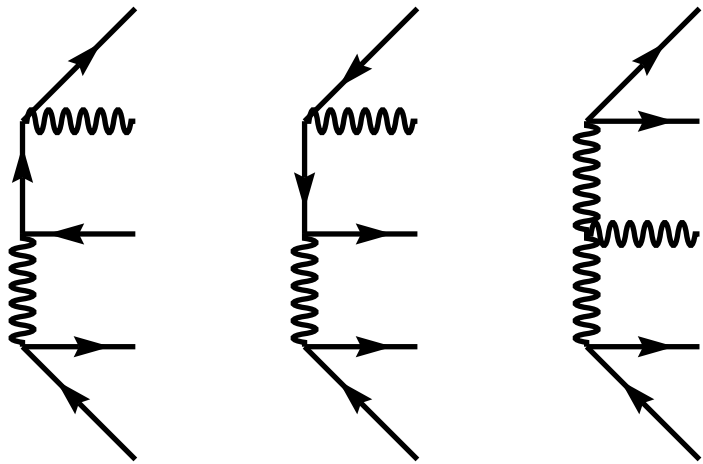
\begin{figure}
\caption{
These generic ``electroweak'' diagrams can interfere with the ``QCD''
diagrams of Fig.\ \protect\ref{Vjjdiags}a.
}
\label{weakdiags}
\end{figure}

In addition to these ``chromodynamic'' or ``QCD'' amplitudes, there
are ``electroweak'' contributions to the process (\ref{processeq}).
For example, replacing the virtual gluon in Fig.\ \ref{Vjjdiags}a 
by an electroweak
boson $V_q$ yields an amplitude of ${\cal O}(e^3)$.  All such
amplitudes can be obtained from the generic diagrams of 
Fig.\ \ref{weakdiags}.  The electroweak cross section is suppressed by a
factor $(\alpha/\alpha_s)^2 \simeq 200$ (for $\alpha_s\simeq 0.1$).
However if $V_q(=W^\pm, Z^0)$ is on shell as can happen in
\begin{eqnarray}
q+\bar{q}\rightarrow & V_q & +V\nonumber \\
&\downarrow&\\
&q+\bar{q}&
\; , \nonumber
\end{eqnarray}
the cross section is enhanced by a factor $(\Gamma_{V_q}/M_{V_q})^2
\simeq 10^3$.
Thus, one might expect the pair-production of on-shell electroweak
bosons to produce an easily observable resonance peak above the QCD
background in the process (\ref{processeq}).  Unfortunately, this
expectation is not borne out by detailed analysis \cite{background}.
There are many more QCD diagrams in Fig.\ \ref{Vjjdiags} than
electroweak diagrams in Fig.\ \ref{weakdiags}.  In particular, gluons
can contribute to the QCD cross section, as can various combinations
of quarks and antiquarks, while on-shell electroweak-boson pair
production can only occur via $q\bar q$ annihilation.  In addition,
the ``abelian'' and ``non-abelian'' amplitudes in 
Fig.\ \ref{weakdiags} tend to contribute with opposite signs since the
renormalizability of the theory requires such cancellations in the
tree amplitudes at high energies.

In Ref.\ \cite{rjg86} it was pointed out that the amplitudes in 
Fig.\ \ref{weakdiags} can interfere with those in Fig.\ \ref{Vjjdiags}a to
give contributions to the cross section of ${\cal
O}(\alpha^2\alpha_s)$.  Once again, the electroweak contributions of
Fig.\ \ref{weakdiags} will be very small unless $V_q$ is almost on
shell when its propagator will behave like
\begin{equation}
{1\over k^2-M^2_{V_q}+i\Gamma_{V_q}M_{V_q}}\simeq{-i\over
\Gamma_{V_q}M_{V_q}}\;.
\label{weakpropeq}
\end{equation}
Comparing this with the propagator of the gluon in 
Fig.\ \ref{Vjjdiags}a which we
can assume has roughly the same momentum
\begin{equation}
{1\over k^2+i\epsilon}\simeq -{1\over M_{V_q}^2}\;,
\label{gluonpropeq}
\end{equation}
we see that the QCD background, the interference contribution, and the
electroweak pair-production cross sections are nominally of relative
magnitudes
\begin{equation}
1\quad:\quad\left({\alpha\over\alpha_s}\right)
\left({M_{V_q}\over\Gamma_{V_q}}\right)
\quad:\quad\left({\alpha\over\alpha_s}\right)^2
\left({M_{V_q}\over\Gamma_{V_q}}\right)^2 \;,
\end{equation}
with $(\alpha/\alpha_s)(M_{V_q}/\Gamma_{V_q})\simeq 2.5$.
The interference contributions might be expected to be suppressed on
account of the following factors: (i) Like the $V-V_q$ pair production
cross section, there are many fewer initial states than contribute to
the QCD background.  (ii) Correlations between initial and final
parton colors, flavors and helicities further restrict the
subprocesses that can receive interference contributions. (iii) Gauge
cancellations between the abelian and non-abelian diagrams in 
Fig.\ \ref{weakdiags}
will also tend to suppress the interference contributions.  There is
one more factor that might, at first sight, actually seem to make the
interference terms vanish, and that is the relative factor of $i$
between Eqs.\ (\ref{weakpropeq}) and (\ref{gluonpropeq}) 
when $V_q$ is exactly on shell: since the
diagrams in Figs.\ \ref{Vjjdiags}a and \ref{weakdiags} 
have the same number of vertices and
propagators, the amplitudes would indeed be $90^\circ$ out of phase
were it not for spinorial factors in the amplitudes that supply a
compensating factor of $i$ that comes from spinor traces such as
\begin{equation}
{\rm Tr}[\not\!k_1\not\!k_2\not\!k_3\not\!k_4\gamma_5]
= 4i\epsilon_{\mu\nu\lambda\sigma}k^\mu_1k^\nu_2
k^\lambda_3k^\sigma_4\;,
\label{traceeq}
\end{equation}
in the spin-averaged cross section.  In fact, the interference cross
sections presented in Sec.\ \ref{subsec:msqds} 
are all proportional to
parity-violating invariants such as that in Eq.\ (\ref{traceeq}).  We
will show that, in contrast to the electroweak pair production cross
section which is parity conserving, the interference terms give rise
to parity-violating asymmetries which might make them observable in
spite of the fact that the magnitudes of the interference cross
sections (like the electroweak pair production cross sections) are
much smaller than the QCD background.

Figure\ \ref{lljjdiags}
shows the three electroweak amplitudes $E_{1-3}$ which produce
an on-shell boson $V_l\equiv V$ shown decaying to a lepton-antilepton
pair with invariant mass $(k_5+k_6)^2=M_{V_l}^2$, and an on-shell
boson $V_q$ which decays to a quark-antiquark pair with invariant mass
$(k_3+k_4)^2 = M_{V_q}^2$.  The quark with momentum $k_1$ and the
antiquark with momentum $k_2$ are to be crossed to the initial state
where they represent the annihilating antiquark and quark
respectively.  The figure also shows the four QCD amplitudes $G_{1-4}$
which are capable of interfering with $E_{1-3}$.  Note that, when the
particles 1 and 2 are crossed to the initial state, the gluon in
amplitudes $G_{1-4}$ has space-like momentum: as in 
Fig.\ \ref{twototwofig}, this is
necessary in order that flavor and color quantum numbers of the
initial and final quarks in $E_{1-3}$ be the same as in $G_{1-4}$.  We
will assume that all quarks are massless.  Helicity, then, is
conserved at each interaction vertex since the electroweak bosons and
the gluons are vector bosons.  Thus the helicities of the initial and
final quarks are the same in the interference, as are the helicities
of the antiquarks.\footnote{
Because initial- and final-state helicities, flavors and colors are
correlated, we expect the interference cross section to be suppressed
by roughly a factor of $2n_c(n_f/2)$ relative to the electroweak
pair-production cross section.}

It is important to remember that the Feynman rules require a relative
minus sign due to Fermi statistics between the amplitudes $E_{1-3}$
and $G_{1-4}$ since the quark lines differ by the interchange of a
pair of labels. This can also be seen by noting that the electroweak
pair-production cross section $\vert E_{1-3}\vert^2$ is got from a
cut-diagram with two quark loops, while the interference 
${\rm Re}(E_{1-3}G_{1-4}^*)$
is got from a cut-diagram which has a single
quark loop.


\subsection{Helicity Amplitudes for $q + \bar q \rightarrow
q + \bar q + V$}

Helicity decomposition techniques\cite{helamps} have proven to be
extremely useful in evaluating tree diagram amplitudes and cross
sections.  If all external particles in a diagram are massless
fermions and all interactions are mediated by vector bosons, the
allowed combinations of helicities of the external particles are
restricted since helicity is conserved at each interaction vertex
along any fermion line.  By judiciously choosing the phases of the
wavefunctions of the external particles, one can obtain very compact
expressions for the independent helicity amplitudes.  A spin-averaged
cross section can then be obtained by squaring these compact
expressions and summing over the allowed helicity combinations.  In
practice, this procedure yields, with considerably less calculational
effort, expressions for cross sections that are more compact than
those obtained by traditional trace-algebra techniques in which the
spin-averaged cross sections are represented by traces of products of
Dirac-matrices.  Helicity amplitude techniques are also extremely
useful in calculating amplitudes involving external massless vector
bosons such as photons and gluons.  They can also be employed if the
external particles are massive, but the expressions obtained are
generally more complex than in the case of massless particles.

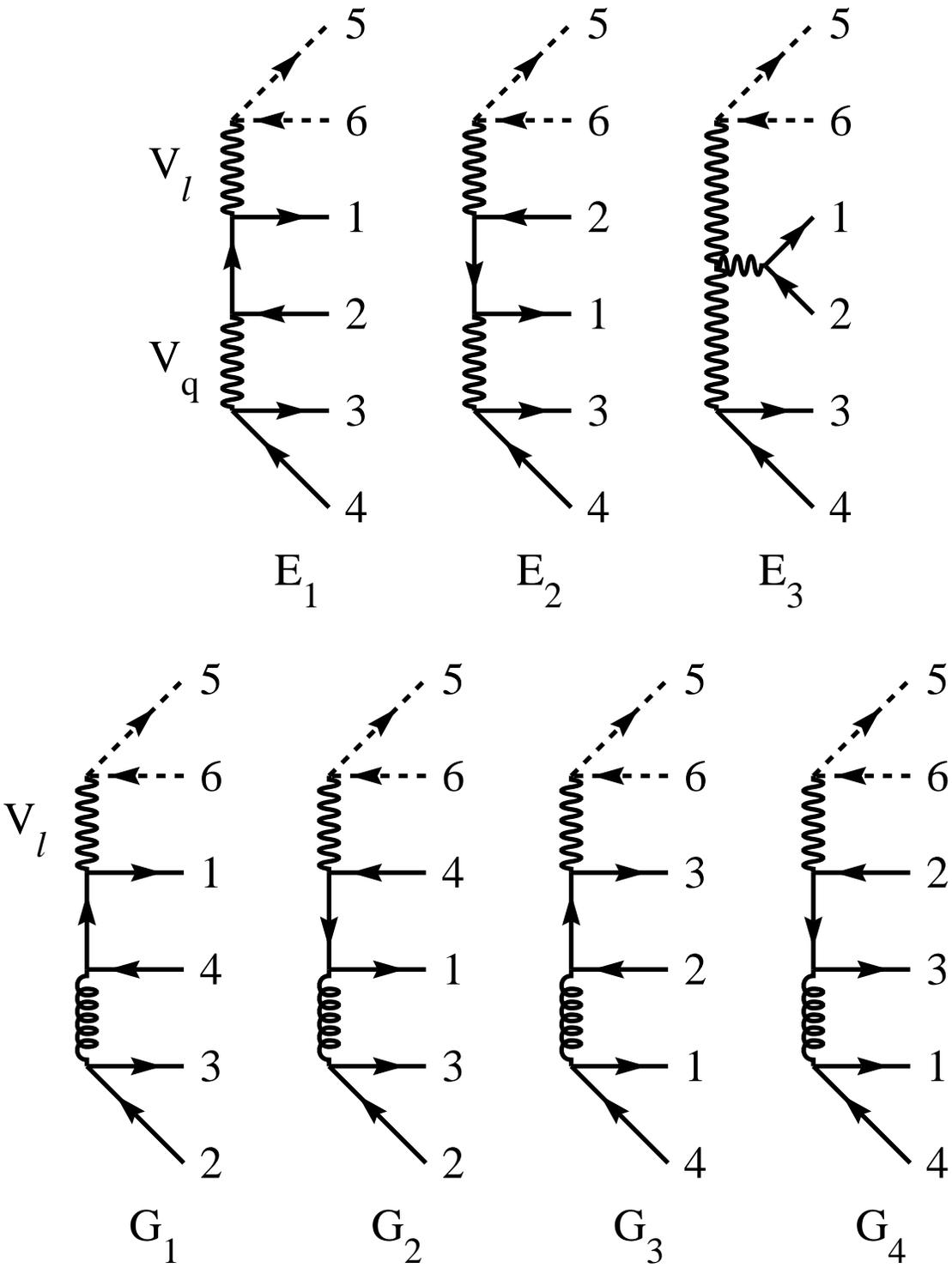
\begin{figure}
\caption{
The dominant diagrams which contribute to chromo-electroweak interference.
$V_l\rightarrow l(k_5)+\bar{l}(k_6)$ with $(k_5+k_6)^2=M_{V_l}^2$, and
$V_q\rightarrow q(k_3)+\bar{q}(k_4)$ with $(k_3+k_4)^2=M_{V_q}^2$.
When crossed to the initial state, $q(k_1)$ and $\bar q(k_2)$
represent an incoming antiquark with momentum $p_1=-k_2$ and an
incoming antiquark with momentum $p_2=-k_1$ respectively.  Other
assignments of particles to the initial state yield amplitudes in
which $V_q$ is far off-shell.
}
\label{lljjdiags}
\end{figure}

In this section, we will present helicity amplitudes for the diagrams
in Fig.\ \ref{lljjdiags}.  
These amplitudes will be used to evaluate interference
contributions to the spin-averaged cross sections in the next section.
Diagrams $E_{1-3}$ were considered in \cite{gunion86}, where
amplitudes and cross sections are presented
for electroweak boson pair production
with both bosons on shell.  To establish notation and conventions, we
list the Feynman rules for vertices at which a quark-antiquark pair
and a gauge boson are produced:
\begin{eqnarray}
W^- &:&\; -ie Q_W U_{u\bar d}\delta_{c\bar c}\gamma^\mu\gamma_L 
    \nonumber \\
W^+ &:&\; -ie Q_W U^*_{d\bar u}\delta_{c\bar c}\gamma^\mu\gamma_L 
    \nonumber \\
Z^0 &:&\; -ie \delta_{f\bar f}\delta_{c\bar c}\gamma^\mu(L_f\gamma_L
                         +R_f\gamma_R)  \\
\gamma &:&\; -ie Q_f\delta_{f\bar f}\delta_{c\bar c} \gamma^\mu 
    \nonumber \\
{\rm Gluon} &:&\; -ig \delta_{f\bar f} 
  \bigl({\lambda_a\over2}\bigr)_{c\bar c} \gamma^\mu\;. \nonumber
\end{eqnarray}
Here, $f=u$ or $d$ is a flavor index, and $u(d)$ are generic labels
for weak $I_z={1\over2}(-{1\over2})$ quark flavors.  $U$ is the
Kobayashi-Maskawa matrix, and with $\theta_W$ being the weak mixing
angle we define
\begin{eqnarray}
&Q_W={1\over\sqrt{2}\sin\theta_W}\;,\; Q_u={2\over3}\;,\;
  Q_d=-{1\over3}\;,\; \tau_u=1\;,\; \tau_d=-1\;,\nonumber \\
&L_f={\tau_f\over\sin(2\theta_W)}-Q_f\tan\theta_W\;,\;
  R_f=-Q_f\tan\theta_W\;.\\ \nonumber
\end{eqnarray}
These definitions can be extended to include couplings to leptons in
an obvious way, i.e., $Q_{e^-}=-1$, $Q_\nu=0$, $\tau_{e^-}=-1$,
$\tau_\nu =+1$, $U_{\nu e^+}=U_{e^-\bar\nu}=1$, etc.  We will often
abbreviate $L_{f_1}$ by $L_1$, $U_{u_3d_{\bar4}}$ by $U_{34}$, etc.,
when the meaning of the abbreviated indices is clear from the context.
The index $c$ is a quark color label, and ${\rm
Tr}(\lambda_a\lambda_a)=4n_cC_F$, with $n_c=3$ and $C_F={4\over3}$.
Finally, $\gamma_L=(1-\gamma_5)/2$ and $\gamma_R=(1+\gamma_5)/2$.  The
Feynman rule for the triple boson vertex is determined by $SU(2)_L$
symmetry.

We next introduce a convenient notation for the wave functions of
massless fermions and spinor products that can be constructed from
these wavefunctions.  An outgoing massless fermion with momentum $k_i$
and helicity $\lambda_i=\pm{1\over2}$, where $i=1,2,3,\dots$, will be
represented by the symbol $\langle {i\pm}\vert$, and an outgoing
antifermion with momentum $k_j$ and helicity $\lambda_j=\mp{1\over2}$
by $\vert {j\pm}\rangle$.  Conventions for crossing and for evaluating
spinor products such as $\langle {i+}|{j-}\rangle$ are explained in
Appendix A.  It is rather remarkable that all of the spinor factors
in the diagrams of Fig.\ \ref{lljjdiags} 
can be expressed in terms of a single
function
\begin{equation}
F_{123456}\equiv 4\,\langle{1-}|3+\rangle\langle{6+}|2-\rangle
\biggl[\langle{4+}|1-\rangle\langle{1-}|5+\rangle +
\langle{4+}|3-\rangle\langle{3-}|5+\rangle \biggr]
\label{Feq}
\end{equation}
of the outgoing momenta $k_i, i=1,\dots,6$, which satisfy momentum
conservation $\sum_{i=1}^6k_i = 0$.  This function was introduced in
\cite{gunion86}: some useful properties connected with it are discussed in
Appendix B.  In the following subsections we present amplitudes for
the sums $E\equiv \sum_{i=1}^3E_i$ and $G\equiv \sum_{i=1}^4G_i$ of
the electroweak and QCD diagrams. These sums are gauge invariant if
the bosons $V_l$ and $V_q$ are on shell.  By conservation of helicity,
$\lambda_2=-\lambda_1$, $\lambda_4= -\lambda_3$,
$\lambda_6=-\lambda_5$, in $E$, and $\lambda_2=-\lambda_3$,
$\lambda_4=-\lambda_1$, $\lambda_6= -\lambda_5$, in $G$: thus, an
amplitude is completely specified by $k_i$ and $\lambda_1$,
$\lambda_3$, and $\lambda_5$, and we will therefore label the sums $E$
and $G$ with a triplet of signs (sign($\lambda_1$), sign($\lambda_3$),
sign($\lambda_5$)).  The on-shell vector bosons involved will be
specified by superscripts as follows: $E^{V_lV_q}$ and $G^{V_l}$.

Finally, we define invariants
\begin{eqnarray}
s_{ij}\equiv(k_i+k_j)^2\quad,&&\quad s_{ijl}\equiv(k_i+k_j+k_l)^2
\;,\nonumber \\
x_{ijlm}&\equiv&\epsilon_{\mu\nu\lambda\sigma}k_i^\mu k_j^\nu
k_l^\lambda k_m^\sigma\;,
\end{eqnarray}
where the metric has signature $(+,-,-,-)$ and $\epsilon_{0123}=+1$.
We also define a function
\begin{equation}
D_{ij}^V\equiv s_{ij}-M_V^2+iM_V\Gamma_V\;,
\end{equation}
which occurs in boson propagator denominators.

\subsubsection{Amplitudes for $V_l=W^\pm$, $V_q=W^\mp$}

Since $W^\pm$ only couple to left handed fermions, the non-zero
helicity amplitudes are $E_{1,2}(-,-,-), E_3(\mp,-,-),
G_{1,2}(-,\mp,-)$, and $G_{3,4}(\mp,-,-)$. However, since
$\lambda_2=-\lambda_1$ and $\lambda_4=-\lambda_3$ in $E_i$ while
$\lambda_2=-\lambda_3$ and $\lambda_4=-\lambda_1$ in $G_i$, only the
amplitudes $E(-,-,-)$ and $G(-,-,-)$ will contribute to
chromo-electroweak interference.  Following \cite{gunion86} we find that the
electroweak amplitude for $V_l=W^-$, $V_q=W^+$ is
\begin{eqnarray}
E^{W^-W^+}&&(-,-,-)
  = {K_eQ_W^2U_{34}U_{65}^*\delta_{12}\over  D^W_{34}D^W_{56}}
  \nonumber\\
&&\times\left[-{(1+\tau_1)Q_W^2\over2s_{234}}F_{125634}
  -{(1-\tau_1)Q_W^2\over2s_{134}}F_{123456}\right.
  \nonumber\\
&&\left.\mbox{}+\left({L_1\cot\theta_W\over D_{12}^Z}
  +{Q_1\over D^\gamma_{12}}\right) (F_{123456}-F_{125634})\right]\;,
\label{EWWeq}
\end{eqnarray}
where
\begin{equation}
K_e=ie^4\delta_{c_1c_2}\delta_{c_3c_4}\;.
\end{equation}
We note that $\delta_{12}\equiv\delta_{f_1f_2}$, i.e., flavors
$f_1=u,d$ are both allowed.  The factor $U_{34}$ implies that $f_3=u$
and $f_4=\bar{d}$.  Further, for the kinematic configurations of
interest, $D^W_{34}=D^W_{56}=i\Gamma_WM_W$, $D_{12}^\gamma=s_{12}$,
and we can replace $D^Z_{12}$ by $s_{12}-M_Z^2$ since
$s_{12}\geq4M_W^2$.  The QCD amplitude is
\begin{eqnarray}
G^{W^-}(-,-,-)&&
  =(-1){K_gQ_W^2U_{65}^*\over D^W_{56}}
   \nonumber\\
&&\times\left[
  {U_{14}\delta_{23}\over s_{23}}\left({F_{125634}\over s_{234}}
  +{F_{341256}\over s_{123}}\right) \right.
  \nonumber\\
&&\left.\mbox{}+{U_{32}\delta_{14}\over s_{14}}
  \left({F_{345612}\over s_{124}}
  +{F_{123456}\over s_{134}}\right)\right]\;,
\label{GWeq}
\end{eqnarray}
where
\begin{equation}
K_g=ie^2g^2\left({\lambda_a\over2}\right)_{c_3c_2}
  \left({\lambda_a\over2}\right)_{c_1c_4}\;.
\end{equation}
The explicit $(-1)$ in this formula represents the relative minus sign
between the electroweak (`annihilation') and QCD (`scattering')
diagrams due to Fermi-Dirac statistics.

The amplitudes for the charge conjugate process $V_l=W^+$, $V_q=W^-$
can be obtained directly from the Feynman rules, or by CP-conjugation
from Eqs.\ (\ref{EWWeq}) and (\ref{GWeq}) using the conventions described in
Appendix A:
\begin{eqnarray}
E^{W^+W^-}(-,-,-)= && {K_eQ_W^2U_{43}^*U_{56}\delta_{12}\over
  D^W_{34}D^W_{56}}
   \nonumber\\
&&\times\left[-{(1+\tau_1)Q_W^2\over2s_{134}}F_{123456}
  -{(1-\tau_1)Q_W^2\over2s_{234}}F_{125634} \right.
  \nonumber \\
&&\left.\mbox{}+\left({L_1\cot\theta_W\over D_{12}^Z}
  +{Q_1\over D^\gamma_{12}}\right)
   (F_{125634}-F_{123456})\right]\;.
\label{EWW-eq}
\end{eqnarray}
We note that Eq.\ (\ref{EWW-eq}) can be obtained from 
Eq.\ (\ref{EWWeq}) either by the formal interchange of labels
$(3\leftrightarrow5,\;4\leftrightarrow6)$, or by the replacements
$\tau_1\rightarrow-\tau_1$, $U\rightarrow U^\dagger$ which interchange
the couplings of $W^+$ and $W^-$ to fermions, and by changing the sign
of the $\bigl({L_1\cot\theta_W/D_{12}^Z} + {Q_1/D^\gamma_{12}}\bigr)$
term, which comes from the three-boson vertex.  Similarly,
\begin{eqnarray}
G^{W^+}(-,-,-)=&&(-1){K_gQ_W^2U_{56}\over D^W_{56}}
  \nonumber\\
&&\times \left[{U_{41}^*\delta_{23}\over s_{23}}
  \left({F_{125634}\over s_{234}}
  +{F_{341256}\over s_{123}}\right) \right.
  \nonumber\\
&&\left.\mbox{}+{U_{23}^*\delta_{14}\over s_{14}}
  \left({F_{345612}\over s_{124}}
  +{F_{123456}\over s_{134}}\right)\right]\;,
\label{GW2eq}
\end{eqnarray}
which can be obtained from Eq.\ (\ref{GWeq}) simply by the replacement
$U\rightarrow U^\dagger$.

\subsubsection{Amplitudes for $V_l=W^\mp$, and $V_q=Z^0$}

In this case, the non-zero electroweak helicity amplitudes are
$E(-,\mp,-)$.  The QCD amplitudes are the same as in the preceding
subsection, and once again, only $E(-,-,-)$ and $G(-,-,-)$ will
contribute to chromo-electroweak interference.
\begin{eqnarray}
E^{W^-Z}(-,-,-)=&&{K_eQ_W^2U_{12}U_{65}^*\delta_{34}L_3\over
  D^Z_{34}D^W_{56}}
  \nonumber\\
&&\times\left[ \left({\cot\theta_W\over D^W_{12}}
    -{L_2\over s_{234}}\right)F_{125634} \right.
  \nonumber\\
&&\left.\mbox{} -\left({\cot\theta_W\over D^W_{12}}
    +{L_1\over s_{134}}\right)F_{123456}\right]\;.
\label{EWZeq}
\end{eqnarray}
In this case, $f_3=u,d$ are both allowed while $f_1=d$.  For the
kinematic configurations of interest, $D_{34}^Z=i\Gamma_ZM_Z$,
$D^W_{56}=i\Gamma_WM_W$, and $D^W_{12}=s_{12}-M_W^2$.  The QCD
amplitude $G^{W^-}$ is given by Eq.\ (\ref{GWeq}) with the flavor indices
taking values appropriate to this process.

The electroweak amplitude for the charge conjugate process is
\begin{eqnarray}
E^{W^+Z}(-,-,-)=&&{K_eQ_W^2U_{21}^*U_{56}\delta_{34}L_3\over
  D^Z_{34}D^W_{56}}
  \nonumber\\
&&\times\left[ \left({\cot\theta_W\over D^W_{12}}
    -{L_1\over s_{134}}\right)F_{123456}\right.
  \nonumber\\
&&\left.\mbox{}-\left({\cot\theta_W\over D^W_{12}}
    +{L_2\over s_{234}}\right)F_{125634}\right]\;,
\label{EWZ2eq}
\end{eqnarray}
which can be obtained from Eq.\ (\ref{EWZeq}) by the replacements
$\cot\theta_W
\rightarrow -\cot\theta_W$ and $U\rightarrow
U^\dagger$. We note that the expressions in square brackets in
Eqs.\ (\ref{EWZeq}) and (\ref{EWZ2eq}) can be expressed by the single formula
\begin{equation}
\left[-\left({\tau_1\cot\theta_W\over D^W_{12}}
    +{L_1\over s_{134}}\right)F_{123456}
  -\left({\tau_2\cot\theta_W\over D^W_{12}}
    +{L_2\over s_{234}}\right)F_{125634}\right]\;.
\end{equation}
$G^{W^+}$ is given by Eq.\ (\ref{GW2eq}).

\subsubsection{Amplitudes for $V_l=Z^0$ and $V_q=W^\pm$}

The non-zero electroweak helicity amplitudes are $E_{1-3}(-,-,\mp)$.
Thus while all helicity combinations are allowed in the QCD
amplitudes, only $G_{1-4}(-,-,\mp)$ will contribute to chromo-electroweak
interference.
\begin{eqnarray}
E^{ZW^+}(-,-,-)=&&{K_eQ_W^2U_{21}^*U_{34}\delta_{56}L_5\over
  D^W_{34}D^Z_{56}}
  \nonumber\\
&&\times\left[ \left({\cot\theta_W\over D^W_{12}}
    -{L_1\over s_{234}}\right)F_{125634}\right.
  \nonumber \\
&&\left.\mbox{} -\left({\cot\theta_W\over D^W_{12}}
    +{L_2\over s_{134}}\right)F_{123456}\right]\;.
\label{EZWeq}
\end{eqnarray}
Note that this amplitude can be obtained from Eq.\ (\ref{EWZ2eq}) by formal
interchange of labels $(3\leftrightarrow5,4\leftrightarrow6)$.  The
amplitude for production of a right-handed lepton pair is
\begin{eqnarray}
E^{ZW^+}(-,-,+)=&&{K_eQ_W^2U_{21}^*U_{34}\delta_{56}R_5\over
  D^W_{34}D^Z_{56}}
  \nonumber \\
&&\times\left[ \left({\cot\theta_W\over D^W_{12}}
    -{L_1\over s_{234}}\right)F_{126534} \right.
 \nonumber\\
&&\left.\mbox{} -\left({\cot\theta_W\over D^W_{12}}
    +{L_2\over s_{134}}\right)F_{123465}\right]\;,
\label{EZW2eq}
\end{eqnarray}
which may be obtained from Eq.\ (\ref{EZWeq}) 
by replacing $L_5$ by $R_5$ and
interchanging labels $(5\leftrightarrow6)$.  The required QCD
amplitudes are
\begin{eqnarray}
G^{Z}(-,-,-)=&&(-1){K_g\delta_{14}\delta_{23}\delta_{56}L_5\over
    D^W_{56}}  
  \nonumber\\
&&\times\left[{L_1\over s_{23}}\left({F_{125634}\over s_{234}}
  +{F_{341256}\over s_{123}}\right)\right.
  \nonumber\\
&&\left.\mbox{}+{L_3\over s_{14}}\left({F_{345612}\over s_{124}}
  +{F_{123456}\over s_{134}}\right)\right]\;,
\label{GZeq}
\end{eqnarray}
\begin{eqnarray}
G^{Z}(-,-,+)=&&(-1){K_g\delta_{14}\delta_{23}\delta_{56}R_5\over
    D^W_{56}}  
  \nonumber\\
&&\times\left[{L_1\over s_{23}}\left({F_{126534}\over s_{234}}
  +{F_{341265}\over s_{123}}\right) \right.
  \nonumber\\
&&\left.\mbox{}+{L_3\over s_{14}}\left({F_{346512}\over s_{124}}
  +{F_{123465}\over s_{134}}\right)\right]\;.
\label{GZ2eq}
\end{eqnarray}
The electroweak amplitudes for the charge conjugate processes are
\begin{eqnarray}
E^{ZW^-}(-,-,-)=&&{K_eQ_W^2U_{12}U_{43}^*\delta_{56}L_5\over
  D^W_{34}D^Z_{56}}
  \nonumber]\\
&&\times\left[- \left({\cot\theta_W\over D^W_{12}}
    +{L_1\over s_{234}}\right)F_{125634} \right.
  \nonumber \\
&&\left.\mbox{} +\left({\cot\theta_W\over D^W_{12}}
    -{L_2\over s_{134}}\right)F_{123456}\right]\;,
\label{EZW3eq}
\end{eqnarray}
which can be got either from Eq.\ (\ref{EWZeq}) by formally interchanging
labels $(3\leftrightarrow5,4\leftrightarrow6)$, or from 
Eq.\ (\ref{EZWeq}) by
letting $\cot\theta_W\rightarrow-\cot\theta_W$ and $U\rightarrow
U^\dagger$; and
\begin{eqnarray}
E^{ZW^-}(-,-,+)=&&{K_eQ_W^2U_{12}U_{43}^*\delta_{56}R_5\over
  D^W_{34}D^Z_{56}}
  \nonumber \\
&&\times\left[- \left({\cot\theta_W\over D^W_{12}}
    +{L_1\over s_{234}}\right)F_{126534} \right.
  \nonumber \\
&&\left.\mbox{} +\left({\cot\theta_W\over D^W_{12}}
    -{L_2\over s_{134}}\right)F_{123465}\right]\;.
\label{EZW4eq}
\end{eqnarray}
We note that the expressions in square brackets in 
Eqs.\ (\ref{EZWeq}) and
(\ref{EZW4eq}) can be expressed by the single formula
\begin{equation}
\left[-\left({\tau_1\cot\theta_W\over D^W_{12}}
    +{L_1\over s_{234}}\right)F_{125634}
  -\left({\tau_2\cot\theta_W\over D^W_{12}}
    +{L_2\over s_{134}}\right)F_{123456}\right]\;. \nonumber
\end{equation}
The QCD amplitudes for the charge conjugate processes are given by
Eqs.\ (\ref{GZeq}) and (\ref{GZ2eq}) with the flavor labels 
appropriately interpreted.

\subsubsection{Amplitudes for $V_l=Z^0$ and $V_q=Z^0$}

In $Z^0$ pair production, all helicity combinations will in general
yield non-zero amplitudes.  However, in the chromo-electroweak
interference terms, the helicities of the initial- and final-state
quarks are constrained to be the same, and we will only present
results for these amplitudes.
\begin{equation}
E^{ZZ}(-,-,-)={K_e\delta_{12}\delta_{34}\delta_{56}L_1^2L_3L_5\over
  D^Z_{34}D^Z_{56}} \left(-{F_{125634}\over s_{234}}
  -{F_{123456}\over s_{134}} \right)\;,
\label{EZZeq}
\end{equation}
\begin{equation}
E^{ZZ}(-,-,+)={K_e\delta_{12}\delta_{34}\delta_{56}L_1^2L_3R_5\over
  D^Z_{34}D^Z_{56}} \left(-{F_{126534}\over s_{234}}
  -{F_{123465}\over s_{134}} \right)\;,
\label{EZZ2eq}
\end{equation}
\begin{equation}
E^{ZZ}(+,+,+)={K_e\delta_{12}\delta_{34}\delta_{56}R_1^2R_3R_5\over
  D^Z_{34}D^Z_{56}} \left({F_{214365}\over s_{234}}
  +{F_{216543}\over s_{134}} \right)\;,
\label{EZZ3eq}
\end{equation}
\begin{equation}
E^{ZZ}(+,+,-)={K_e\delta_{12}\delta_{34}\delta_{56}R_1^2R_3L_5\over
  D^Z_{34}D^Z_{56}} \left({F_{214356}\over s_{234}}
  +{F_{215643}\over s_{134}} \right)\;.
\label{EZZ4eq}
\end{equation}
Note that Eq.\ (\ref{EZZ3eq}) can be obtained from 
Eq.\ (\ref{EZZeq}) by changing
left-handed $Z^0$ couplings to right, interchanging labels
$(1\leftrightarrow2, 3\leftrightarrow4, 5\leftrightarrow6)$, and
appending a minus sign.\footnote{
This minus sign is omitted in Eq.~(16) of \cite{gunion86}.  
The relative phase
of amplitudes with different helicities is of course arbitrary: the
negative sign in this case is consistent with a naive application of
the Feynman rules, and with the conventions for charge conjugation
discussed in Appendix A; it arises from charge conjugating the virtual
quark in Fig.\ \ref{lljjdiags}.}

The QCD amplitudes for $Z^0$ production with negative quark helicities
were given in Eqs.\ (\ref{GZeq}) and (\ref{GZ2eq}).  In addition, we need
\begin{eqnarray}
G^{Z}(+,+,+)=&&(-1){K_g\delta_{14}\delta_{23}\delta_{56}R_5\over
    D^W_{56}}  
  \nonumber \\
&&\times\left[-{R_1\over s_{23}}\left({F_{214365}\over s_{234}}
  +{F_{436521}\over s_{123}}\right) \right.
  \nonumber \\
&&\left.\mbox{}-{R_3\over s_{14}}\left({F_{432165}\over s_{124}}
  +{F_{216543}\over s_{134}}\right)\right]\;,
\end{eqnarray}
\begin{eqnarray}
G^{Z}(+,+,-)=&&(-1){K_g\delta_{14}\delta_{23}\delta_{56}L_5\over
    D^W_{56}}  
  \nonumber \\
&&\times\left[-{R_1\over s_{23}}\left({F_{214356}\over s_{234}}
  +{F_{435621}\over s_{123}}\right) \right.
  \nonumber \\
&&\left.\mbox{}-{R_3\over s_{14}}\left({F_{432156}\over s_{124}}
  +{F_{215643}\over s_{134}}\right)\right]\;.
\end{eqnarray}
These two amplitudes may be obtained from Eqs.\ (\ref{GZeq}) 
and (\ref{GZ2eq}) by
changing left- to right-handed $Z$ couplings, interchanging labels
$(1\leftrightarrow2, 3\leftrightarrow4, 5\leftrightarrow6)$, and
appending a minus sign.  The interchange of labels
$(1\leftrightarrow4,3\leftrightarrow2, 5\leftrightarrow6)$ with the
change of couplings and an overall minus sign also produces the
desired result.

\subsubsection{Amplitudes for $V_l=\gamma$ and $V_q=W^\pm,Z^0$}

The amplitudes for production of a photon (either real or virtual) plus
two jets can in principle be obtained from the corresponding amplitudes
for production of a $Z^0$ plus two jets by setting $M_Z=0$ and
$R_5=L_5=1$. In the case of a real photon, the limit $M_Z\rightarrow 0$
is rather subtle since the photon has zero width, and it is simpler to
derive the amplitudes from scratch.  The amplitude for production of a
photon that is off-shell by an amount $Q^2$ is suppressed relative that
for an  on-shell $Z^0$ by a factor of $\Gamma_ZM_Z/Q^2$:  thus, photons
with $Q^2 \geq 10\;\hbox{GeV}$ are not likely to be phenomenologically
very interesting, and we will therefore confine our attention to real
photons.  Following \cite{gunion86} we present 
the amplitudes in terms of the
functions\footnote{
We note that $H_{12345}=-C(1,2,3,4,5)=-(\langle1-|3+\rangle)^2\delta$
where $C$ and $\delta$ are defined in Eqs.~(19,20) of \cite{gunion86}.
$H^+=H^*$ if $k_i^0 > 0,\;i=1,\dots,5$, see Appendix A.}
\begin{eqnarray}
H_{12345}&=& 2\sqrt2{\langle1-|3+\rangle^2\langle3+|4-\rangle
  \over\langle1-|5+\rangle\langle2-|5+\rangle} \;, 
  \nonumber \\
H_{12345}^+&=& 2\sqrt2{\langle3+|1-\rangle^2\langle4-|3+\rangle
  \over\langle5+|1-\rangle\langle5+|2-\rangle} \;,
\label{Hdefeq}
\end{eqnarray}
of the outgoing momenta $k_i,\,i=1,\dots,5$, which satisfy momentum
conservation $\sum_{i=1}^5 k_i = 0$.  Here, $k_5$ is the momentum of
the real photon (compare Eq.\ (\ref{Feq}) {\it et seq}.). Amplitudes will be
labeled by (sign($\lambda_1$),sign($\lambda_3$),sign($\lambda_5$)),
where $\lambda_5$ is now the helicity of the photon
($\lambda_5=\pm1$).

For $V_q=W^+$, the non-zero electroweak helicity amplitudes are
\begin{equation}
E^{\gamma W^+}(-,-,+)={K_e^\gamma Q_W^2U_{21}^*U_{34}\over D^W_{34}}
   \left(Q_2-\tau_1{s_{25}\over D^W_{12}}\right) H_{12345}\;,
\label{EAWeq}
\end{equation}
and
\begin{equation}
E^{\gamma W^+}(-,-,-)={K_e^\gamma Q_W^2U_{21}^*U_{34}\over D^W_{34}}
  \left(Q_1-\tau_2{s_{15}\over D^W_{12}}\right) H_{21435}^+\;,
\label{EAW2eq}
\end{equation}
where
\begin{equation}
K_e^\gamma \equiv ie^3\delta_{c_1c_2}\delta_{c_3c_4}\;. \nonumber
\end{equation}
We note that for a weak-isospin doublet $\Delta Q=\Delta\tau/2$
and $Q_1-Q_2=\tau_1=-\tau_2$, so
that when the $W$ is on shell, i.e., when $s_{34}=M_W^2$,
\begin{equation}
Q_2-\tau_1{s_{25}\over s_{12}-M_W^2}=Q_1-\tau_2{s_{15}\over
  s_{12}-M_W^2}\;.
\end{equation}
The non-zero electroweak amplitudes for the case $V_q=W^-$ are
\begin{equation}
E^{\gamma W^-}(-,-,+)={K_e^\gamma Q_W^2U_{12}U_{43}^*\over D^W_{34}}
   \left(Q_2-\tau_1{s_{25}\over D^W_{12}}\right) H_{12345}\;,
\end{equation}
and
\begin{equation}
E^{\gamma W^-}(-,-,-)={K_e^\gamma Q_W^2U_{12}U_{43}^*\over D^W_{34}}
  \left(Q_1-\tau_2{s_{15}\over D^W_{12}}\right) H_{21435}^+\;,
\end{equation}
which are identical in form to Eqs.\ (\ref{EAWeq}) and 
(\ref{EAW2eq}) except that $U$
is replaced by $U^\dagger$.

For the case that $V_q=Z^0$, all helicity combinations yield
non-vanishing electroweak amplitudes.  However, only amplitudes with
the same helicities for the quarks (i.e., $\lambda_1=\lambda_3$)
contribute to chromo-electroweak interference.  These amplitudes are
\begin{equation}
E^{\gamma Z}(+,+,+)={K_e^\gamma\delta_{12}\delta_{34}Q_1\over
  D^Z_{34}}R_1R_3H_{21435}\;,
\end{equation}
\begin{equation}
E^{\gamma Z}(+,+,-)={K_e^\gamma\delta_{12}\delta_{34}Q_1\over
  D^Z_{34}}R_1R_3H_{12345}^+\;,
\end{equation}
\begin{equation}
E^{\gamma Z}(-,-,+)={K_e^\gamma\delta_{12}\delta_{34}Q_1\over
  D^Z_{34}}L_1L_3H_{12345}\;,
\end{equation}
\begin{equation}
E^{\gamma Z}(-,-,-)={K_e^\gamma\delta_{12}\delta_{34}Q_1\over
  D^Z_{34}}L_1L_3H_{21435}^+\;.
\label{EAZ4eq}
\end{equation}

The QCD amplitudes that can interfere with the electroweak amplitudes
given in Eqs.\ (\ref{EAWeq}) through (\ref{EAZ4eq}) are
\begin{equation}
G^\gamma(+,+,+)=(-1)K^\gamma_g\delta_{14}\delta_{23}\left(-{Q_1\over
   s_{23}}H_{41235}-{Q_3\over s_{14}}H_{23415}\right)\;,
\end{equation}
\begin{equation}
G^\gamma(+,+,-)=(-1)K^\gamma_g\delta_{14}\delta_{23}\left({Q_1\over
   s_{23}}H_{14325}^++{Q_3\over s_{14}}H_{32145}^+\right)\;,
\end{equation}
\begin{equation}
G^\gamma(-,-,+)=(-1)K^\gamma_g\delta_{14}\delta_{23}\left({Q_1\over
   s_{23}}H_{14325}+{Q_3\over s_{14}}H_{32145}\right)\;, 
\end{equation}
\begin{equation}
G^\gamma(-,-,-)=(-1)K^\gamma_g\delta_{14}\delta_{23}\left(-{Q_1\over
   s_{23}}H_{41235}^+-{Q_3\over s_{14}}H_{23415}^+\right)\;,
\label{GA4eq}
\end{equation}
where
\begin{equation}
K^\gamma_g\equiv ieg^2\left({\lambda_a\over2}\right)_{c_3c_2}
   \left({\lambda_a\over2}\right)_{c_1c_4}\;.
\end{equation}


\subsection{Squared Amplitudes:  Chromo-electroweak Interference Terms}
\label{subsec:msqds}

The amplitudes presented in the preceding section can easily be
evaluated numerically and the resulting complex numbers squared to
obtain a cross section.  We will, however, square the helicity
amplitudes analytically and present formulas for the chromo-electroweak
interference contributions to the cross section.  These formulas turn
out to be as simple and compact as the helicity amplitudes themselves,
and their analytic forms explicitly exhibit their parity-violating
nature and are therefore quite instructive.

The interference contributions will all be linear functions of the
invariants 
\begin{equation}
x_{ijlm} \equiv \epsilon_{\mu\nu\lambda\sigma}k_i^\mu k_j^\nu k_l
  ^\lambda k_m^\sigma\; ,
\end{equation}
where $\epsilon_{\mu\nu\lambda\sigma}$ is the totally antisymmetic
Levi-Civita tensor with $\epsilon_{0123}=+1$, and $k_l$ are
four-momenta.  In processes of the type $q_1\bar q_2\rightarrow
(V_q\rightarrow q_3\bar q_4) + V_l$, where $V_l$ is detected as a real
particle, there are only four independent momenta and 
there is therefore
essentially only one parity-odd invariant $x_{1234}$.  In this case,
the use of traditional trace-algebra techniques to compute
spin-averaged squared matrix elements leads to essentially the same
analytic forms as does use of the helicity amplitude methods.
However, if $V_l\rightarrow l_5\bar l_6$, there are five independent
momenta and hence five essentially different parity-odd invariants
(e.g., $x_{1234},\,x_{1235},\,x_{1245},\,x_{1345},\,x_{2345})$.  These
five invariants are not functionally independent however: for a
process involving $n$ particles, the number of functionally
independent invariants is $3n-10$, and this number includes the scalar
products $k_i\cdot k_j$.\footnote{
There is in addition one discrete invariant which can take two 
values and which describes the handedness of the coordinate system used.}
For a $2\rightarrow4$
process, only 8 of the 15 invariants $s_{ij}$ and $x_{ijlm}$ are
functionally independent, and the relations between them are
nonlinear and not trivial.  In this case, there is a dramatic
difference between the use of trace-algebra and helicity-amplitude
techniques: straightforward application of the former leads to
interference contributions that involve in excess of a hundred terms
linear in the invariants $x_{ijlm}$, while squaring the helicity
amplitudes generates on the order of ten such terms!  In addition, it
is very difficult to show analytically that the expressions obtained
using the two methods are in fact identical (though their equality can
be checked numerically).\footnote{
This is because $x_{ijlm} = \pm\left[-G\pmatrix{k_ik_jk_lk_m\cr
k_ik_jk_lk_m\cr}\right]^{1/2}$ where $G$ is a Gram determinant 
(see e.g., \cite{eden66}).
Thus, linear relations between the different $x$'s have
coefficients that are cubic in the invariants $s_{ij}$.}
The calculation of the 4-particle
interference contributions thus provides a rather striking
demonstration of the advantages of using helicity amplitude
techniques.

\subsubsection{Four particle final states: $V_q\rightarrow q_3\bar
q_4$, $V_l\rightarrow l_5\bar l_6$}

In this section, we will present results for the chromo-electroweak
interference contributions ${\cal J}_4$ to the helicity- and
color-averaged squared amplitudes $(E+G)(E^++G^+)$ for fixed flavors,
i.e.,
\begin{equation}
{\cal J}_4^{V_lV_q} \equiv {1\over4n_c^2}\sum_{\rm colors}\sum_{\rm
  helicities}{\rm Re}\left[E^{V_lV_q}\left(G^{V_lV_q}\right)^+
  +\left(E^{V_lV_q}\right)^+G^{V_lV_q}\right]\;.
\label{JVVeq}
\end{equation}
The operation denoted by the superscript $+$ is the crossing invariant
extension of complex conjugation of amplitudes involving particles
with negative as well as physical energies as explained in Appendix A.
These contributions are most compactly expressed in terms of a set of
auxiliary functions of the external momenta which we define as
follows:
\begin{equation}
I_{123456} \equiv F_{123456}F_{123456}^+\;,
\label{Idefeq}
\end{equation}
\begin{equation}
J_{123456} \equiv F_{123456}F_{125634}^+\;,
\label{Jdefeq}
\end{equation}
\begin{equation}
K_{123456} \equiv F_{123456}F_{341256}^+\;,
\label{Kdefeq}
\end{equation}
\begin{equation}
X_{123456} \equiv {1\over s_{14}}\left({I_{123456}\over s_{134}}
   + {J_{321456}\over s_{124}} \right) \;,
\label{Xdefeq}
\end{equation}
\begin{equation}
Y_{123456} \equiv {1\over s_{23}}\left({J_{123456}\over s_{234}}
   + {K_{123456}\over s_{123}} \right) \;.
\label{Ydefeq}
\end{equation}
In fact, only the imaginary parts of the functions $X$ and $Y$ will
contribute to the interference terms.  This can be seen, for example,
by writing out the squared matrix element explicitly in the case of
$W^+W^-$ production using Eqs.\ (\ref{EWWeq}) and (\ref{GWeq}):
\begin{eqnarray}
\lefteqn{E^{W^-W^+}(-,-,-)\left[G^{W^-}(-,-,-)\right]^+= (-1)
  {K_eK_g^*Q_W^4|U_{34}|^2|U_{65}|^2\delta_{12}\over D_{34}^W|D_{56}^W|^2}}
  \quad\nonumber \\
&&\times\left[\left({L_1\cot\theta_W\over s_{12}-M_Z^2}+{Q_1\over s_{12}}
  +{(\tau_1-1)Q_W^2\over 2s_{134}}\right)F_{123456} \right.
  \nonumber \\
&&\left.\mbox{}-\left({L_1\cot\theta_W\over s_{12}-M_Z^2}+{Q_1\over s_{12}}
  +{(\tau_1+1)Q_W^2\over 2s_{234}}\right)F_{125634}\right]
  \nonumber \\
&&\times\left[{\delta_{23}\over s_{23}}\left({F_{125634}^+\over s_{234}}
  +{F_{341256}^+\over s_{123}}\right) + {\delta_{14}\over s_{14}}\left( 
  {F_{345612}^+\over s_{124}} + {F_{123456}^+\over
s_{134}}\right)\right]
  \nonumber \\
= && {K_eK_g^*Q_W^4|U_{34}|^2|U_{65}|^2\delta_{12}\over D_{34}^W|D_{56}^W|^2}
 \nonumber \\
&&\times\left[\left({L_1\cot\theta_W\over s_{12}-M_Z^2}+{Q_1\over s_{12}}
  +{(\tau_1+1)Q_W^2\over 2s_{234}}\right) \left(\delta_{23}X_{214365}^+
  +\delta_{14}Y_{214365}^+\right) \right.
  \nonumber \\
&&\left.\mbox{}-\left({L_1\cot\theta_W\over s_{12}-M_Z^2}+{Q_1\over s_{12}}
  +{(\tau_1-1)Q_W^2\over 2s_{134}}\right) \left(\delta_{23}Y_{123456}
  +\delta_{14}X_{123456} \right) \right]\;.
\end{eqnarray}
This rather compact form was obtained by using the symmetry properties
of the functions $I$, $J$ and $K$ discussed in Appendix B---see
Eqs.\ (\ref{Ipropseq},\ref{Jpropseq},\ref{Kpropseq}).  
We now note that $D^W_{12}=s_{34}-M_W^2 +i\Gamma_WM_W$ is
imaginary on shell, that is, when $s_{34}=M_W^2$.  In \cite{rjg86}, we
integrated the cross section over a small interval $(M_W-\Delta)^2
\leq s_{34} \leq (M_W+\Delta)^2$, with $\Delta \simeq
\Gamma_W$.  However, in this paper, we will use the narrow width
approximation and replace 
\begin{equation}
{1\over D_{34}^W} \longrightarrow
-i\pi\delta(s_{34}-M_W^2)\;. 
\label{narrowWidthEq}
\end{equation}
The two procedures yield roughly the same numerical results.  The
narrow width approximation is, at least theoretically if not also
numerically, the more consistent of the two procedures since the
diagrams we include are gauge invariant only when $V_q$ ($W^+$ in this
case) is exactly on shell.  We also use a narrow width approximation
for $V_l$ ($W^+$ in this case)
\begin{equation}
{1\over |D_{56}^W|^2} \longrightarrow {\pi\over \Gamma_WM_W}
  \delta(s_{56}-M_w^2)\;. 
\end{equation}
Performing the color and helicity final state sums and initial state
averages, we obtain
\begin{eqnarray}
\lefteqn{{\cal J}^{W^-W^+}_4 = {\cal K}^{WW}_4 Q_W^4
  \delta_{12} |U_{34}|^2 |U_{65}|^2 }
  \quad\nonumber \\
&&\times\left[\left({L_1\cot\theta_W\over s_{12}-M_Z^2}+{Q_1\over s_{12}}
  +{(\tau_1+1)Q_W^2\over 2s_{234}}\right) 
  \left(\delta_{23}{\rm Im} X_{214365}^+
  +\delta_{14}{\rm Im} Y_{214365}^+\right) \right.
  \nonumber \\
&&\left.\mbox{}-\left({L_1\cot\theta_W\over s_{12}-M_Z^2}+{Q_1\over s_{12}}
  +{(\tau_1-1)Q_W^2\over 2s_{134}}\right) 
  \left(\delta_{23}{\rm Im} Y_{123456}
  +\delta_{14}{\rm Im} X_{123456} \right) \right]\;,
\label{JWWeq}
\end{eqnarray}
where,
\begin{equation}
{\cal K}^{V_lV_q}_4 = {2^7\pi^6\alpha^3\alpha_s C_F\over n_c}
  {\delta(s_{34}-M_{V_q}^2)\delta(s_{56}-M_{V_l}^2)\over\Gamma_{V_l}
  M_{V_l}} \; .
\end{equation}
Explicit formulas for the imaginary parts
of the functions $X$ and $Y$
are derived in Appendix B:
\begin{eqnarray}
{\rm Im} X_{123456} = && 32 {s_{135}\over s_{14}s_{124}}
  \left[(s_{15}+s_{35})x_{1234}+s_{13}x_{1245}+s_{24}x_{1345}\right]\;, 
\label{ImXeq}\\
{\rm Im} Y_{123456} = && 32 {s_{135}\over s_{23}} \left[{s_{13}\over
  s_{123}}(x_{1245}-x_{2345}) \right. \nonumber\\
&&\left.\mbox{} -{1\over s_{234}} \left\{
  (s_{15}+s_{35})x_{1234}+s_{13}x_{2345}+s_{24}x_{1345}\right\}\right]
  \;. 
\label{ImYeq}
\end{eqnarray}
For the charge conjugate process, that is for $V_q=W^-\rightarrow
d\bar{u}$, $V_l=W^+\rightarrow \nu\bar{l}$,
\begin{eqnarray}
\lefteqn{{\cal J}^{W^+W^-}_4 = {\cal K}^{WW}_4 Q_W^4
  \delta_{12} |U_{43}|^2 |U_{56}|^2}
  \quad\nonumber \\
&&\times\left[\left({L_1\cot\theta_W\over s_{12}-M_Z^2}+{Q_1\over s_{12}}
  +{(\tau_1+1)Q_W^2\over 2s_{134}}\right) 
  \left(\delta_{14}{\rm Im} X_{123456}
  +\delta_{23}{\rm Im} Y_{123456}\right) \right.
  \nonumber\\
&&\left.\mbox{}-\left({L_1\cot\theta_W\over s_{12}-M_Z^2}+{Q_1\over s_{12}}
  +{(\tau_1-1)Q_W^2\over 2s_{234}}\right) \left(\delta_{14}{\rm Im} 
    Y_{214365}^+ +\delta_{23}{\rm Im} X_{214365}^+\right) \right]\;.
\label{JWW2eq}
\end{eqnarray}
Note that Eq.\ (\ref{JWWeq}) and Eq.\ (\ref{JWW2eq}) 
are related by the interchanges
$(1\leftrightarrow2,\;3\leftrightarrow4,\;5\leftrightarrow6,\;
X\leftrightarrow X^+,\;Y\leftrightarrow Y^+)$, or alternatively by the
replacements $U\rightarrow U^\dagger$, $\tau_1\rightarrow -\tau_1$,
$\cot\theta_W\rightarrow-\cot\theta_W$, and $Q_1\rightarrow-Q_1$.  

For the case $V_q=Z^0\rightarrow q_3\bar q_4$, $V_l=W^-\rightarrow l_5
\bar\nu_6$,
\begin{eqnarray}
\lefteqn{{\cal J}^{W^-Z}_4 = {\cal K}^{WZ}_4 Q_W^4
  |U_{12}|^2 \delta_{34} |U_{65}|^2 L_3}
  \quad\nonumber \\
&&\times \left[\left({\tau_1\cot\theta_W\over s_{12}-M_W^2}
  +{L_1\over s_{134}}\right) \left(\delta_{14}{\rm Im} X_{123456}
  +\delta_{23}{\rm Im} Y_{123456} \right) \right.
  \nonumber \\
&&\left.\mbox{}+\left({\tau_2\cot\theta_W\over s_{12}-M_W^2}
  +{L_2\over s_{234}}\right) \left(\delta_{14}{\rm Im} 
    Y_{214365}^+ +\delta_{23}{\rm Im} X_{214365}^+ \right) \right]\;,
\label{JWZeq}
\end{eqnarray}
while for the charge conjugate process, that is for $V_l=W^+\rightarrow
\nu_5\bar{l}_6$,
\begin{eqnarray}
\lefteqn{{\cal J}^{W^+Z}_4 = {\cal K}^{WZ}_4 Q_W^4
  |U_{21}|^2 \delta_{34} |U_{56}|^2 L_3}
  \quad\nonumber \\
&&\times \left[\left({\tau_1\cot\theta_W\over s_{12}-M_W^2}
  +{L_1\over s_{134}}\right) \left(\delta_{14}{\rm Im} X_{123456}
  +\delta_{23}{\rm Im} Y_{123456} \right) \right.
  \nonumber \\
&&\left.\mbox{}+\left({\tau_2\cot\theta_W\over s_{12}-M_W^2}
  +{L_2\over s_{234}}\right) \left(\delta_{14}{\rm Im} 
    Y_{214365}^+ +\delta_{23}{\rm Im} X_{214365}^+ \right) \right]\;.
\label{JWZ2eq}
\end{eqnarray}
Eqs.\ (\ref{JWZeq}) and (\ref{JWZ2eq}) are related by the interchanges
$(1\leftrightarrow2,\;3\leftrightarrow4,\;5\leftrightarrow6,\;
X\leftrightarrow X^+,\;Y\leftrightarrow Y^+)$, or alternatively by the
replacement $U\rightarrow U^\dagger$.  

For the case $V_q=W^+\rightarrow u_3\bar d_4$, $V_l=Z^0\rightarrow l_5
\bar l_6$,
\begin{eqnarray}
\lefteqn{{\cal J}^{ZW^+}_4 = {\cal K}^{ZW}_4 Q_W^2
  |U_{21}|^2 \delta_{14} \delta_{23} \delta_{56} }
  \quad\nonumber \\
&&\times \left[L_5^2\left\{\left({\tau_2\cot\theta_W\over s_{12}-M_W^2}
  +{L_2\over s_{134}}\right) \left(L_2{\rm Im} X_{123456}
  +L_1{\rm Im} Y_{123456} \right) \right. \right.
  \nonumber \\
&&\left.\mbox{}+\left({\tau_1\cot\theta_W\over s_{12}-M_W^2}
  +{L_1\over s_{234}}\right) \left(
    L_2{\rm Im} Y_{214365}^+
  +L_1{\rm Im} X_{214365}^+ \right) \right\}
  \nonumber \\
&&\mbox{}\left. +R_5^2 \left\{5\leftrightarrow6
  \vphantom{{L_2\over s_{134}}}\right\}\right]\;,
\end{eqnarray}
while for the charge conjugate process,
\begin{eqnarray}
\lefteqn{{\cal J}^{ZW^-}_4 = {\cal K}^{ZW}_4 Q_W^2
  |U_{12}|^2 \delta_{14} \delta_{23} \delta_{56} }
  \quad\nonumber \\
&&\times \left[L_5^2\left\{\left({\tau_1\cot\theta_W\over s_{12}-M_W^2}
  +{L_1\over s_{234}}\right) \left(L_2{\rm Im} Y_{214365}^+
  +L_1{\rm Im} X_{214365}^+ \right) \right. \right.
  \nonumber \\
&&\left.\mbox{}+\left({\tau_2\cot\theta_W\over s_{12}-M_W^2}
  +{L_2\over s_{134}}\right) \left(L_2{\rm Im} 
    X_{123456} +L_1{\rm Im} Y_{123456} \right) \right\}
  \nonumber \\
&&\mbox{}\left. +R_5^2 \left\{5\leftrightarrow6
  \vphantom{{L_2\over s_{134}}}\right\}\right]\;,
\end{eqnarray}
which can be obtained from Eq.\ (\ref{JWZ2eq}) by the interchanges
$(1\leftrightarrow2,\;3\leftrightarrow4,\;5\leftrightarrow6,\;
X\leftrightarrow X^+,\;Y\leftrightarrow Y^+)$, or by letting $U
\rightarrow U^\dagger$.
The two expressions are in fact formally identical, but they are not
equal because the flavors involved are different in the two cases.

Finally, for the case $V_q=Z^0\rightarrow q_3\bar{q}_4$, $V_l=Z^0
\rightarrow l_5\bar{l}_6$,
\begin{eqnarray}
\lefteqn{{\cal J}^{ZZ}_4 = {\cal K}^{ZZ}_4 \delta_{12}
  \delta_{14} \delta_{23} \left[\vphantom{{1\over s_{134}}} \left(
  L_1^4L_5^2-R_1^4R_5^2\right) \right. }
  \quad\nonumber \\
&&\times \left\{{1\over s_{134}}({\rm Im} X_{123456}+{\rm Im} Y_{123456})
   +{1\over s_{234}}({\rm Im} X_{214365}^++{\rm Im} Y_{214365}^+) \right\}
  \nonumber \\
&&\mbox{}\left. +\left(L_1^4R_5^2-R_1^4L_5^2\right)
  \left\{5\leftrightarrow6\vphantom{{1\over s_{134}}} \right\} \right] \;.
\label{JZZeq}
\end{eqnarray}

\subsubsection{Three particle final states $(V_q\rightarrow
q_3\bar{q}_4)+(V_l=W^\pm$ or $Z^0)$}

In the preceding subsection we presented chromo-electroweak interference
contributions to the cross section for the production of two
electroweak bosons, one of which $V_q$ decays to a quark-antiquark
pair, and the other $V_l$ to a lepton pair, and in which we used the
narrow width approximation for the propagators of both bosons.  In
this subsection we will consider production of a real $V_l$.  If $V_l$
is a real photon, it can be detected experimentally; in the cases
$V_l=W^\pm,Z^0$, which might be detected for example through
particular leptonic decay channels, our cross sections must be
multiplied by the appropriate branching ratio.

The case of a real $V_l=W^\pm,Z^0$ can be treated by introducing
helicity eigenfunctions for the massive vector boson and computing the
required helicity amplitudes.  It is simpler, however, to integrate
the expressions for the chromo-electroweak interference cross sections
given in the preceding subsection over the phase space of the final
leptonic decay products with the momentum of $V_l$ held fixed, and
divide the result by the leptonic branching ratio.

We wish to integrate the squared amplitude over the $k_5k_6$ phase
space holding $k_5+k_6$ fixed.
If the narrow width approximation
\begin{equation}
{1\over\left|D^{V_l}_{56}\right|^2 }
  = {\pi\delta(s_{56}-M_{V_l}^2)\over \Gamma_{V_l}  M_{V_l}}\;,
\end{equation}
is used for the $V_l$ propagator, the $k_5k_6$ invariant phase space
can be converted to the appropriate phase space for a real $V_l$ as
follows:
\begin{equation}
\int{d^3k_5d^3k_6\over (2\pi)^6 4k_5^0k_6^0}{\delta^4(\sum_{i=1}^6
  k_i)\over|D_{56}^{V_l}|^2}={1\over 16\pi M_{V_l}\Gamma_{V_l}}
  \int{d^3q\over (2\pi)^3 2 q^0} \delta^4\left(q+\sum_{i=1}^4 k_i\right)\;.
\label{56eq}
\end{equation}
The 4-particle squared amplitudes presented in the preceding
subsection were for a particular decay channel $V_l\rightarrow
l_5\bar{l}_6$.  If the coupling in this channel is
$-ie\gamma^\mu({\cal L}\gamma_L +{\cal R}\gamma_R)$, the partial width
\begin{equation}
\Gamma_{V_l}^{l\bar l} = {e^2(|{\cal L}|^2+|{\cal R}|^2)M_{V_l}
  \over 24\pi}\;,
\end{equation}
and we need to divide the 4-particle squared amplitudes by the
branching ratio $\Gamma_{V_l}^{l\bar l}/\Gamma_{V_l}$.  The 4-particle
squared amplitudes are polynomials in $k_5$ when $q=k_5+k_6$ is held
fixed.  The averages over the decay phase-space can be effected by
using the formulas
\begin{eqnarray}
\int_{5,6} &\equiv & \int{d^3k_5d^3k_6\over (2\pi)^6
  4k_5^0k_6^0} (2\pi)^4\delta^4(q-k_5-k_6) \nonumber \\
&=& {1\over 8\pi}\;, \cr
  \int_{5,6} k_5^\mu &=& {1\over2}q^\mu \int_{5,6}\;, \nonumber \\
\int_{5,6} k_5^\mu k_5^\nu &=& \left({1\over3}q^\mu q^\nu -{1\over12}
  g^{\mu\nu}q^2\right) \int_{5,6}\;,
\label{56aveq}
\end{eqnarray}
etc.  From Eqs.\ (\ref{56eq}) through (\ref{56aveq}) 
it is easy to see that the 3-particle
squared amplitudes can be obtained from the corresponding 4-particle
squared amplitudes given in the preceding subsection by the
replacements
\begin{eqnarray}
{1\over |D^{V_l}_{56}|^2} \approx{\pi\delta(s_{56}-M_{V_l}^2)
  \over \Gamma_{V_l}M_{V_l}} 
&&\rightarrow   \left({1\over 16\pi\Gamma_{V_L}
  M_{V_l}}\right)\left({\Gamma_{V_l}\over \Gamma_{V_l}^{l\bar l}}
  \right) \nonumber \\
&&= {3\over 2 e^2(|{\cal L}|^2+ |{\cal R}|^2)}\;,\nonumber \\
  k_5^\mu &\rightarrow & {1\over 2} q^\mu \;, \nonumber \\
  k_5^\mu k_6^\nu &\rightarrow & {1\over 3} q^\mu q^\nu - {1\over12}
  g^{\mu \nu}q^2\;.
\end{eqnarray}
In particular, Eqs.\ (\ref{56aveq}) require the following replacements
for Eqs.\ (\ref{ImXeq},\ref{ImYeq}):
\begin{equation}
{\rm Im} X_{123456} \rightarrow {32\over3} M_{V_l}^2 {x_{1234} 
  \over s_{14} s_{124}}(s_{12}+s_{34}+s_{14}+s_{23})\;,
\end{equation}
\begin{equation}
{\rm Im} Y_{123456} \rightarrow -{32\over3} M_{V_l}^2{x_{1234}\over
   s_{23} s_{234}}(s_{12}+s_{34}+s_{14}+s_{23})\;. 
\end{equation}
Let us define in analogy with Eq.\ (\ref{JVVeq})
\begin{equation}
{\cal J}_3^{V_lV_q} \equiv {1\over4n_c^2}\sum_{\rm colors}\sum_{\rm
  helicities}{\rm Re}\left[E^{V_lV_q}\left(G^{V_lV_q}\right)^+
  +\left(E^{V_lV_q}\right)^+G^{V_lV_q}\right]\;.
\label{J3VVeq}
\end{equation}
We also define
\begin{equation}
{\cal K}_3^{V_q} \equiv {2^9\pi^4\alpha^2\alpha_s C_F\over n_c}
  \delta(s_{34} - M_{V_q}^2)\;. 
\label{K3Veq}
\end{equation}
Then, from Eqs.\ (\ref{JWWeq})-(\ref{JZZeq}) we obtain the 3-particle squared
amplitudes listed below:
\begin{eqnarray}
\lefteqn{{\cal J}_3^{W^-W^+} ={\cal K}_3^{W}Q_W^2\delta_{12}
  |U_{34}|^2 \left(s_{12}+s_{34}+s_{14}+s_{23}\right)  x_{1234}}
  \quad \nonumber \\
&&\times\left[ \left({L_1\cot\theta_W\over s_{12}-M_Z^2}+{Q_1\over s_{12}}
  +{(\tau_1+1)Q_W^2\over 2s_{234}} \right) \left({\delta_{14}\over
  s_{14}s_{134}}-{\delta_{23}\over s_{23}s_{123}}\right) \right.
  \nonumber \\
&&\left.\mbox{} -\left({L_1\cot\theta_W\over s_{12}-M_Z^2}+{Q_1\over s_{12}}
  +{(\tau_1-1)Q_W^2\over 2s_{134}} \right) \left({\delta_{14}\over
  s_{14}s_{124}}-{\delta_{23}\over s_{23}s_{234}}\right)\right]
  \;, 
\label{J3WWeq}
\end{eqnarray}
\begin{eqnarray}
\lefteqn{{\cal J}_3^{W^+W^-} ={\cal K}_3^{W}Q_W^2\delta_{12}
  |U_{43}|^2 \left(s_{12}+s_{34}+s_{14}+s_{23}\right)  x_{1234} }
  \quad \nonumber \\
&&\times\left[ \left({L_1\cot\theta_W\over s_{12}-M_Z^2}+{Q_1\over s_{12}}
  +{(\tau_1+1)Q_W^2\over 2s_{134}} \right) \left({\delta_{14}\over
  s_{14}s_{124}}-{\delta_{23}\over s_{23}s_{234}}\right) \right.
  \nonumber \\
&&\left.\mbox{}-\left({L_1\cot\theta_W\over s_{12}-M_Z^2}+{Q_1\over s_{12}}
  +{(\tau_1-1)Q_W^2\over 2s_{234}} \right) \left({\delta_{14}\over
  s_{14}s_{134}}-{\delta_{23}\over s_{23}s_{123}}\right)\right]
  \nonumber \\
&&=\left. -{\cal J}_3^{W^-W^+}\right|_{1\leftrightarrow2,3\leftrightarrow4} \;,
\end{eqnarray}
\begin{eqnarray}
\lefteqn{{\cal J}_3^{W^-Z} ={\cal K}_3^{Z}Q_W^2\delta_{34}
  |U_{12}|^2 L_3\left(s_{12}+s_{34}+s_{14}+s_{23}\right)  x_{1234} }
  \quad \nonumber \\
&&\times\left[ \left({\tau_1\cot\theta_W\over s_{12}-M_W^2}
  +{L_1\over s_{134}}\right)\left({\delta_{14}\over s_{14}
  s_{124}}-{\delta_{23}\over s_{23} s_{234}}\right) \right.
  \nonumber \\
&&\left.\mbox{}+\left({\tau_2\cot\theta_W\over s_{12}-M_W^2}
  +{L_2\over s_{234}} \right)\left({\delta_{14}\over s_{14}
  s_{134}}-{\delta_{23}\over s_{23}
  s_{123}}\right)\right]\;,
\label{J3WZeq}
\end{eqnarray}
\begin{eqnarray}
\lefteqn{{\cal J}_3^{W^+Z} ={\cal K}_3^{Z}Q_W^2\delta_{34}
  |U_{21}|^2 L_3\left(s_{12}+s_{34}+s_{14}+s_{23}\right)  x_{1234} }
  \quad \nonumber \\
&&\times\left[ \left({\tau_1\cot\theta_W\over 
  s_{12}-M_W^2}+{L_1\over s_{134}}
  \right)\left({\delta_{14}\over s_{14}s_{124}}-
  {\delta_{23}\over s_{23} s_{234}}\right) \right.
  \nonumber \\
&&\left.\mbox{}+\left({\tau_2\cot\theta_W\over s_{12}-M_W^2}
+{L_2\over s_{234}} \right)\left({\delta_{14}\over s_{14}s_{134}}-
  {\delta_{23}\over s_{23} s_{123}}\right)\right] 
  \nonumber \\
&& =\left. -{\cal J}_3^{W^-Z}\right|_{1\leftrightarrow2,3\leftrightarrow4} \;.
\label{J3WZ2eq}
\end{eqnarray}
We note that the equations (\ref{J3WZeq}) and (\ref{J3WZ2eq}) 
are formally identical
(except for the change $U\rightarrow U^\dagger$), but the two
expressions are not equal because the flavors involved are different.
The same remark holds for the following two equations:
\begin{eqnarray}
\lefteqn{{\cal J}_3^{ZW^+} ={\cal K}_3^{W}Q_W^2\delta_{14}
  \delta_{23}|U_{21}|^2 \bigl(s_{12}+
  s_{34}+s_{14}+s_{23}\bigr)  x_{1234} }
  \quad \nonumber \\
&&\times\left[ \left({\tau_1\cot\theta_W\over 
  s_{12}-M_W^2}+{L_1\over s_{234}}
  \right)\left({L_2\over s_{14}s_{134}}-
  {L_1\over s_{23} s_{123}}\right) \right.
  \nonumber \\
&&\left.\mbox{}+\left({\tau_2\cot\theta_W\over s_{12}-M_W^2}
  +{L_2\over s_{134}} \right)\left({L_2\over s_{14}s_{124}}-
  {L_1\over s_{23} s_{234}}\right)\right]\;,
\end{eqnarray}
\begin{eqnarray}
\lefteqn{{\cal J}_3^{ZW^-} ={\cal K}_3^{W}Q_W^2\delta_{14}
  \delta_{23}|U_{12}|^2 \left(s_{12}+s_{34}+s_{14}
  +s_{23}\right)  x_{1234} }
  \quad \nonumber \\
&&\times\left[ \left({\tau_1\cot\theta_W\over 
  s_{12}-M_W^2}+{L_1\over s_{234}}
  \right)\left({L_2\over s_{14}s_{134}}-
  {L_1\over s_{23} s_{123}}\right) \right.
  \nonumber \\
&&\left.\mbox{}+\left({\tau_2\cot\theta_W\over 
  s_{12}-M_W^2}+{L_2\over s_{134}}
  \right)\left({L_2\over s_{14}s_{124}}-
  {L_1\over s_{23} s_{234}}\right)  \right] 
  \nonumber \\
&& =\left. -{\cal J}_3^{ZW^+}\right|_{1\leftrightarrow2,3\leftrightarrow4}
    \;.
\end{eqnarray}
Finally,
\begin{eqnarray}
\lefteqn{{\cal J}_3^{ZZ} ={\cal K}_3^{Z}\delta_{12}
  \delta_{14} \delta_{23} \left(L_1^4-R_1^4\right)
  \left(s_{12}+s_{34}+s_{14}+s_{23}\right)  x_{1234} }
  \quad \nonumber \\
&&\times\left[{1\over s_{134}}\left({1\over s_{14} s_{124}}-{1\over s_{23}
  s_{234}} \right) + {1\over s_{234}}\left({1\over s_{14} s_{134}}
  -{1\over s_{23}  s_{123}} \right)\right] \;.
\end{eqnarray}

\subsubsection{Three particle final states: $(V_q\rightarrow
q_3\bar q_4) + \gamma $}

In this subsection we present formulas for the chromo-electroweak
contributions ${\cal J}_3^{\gamma V_q}$ defined in 
Eq.\ (\ref{J3VVeq}) to the
squares of the amplitudes for real photon production which were
presented in the preceding subsection.  
The squared amplitudes can be compactly written
in terms of the functions
\begin{equation}
X_{12345}^\gamma \equiv {H_{12345}H_{32145}^+\over s_{14}}\;,
\label{Xpheq}
\end{equation}
\begin{equation}
Y_{12345}^\gamma \equiv {H_{12345}H_{14325}^+\over s_{23}}\;.
\end{equation}
Only the imaginary parts of these functions contribute to
${\cal J}_3^{\gamma V_q}$; in Appendix B we show that
\begin{equation}
{\rm Im} X_{12345}^\gamma = 16{s_{13}^2\over s_{14}s_{15}s_{25}
  s_{35}} x_{1234} \;,
\end{equation}
\begin{equation}
{\rm Im} Y_{12345}^\gamma = 16{s_{13}^2\over s_{23}s_{15}s_{25}
  s_{45}} x_{1234} \;.
\label{ImYpheq}
\end{equation}
From Eqs.\ (\ref{EAWeq}) to (\ref{GA4eq}), we obtain
\begin{eqnarray}
{\cal J}_3^{\gamma W^+} = && {\cal K}_3^{W}Q_W^2|U_{21}|^2
  \delta_{14}\delta_{23} (s_{13}^2-s_{24}^2) x_{1234} 
\nonumber \\
&& \times {1\over s_{15}} \left({\tau_2\over s_{12}-M_W^2}
  +{Q_2\over s_{25}}\right) \left({Q_1\over s_{23} s_{45}} +
  {Q_2\over s_{14} s_{35}} \right) \;,
\end{eqnarray}
\begin{eqnarray}
{\cal J}_3^{\gamma W^-}= && {\cal K}_3^{W}Q_W^2|U_{12}|^2
  \delta_{14}\delta_{23} (s_{13}^2-s_{24}^2) x_{1234} 
\nonumber \\
&& \times {1\over s_{25}} \left({\tau_1\over s_{12}-M_W^2}
  +{Q_1\over s_{15}}\right) \left({Q_1\over s_{23} s_{45}} +
  {Q_2\over s_{14} s_{35}} \right)
\nonumber \\
  =&&\left. -{\cal J}_3^{\gamma W^+}
  \right|_{1\leftrightarrow2,3\leftrightarrow4} 
  \;, 
\end{eqnarray}
\begin{eqnarray}
{\cal J}_3^{\gamma Z} = && {\cal K}_3^{Z}Q_1^2\delta_{12}
  \delta_{14} \delta_{23} \bigl(L_1^2-R_1^2\bigr) (s_{13}^2-s_{24}^2)
   x_{1234} \nonumber \\
&& \times {1\over s_{15} s_{25}}
  \left({1\over s_{23} s_{45}}+{1\over s_{14} s_{35}}\right)\;.
\end{eqnarray}
%

\section{Parity-violating Asymmetries}

\subsection{General Definition}

The chromo-electroweak interference contributions to the cross section
for producing a vector boson plus two jets which were presented above
are odd under space-reflection.  A simple way of defining experimental
observables that are sensitive to this parity-odd character of the
interference terms is as follows: Imagine that the incident beams lie
in the plane of a mirror.  If the incident beams are not polarized,
the initial state is invariant under reflection in this mirror.  We
will also assume that the spins of the final state particles are not
detected, i.e., particles are identified by their momenta and internal
quantum numbers only.  A parity-odd contribution to the cross section
can make the probabilities for observing a particular event (i.e., a
particular configuration of final state particles) and its mirror
image different from one another.  Since the events are continuously
distributed in phase space, the likelihood of finding an event and its
geometrical mirror image in any finite sample of events is vanishingly
small.  To decide experimentally whether or not there is an asymmetry
with respect to mirror reflection in the event sample, one must count
the number of events that fall in some region of phase space (which we
will call a ``bin'') and the number of events that fall in the mirror
image of this region (which we will the ``image bin'') and then decide
whether or not there is a statistically significant difference between
these two numbers.  This difference can be compared with a
``parity-violating asymmetry'' which we define as follows:
\begin{equation}
a^{\rm pv}({\rm bin}) = \int\limits_{\rm bin}\,d\sigma \quad-\quad
  \int\limits_{\rm image\,bin}d\sigma\;.
\label{apveq}
\end{equation}
The integral over the bin includes an implicit sum over all final
state quantum numbers that are not observed, i.e., color quantum
numbers and any spin or flavor quantum numbers that cannot be
experimentally measured.  Note that the parity-even terms in the
differential cross section $d\sigma$ do not contribute to $a^{\rm
pv}$, which therefore provides a direct measure of the parity-odd
terms.

The asymmetry $a^{\rm pv}({\rm bin})$ will obviously depend on the
location of the chosen bin in phase space and on its size and shape.
It will be small if the bin is very small.  It will obviously be zero
if the bin consists of the whole phase space, or if the bin is
invariant under mirror reflection.  Note also that events which lie in
any portion of a bin that overlaps with a portion of the image bin
will not contribute to $a^{\rm pv}$.  While such ``symmetric'' events
do not contribute to the size of the asymmetry, they will tend to
exaggerate its statistical significance. We will therefore restrict
the chosen bins to those which do not overlap with their mirror
images.

A set of bins will have to be judiciously chosen to maximize the
observed effects of parity violation, i.e., to yield the largest
cumulative asymmetry.  Thus we also define a cumulative
parity-violating asymmetry as follows:
\begin{equation}
A^{\rm pv} = \sum_{\rm bins} \mid a^{\rm pv}({\rm bin})\mid\;,
\label{Aeq}
\end{equation}
where the sum is taken over bins that do not overlap with one another.
To avoid counting events more than once we will also demand that the
bins be chosen so that no bin in this sum overlaps with the mirror
image of any other bin.  $A^{\rm pv}$ will obviously be largest if
$a^{\rm pv}$ can be measured in a large number of small bins.  The
size of the total event sample will put a lower limit on the number of
bins and their sizes since the asymmetry in each bin-image pair must
be statistically significant.

A theoretical upper bound on $A^{\rm pv}$ is given by
\begin{equation}
A^{\rm pv}_{\max} = \sum_{\rm bins} \lim_{\rm bin\,size\rightarrow 0}
 \mid a^{\rm pv}({\rm bin})\mid\;.
\label{Amaxeq}
\end{equation}
If ${d\sigma/d\Omega}$ is the fully differential cross section
(which includes a sum over unobserved quantum numbers),
\begin{equation}
A^{\rm pv}_{\max} = {1\over2}\int\limits_{\rm <cuts>} d\Omega\;\left|
  \left({d\sigma\over d\Omega}\right)-\left({d\sigma\over d\Omega}
  \right)_{\rm mirror\,image} \right| \;.
\label{Amax2eq}
\end{equation}
The factor of $1\over2$ ensures that no event is counted twice.  The
region of integration is taken to be all of the phase space 
that satisfies cuts that
are mandated by experimental considerations, and by the need to
exclude regions where the theoretical expressions for
${d\sigma/d\Omega}$ become unreliable due to non-perturbative effects.
The cuts will be taken to be invariant under mirror-reflection: this
assumption will not unduly restrict the event sample in a
colliding-beam experiment with a $4\pi$ or a cylindrically symmetrical
detector.

\subsection{Asymmetries in Electroweak Boson + 2 Jet Production }

In this subsection we will apply the general definition of
parity-violating asymmetries introduced above to the production of an
electroweak boson and two jets.  We first describe the kinematics
appropriate to the three-particle final state at the parton level.  We
will then define and present results for asymmetries at the parton and
the hadron levels.

\subsubsection{Kinematics }

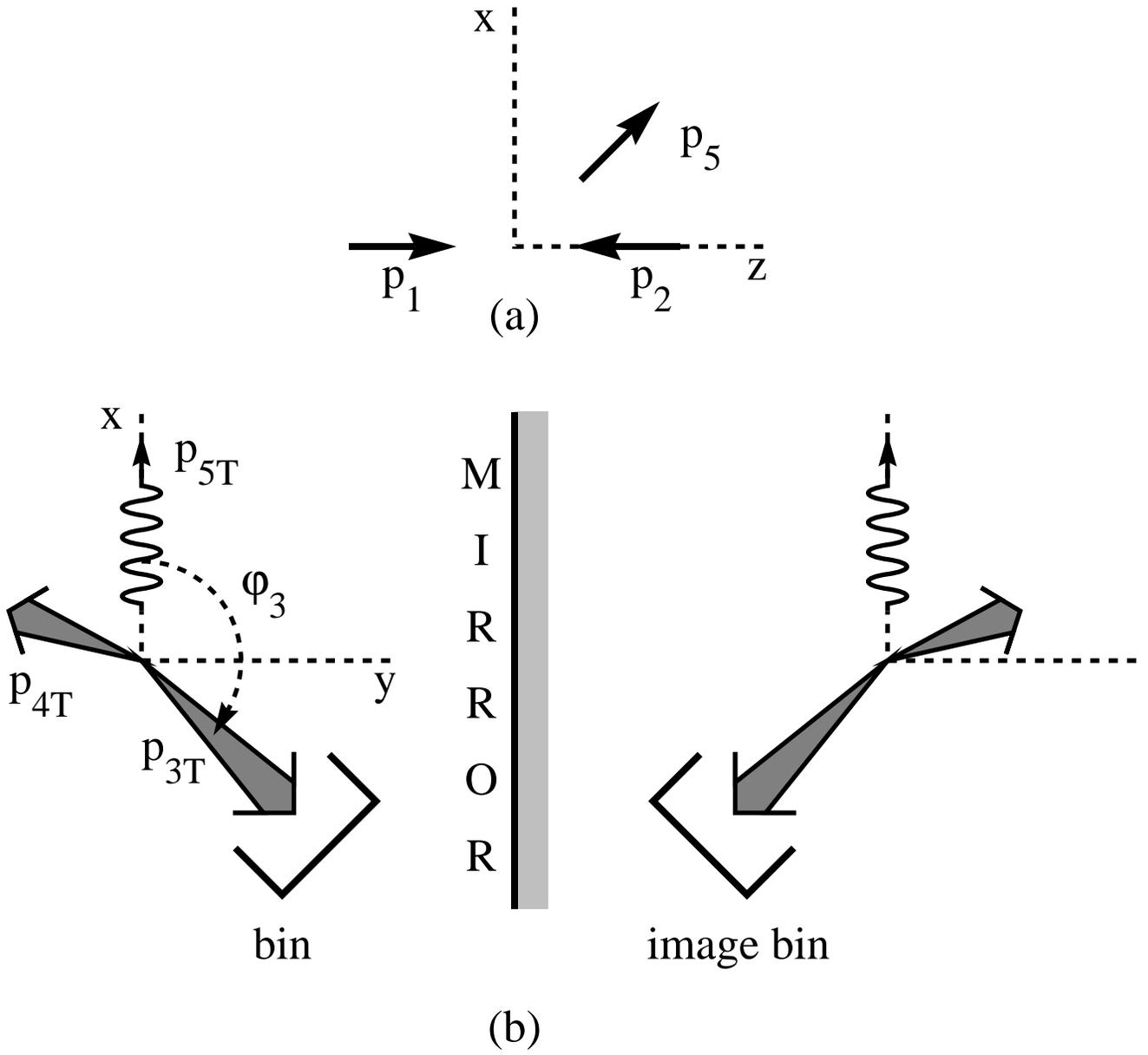
\begin{figure}
\caption{
(a) Momenta in the $x{-}z$ plane.  $p_1$ is the momentum of the quark
(parton-level asymmetry), the proton ($p\bar p$ collisions), or one of
the protons chosen arbitrarily ($pp$ collisions).  $p_5$ is the
momentum of $V=W^\pm,Z^0$ (3-particle final state) or of the lepton
(4-particle final state).  A right-handed coordinate system is defined
such that $p_5^x>0$ and $p_5^y=0$. (b) Momenta in the $x{-}y$ plane.
$p_3$ and $p_4$ are the momenta of the two jet partons.  Parity is
violated if the event shown on the left and its mirror image shown on
the right occur with different probabilities.  In the case of the
4-particle final state, the azimuthal angle $\phi_6$ of the antilepton
may also be used to define a bin.
}
\label{pvfig}
\end{figure}

We assume that the colliding beams are collinear and define a $z$
axis.  The colliding partons are assumed to have momenta $p_1$ and
$p_2$, with $\vec p_1$ in the positive $z$-direction.  We assume that
the electroweak boson has momentum $q$, and choose the $x$-axis so
that $\vec q$ lies in the $x{-}z$ plane and $q^x>0$.  
(We assume that an on-shell $W$ or $Z$ is detected in its leptonic decay mode,
and that its momentum has been reconstructed from measurement of the
decay-lepton momenta.)
The $y$ axis
is chosen so that the coordinate system is right-handed as illustrated
in Fig.\ \ref{pvfig}.  The momenta of the two final state partons
(jets) will be denoted by $p_3$ and $p_4$.

If ${\cal J}_3$ is a contribution to a spin and helicity averaged
squared matrix element, the corresponding contribution to the
differential cross section at the parton level is
\begin{equation}
d\hat\sigma = {1\over2s_{12}} (2\pi)^4\delta^4(p_1+p_2-q-p_3-p_4)
  {\cal J}_3 {\cal B}_{\rm lep}
  {d^3p_3d^3p_4d^3q\over(2\pi)^9 8 p_3^0p_4^0q^0}\;.
\label{sigmahateq}
\end{equation}
In this equation we have explicitly included a factor of the 
leptonic branching ratio ${\cal B}_{\rm lep}$ of the electroweak boson
(i.e., ${\cal B}_{\rm lep}=\Gamma(W\rightarrow e\nu)/\Gamma_{\rm tot}$ 
for $W^\pm$, $\Gamma(Z\rightarrow e^+e^-)/\Gamma_{\rm tot}$ for $Z^0$, and
1 for a real photon).
We note that the subscripts labeling the parton momenta $p_i$ stand
for particular points in momentum space and not for parton quantum
numbers.  To be more precise, let us write the parton level reaction
as follows:
\begin{equation}
a_1(p_1)+a_2(p_2)\rightarrow a_3(p_3)+a_4(p_4)+V_l(q)\;,
\label{aaaaVeq}
\end{equation}
where $a_i$ stand for parton flavors.  The quantities ${\cal J}_3$
defined in Eq.\ (\ref{J3VVeq}) were 
functions of momenta $k_i,\;i=1,\dots4$ all
of which were taken to be outgoing, and they also depended on the
flavors $f_i$ of the quarks or antiquarks which carried these momenta.
We express this dependence explicitly as follows:
\begin{equation}
{\cal J}_3^{V_lV_q}\bigl(f_1(k_1),f_2(k_2),f_3(k_3),f_4(k_4)\bigr)
  \;.
\end{equation}
Partons $f_1$ and $f_2$ must be crossed to the initial state, and then
either one of them can be identified with the beam parton $a_1(p_1)$.
Parton $f_3$ can then be assigned either a momentum $p_3$ or a
momentum $p_4$. For example if $V_l=Z^0,\,V_q=W^+$, we have the
following possibilities:
\begin{eqnarray}
u(p_1)+\bar d(p_2)\rightarrow u(p_3)+\bar d(p_4)+Z^0 & \Rightarrow &
{\cal J}_3^{ZW^+}(d(-p_2),\bar u(-p_1),u(p_3),\bar d(p_4)) \;,
\nonumber \\
\bar d(p_1)+u(p_2)\rightarrow u(p_3)+\bar d(p_4)+Z^0 & \Rightarrow &
{\cal J}_3^{ZW^+}(d(-p_1),\bar u(-p_2),u(p_3),\bar d(p_4)) \;,
\nonumber \\
u(p_1)+\bar d(p_2)\rightarrow \bar d(p_3)+u(p_4)+Z^0 & \Rightarrow &
{\cal J}_3^{ZW^+}(d(-p_2),\bar u(-p_1),u(p_4),\bar d(p_3)) \;,
\nonumber \\
\bar d(p_1)+u(p_2)\rightarrow \bar d(p_3)+u(p_4)+Z^0 & \Rightarrow &
{\cal J}_3^{ZW^+}(d(-p_1),\bar u(-p_2),u(p_4),\bar d(p_3)) \;.
\end{eqnarray}
In these expressions, $u$ and $d$ stand for generic weak
$I_z=\pm{1\over2}$ flavors, i.e., ``{\it u}\thinspace''$=u,c,\dots$
and ``{\it d}\thinspace''$=d,s,b,\dots$, etc.  In this paper we will
include the contributions of the ``lighter'' flavors $u,d,s,c,b$ only in
making numerical predictions.

Since we have used the narrow width approximation for $V_q$, see
Eq.\ (\ref{narrowWidthEq}), let us write 
\begin{equation}
{\cal J}_3=\hat{\cal J}_3\delta(s_{34}-M_{V_q}^2)\;.
\label{J3hateq}
\end{equation}
Because the initial partons are assumed unpolarized, ${\cal J}_3$ is
invariant under rotations about the $z$-axis.  The 3-particle phase
space in Eq.\ (\ref{sigmahateq}) is thus effectively 3-dimensional.
We will choose the rapidity $y_3$ and azimuthal angle $\phi_3$ of the
parton with momentum $p_3$ and the rapidity $y_q$ of $V_l$ as the
three independent phase space variables.  The rapidities are invariant
under reflection in the $x{-}z$ plane, while $\phi_3\rightarrow
2\pi-\phi_3$.  Performing the trivial integrations we obtain
\begin{equation}
{d\hat\sigma\over dy_qdy_3d\phi_3}=
  {p_{3T}^2E_{qT}^2\hat{\cal J}_3{\cal B}_{\rm lep}
  \over 2^7\pi^4 s_{12}s_{34}(s_{12}+M_{V_l}^2-s_{34})}\;,
\label{diffxceq}
\end{equation}
where
\begin{eqnarray*}
&&E_{qT}\equiv \sqrt{q_x^2+q_y^2+M_{V_l}^2}={s_{12}+M_{V_l}^2-M_{V_q}^2
  \over 2\sqrt{s_{12}}\cosh(y_{12}-y_q)}\;,\\
&&p_{3T}\equiv \sqrt{p_{3x}^2+p_{3y}^2}={M_{V_q}^2\over 2\bigl[
  \sqrt{s_{12}}\cosh(y_{12}-y_3)-E_{qT}\cosh(y_q-y_3)+q_T\cos\phi_3
  \bigr]}\;,\\
&&q_T\equiv \sqrt{q_x^2+q_y^2}\;,\qquad y_{12}\equiv{1\over2}\ln
  \left({p_1^0+p_2^0+p_1^z+p_2^z
  \over p_1^0+p_2^0-p_1^z-p_2^z}\right)\;.
\end{eqnarray*}
The parton subprocess (\ref{aaaaVeq}) will contribute to the cross
section for the hadronic process $h_1(P_1)+h_2(P_2)\rightarrow
V_l+2$ Jets:
\begin{equation}
{d\sigma\over d\tau\,dy_{12}dy_qdy_3d\phi_3} = \sum_{a_1a_2}f_{a_1/
  h_1}(x_1,Q^2)f_{a_2/h_2}(x_2,Q^2){d\hat\sigma\over dy_qdy_3d\phi_3}
  \;,
\label{hadxceq}
\end{equation}
where
\begin{equation}
\tau=x_1x_2={s_{12}\over s}\;,\quad s=(P_1+P_2)^2\;,\quad
 y_{12}={1\over2}\ln\left({x_1\over x_2}\right)\;,
\end{equation}
and $f_{a/h}(x,Q^2)$ is the density of partons of flavor $a$ and
momentum fraction $x$ in the hadron $h$. In this paper we will use the
structure function parametrizations of Martin, Roberts and
Stirling \cite{MRSB} (set MRSB).
The momentum
scale $Q$ of the parton densities is not precisely determined since we
have not computed the next-to-leading corrections to the cross
section, so we will take $Q^2=s_{12}/4$ in obtaining numerical
predictions.

We next specify a set of cuts that we will use when we integrate the
differential cross section to obtain the asymmetries defined in 
Eqs.\ (\ref{apveq}) through (\ref{Amax2eq}).  
We will apply a single representative set of cuts on the momenta of the final
state particles for most of the numerical results presented in this paper.
These cuts are defined in terms of the transverse momenta $p_{iT}$,
pseudorapidities $\eta_i=-\ln\tan(\theta_i/2)$ 
and azimuthal angles $\phi_i$, in the laboratory frame.
The index $i$ runs over $V_l$, $j_1$ and $j_2$ for three-particle final
states, and over $l$, $\bar l$, $j_1$ and $j_2$ for four-particle final
states.
We choose
\begin{eqnarray}
{\rm (i)}\quad && p_{iT} \geq 10{\rm\ GeV}\;, \nonumber\\
{\rm (ii)}\quad && \left|\eta_i\right| \leq 2.5 \;, \\
\label{cutseq}
{\rm (iii)}\quad && \left[(\eta_i-\eta_j)^2+(\phi_i-\phi_j)^2\right] 
	\geq 0.7 \;.\nonumber
\end{eqnarray}
These cuts are meant to ensure that the jets are well
resolved experimentally, and to exclude regions of phase space where
the theoretical predictions are likely to become unreliable due to
infrared and collinear singularities. 
We will use the Monte Carlo program {\sc vegas}\ \cite{vegas}
to perform the phase space integrations.

\subsubsection{Parton-level Asymmetries}

To illustrate the nature and magnitude of the
parity-violating effects, we first present results for the
differential asymmetry $da^{\rm pv}/d\phi_3$ that would be observed in
a quark-antiquark collisions if different quark flavors could be
distinguished from one another.  We assign the incoming quark the
momentum $p_1$ and the outgoing quark the momentum $p_3$.  According
to Eq.\ (\ref{apveq}), the differential asymmetry represents the excess of
events/degree in which the final state quark is observed at an
azimuthal angle $\phi_3$ over those in which it is observed at an
angle $2\pi\!-\!\phi_3$.  The asymmetry at an angle
$\phi_3\geq180^\circ$ is the negative of that at the complementary
angle $2\pi\!-\!\phi_3$.  

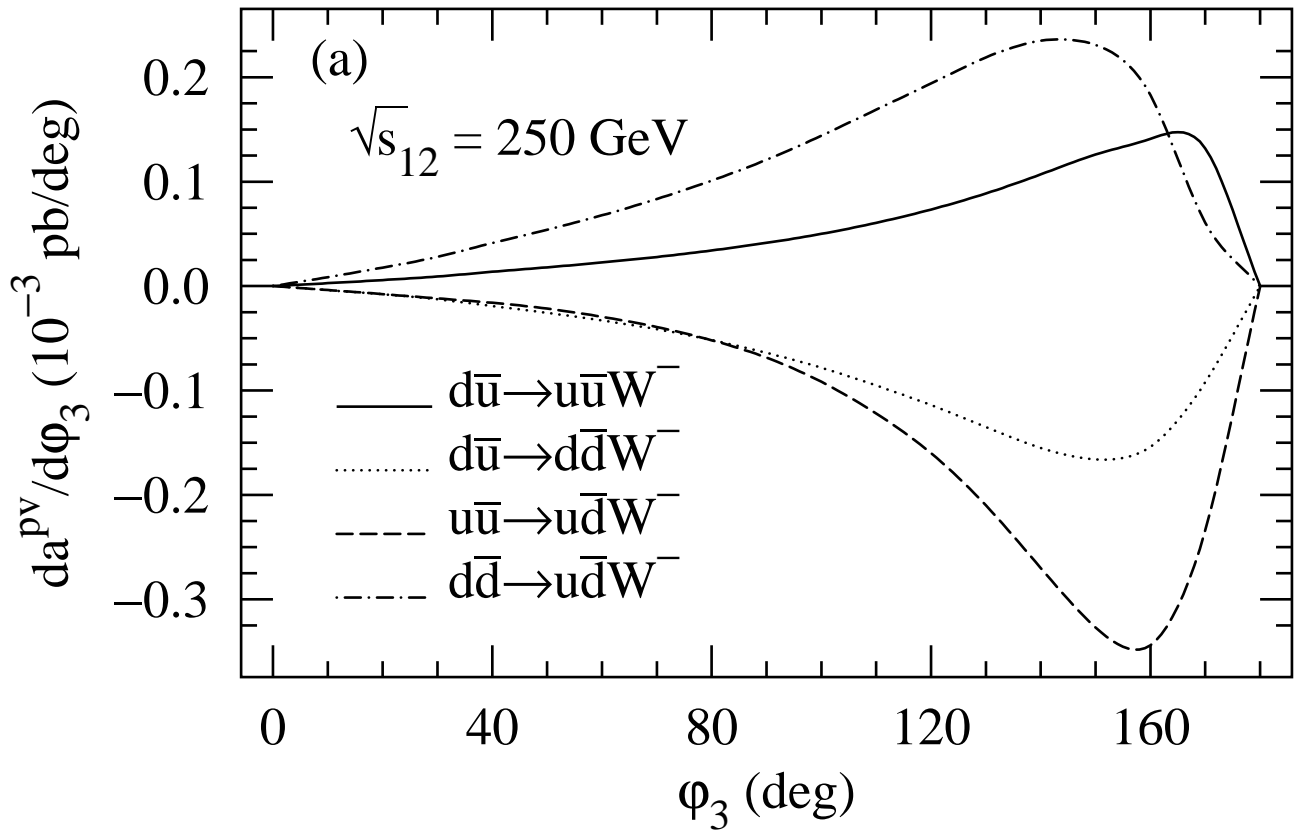
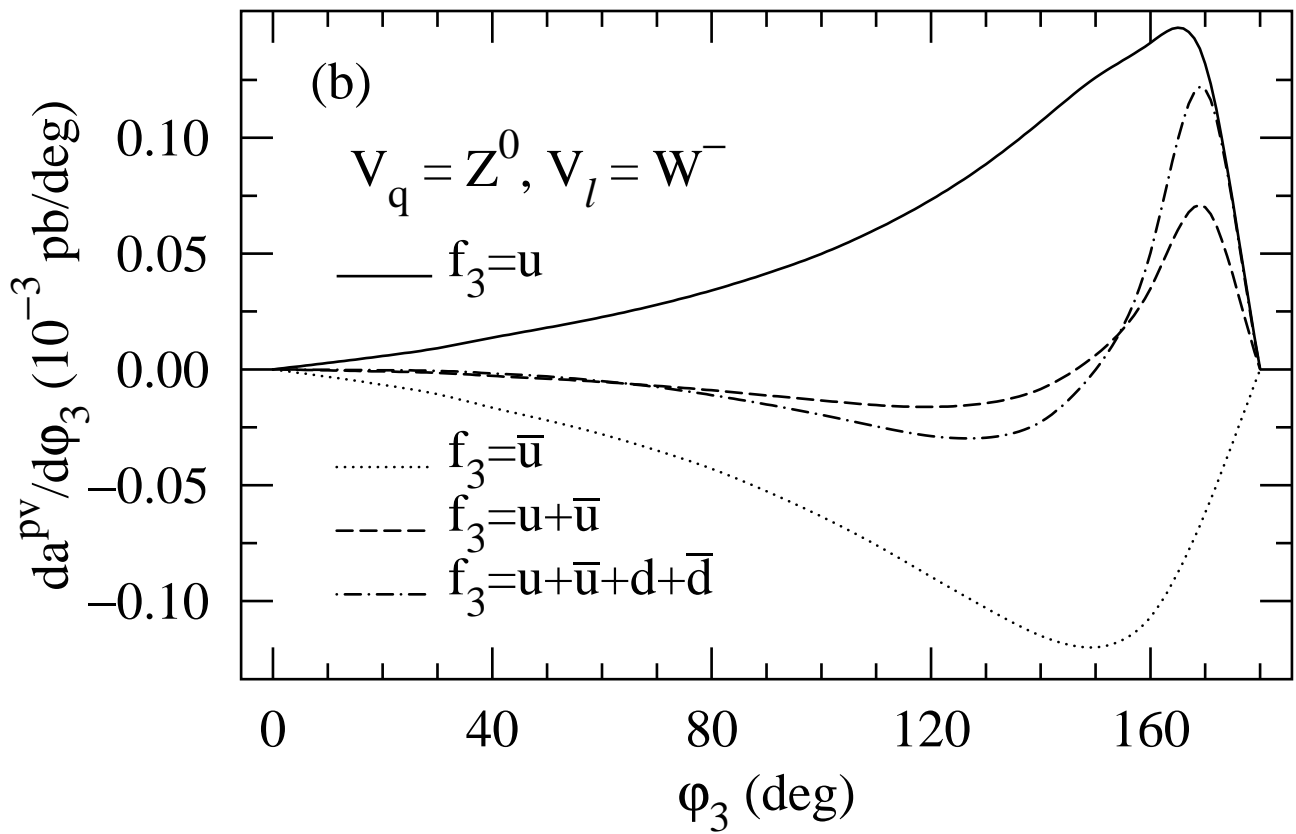
\begin{figure}
\caption{
Differential parton-level asymmetries in the process $q\bar q
\rightarrow W^-q\bar q$. 
(a) Contributions from 2 subprocesses with $V_q=W^+$ and 2 with
$V_q=Z^0$.
$\phi_3$ is the azimuthal angle of the quark in the final state.
(b) Contributions from subprocesses with $V_q=Z^0$ decaying to
a $q\bar q$ with flavor $f_3$, and 4 different assumptions concerning
the distinguishability of the parton observed at azimuth $\phi_3$.
}
\label{daWfig}
\end{figure}

Fig.\ \ref{daWfig}a shows differential asymmetries for
the case $V_l=W^-$.  There are four possible subprocesses, two with
$V_q=Z^0$ and two with $V_q=W^+$.  
The asymmetry is largest when the quark is emitted almost diametrically
opposite in azimuth to the $W$.  It is zero when $\phi_3=0$ or
180$^\circ$ because the 5 particles involved in the collision are then
coplanar, and the parity-violating invariant $x_{1234}$ vanishes.
In Fig.\ \ref{daWfig}b we present asymmetries for the two processes with
$V_q=Z^0$, with various assumptions.  The solid curve, reproduced from 
Fig.\ \ref{daWfig}a, is for the
subprocess $d\bar{u}\rightarrow u\bar{u}W^-$ with the $u$ quark
observed at angle $\phi_3$.  The dotted curve is for observation of
the $\bar{u}$ at angle $\phi_3$.  If one is unable to distinguish
quarks from antiquarks, one must add the corresponding cross sections:
this is shown by the dashed curve.  The subprocess
$d\bar{u}\rightarrow d\bar{d}W^-$ can also contribute if one cannot
distinguish $u$ from $d$ quark jets: the dot-dashed curve shows the sum
of the asymmetries in which the jet observed at angle $\phi_3$ can
come from $u$, $\bar{u}$, $d$ or $\bar{d}$.  The figure shows that the
inability to distinguish flavors experimentally 
in a jet may reduce the asymmetry, but
does not necessarily reduce it to zero.

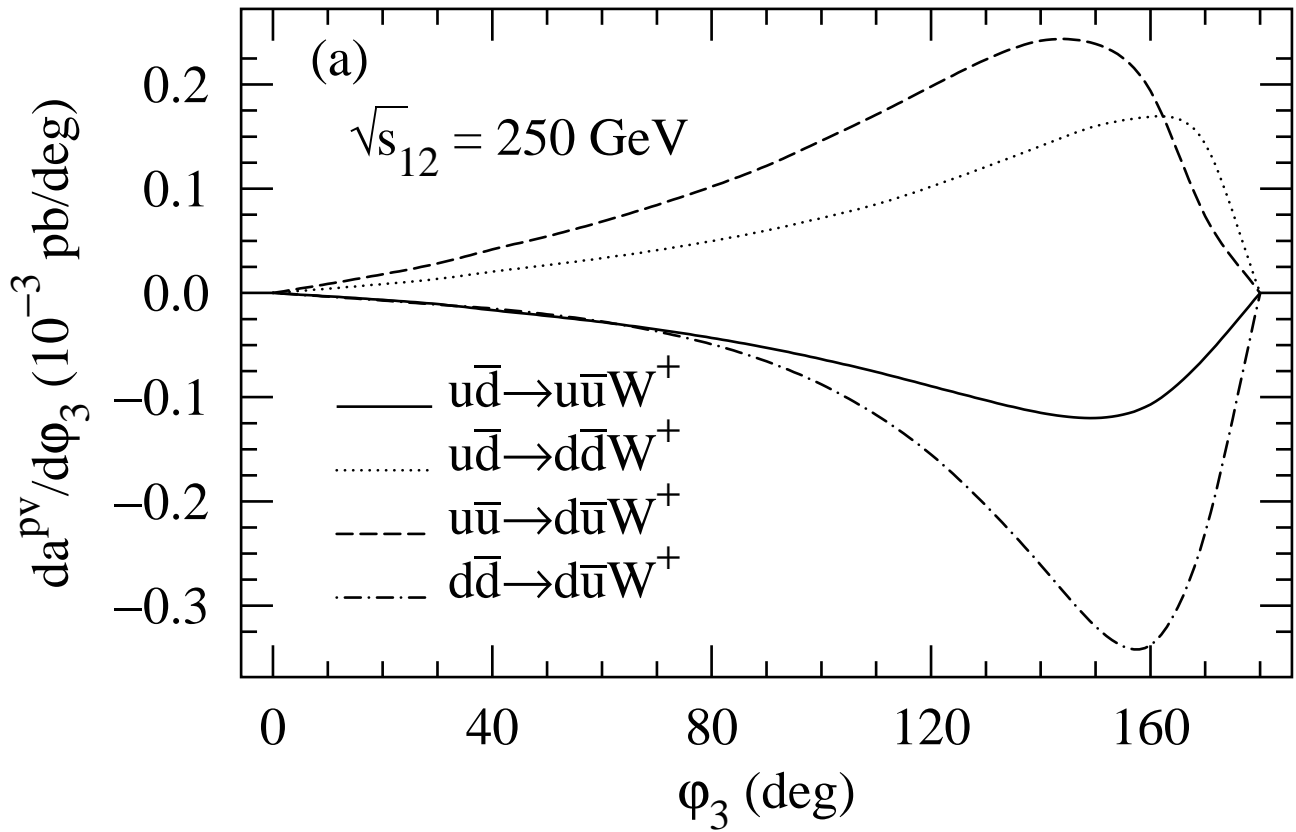
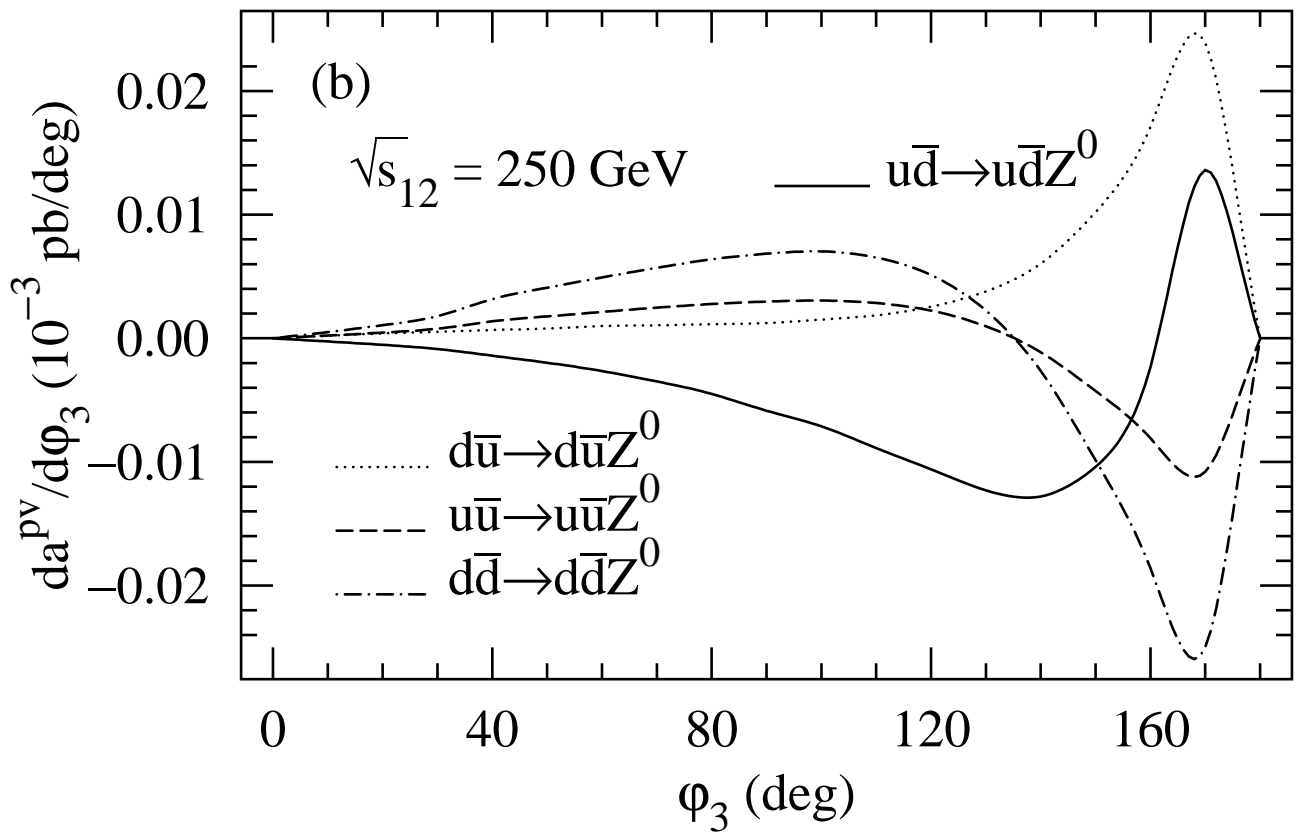
\begin{figure}
\caption{
Differential asymmetries of the type in Fig.\ \protect\ref{daWfig}a,
for (a) $W+$ production, and (b) $Z^0$ production.
}
\label{daWZfig}
\end{figure}

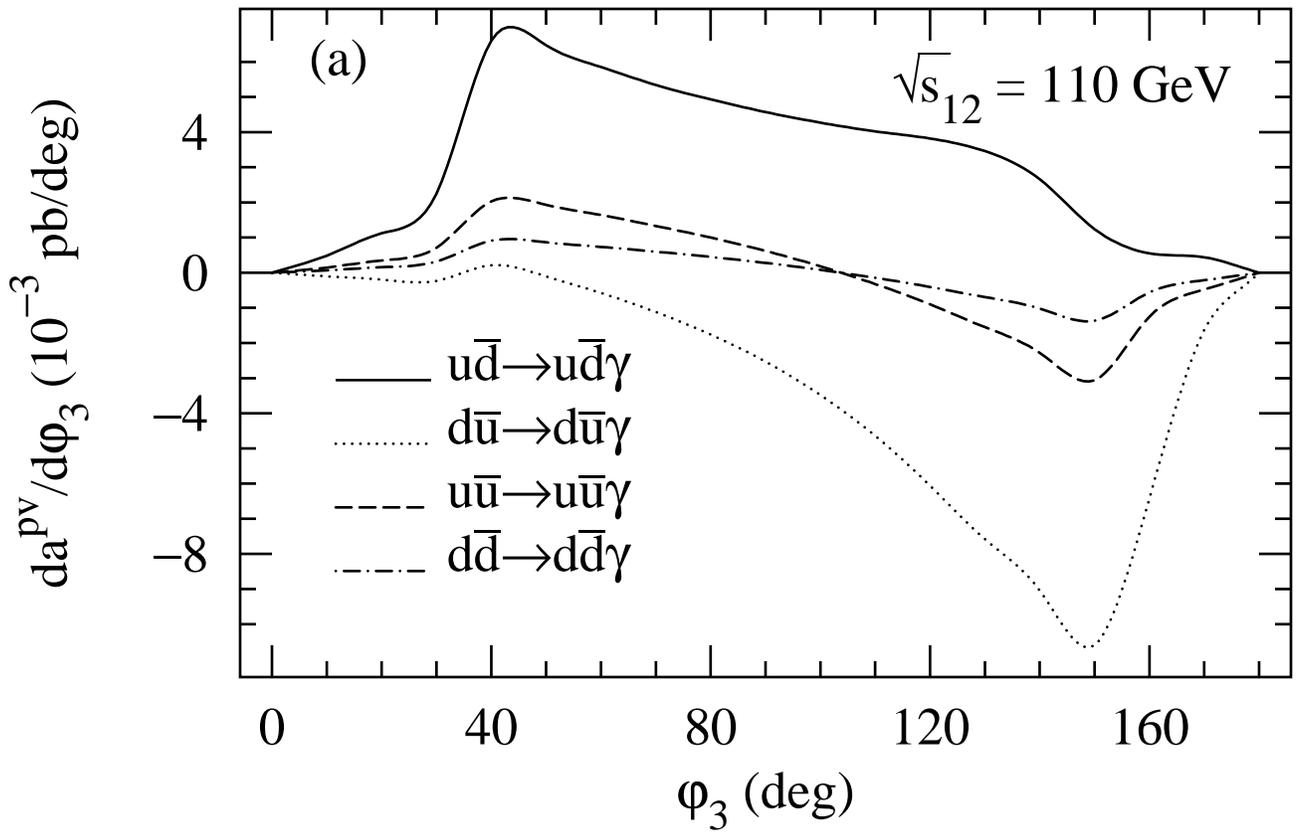
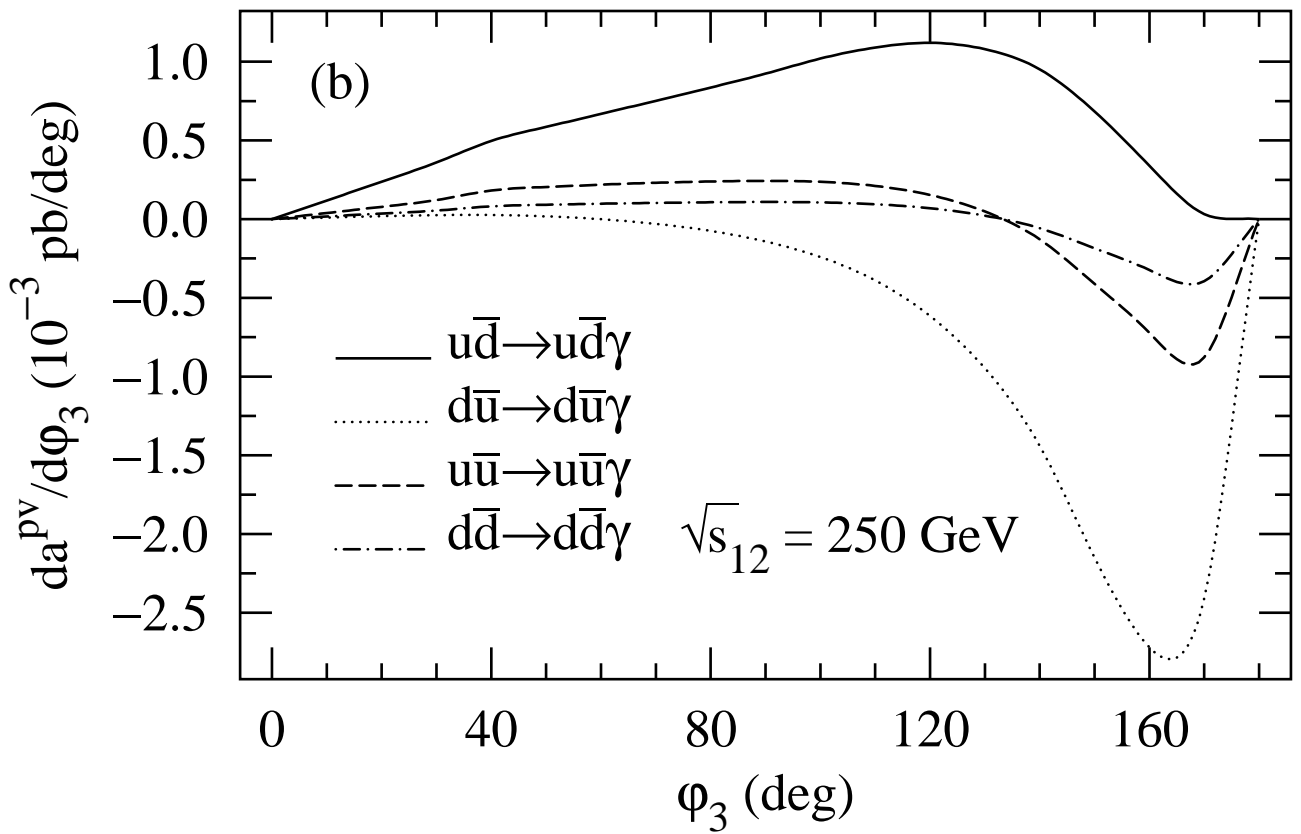
\begin{figure}
\caption{
Differential asymmetries of the type in Fig.\ \protect\ref{daWfig}a,
for real photon production (a) at $\protect\sqrt{s_{12}}=110$ GeV just above
threshold for the process, and (b) at $\protect\sqrt{s_{12}}=250$ GeV.
}
\label{daphfig}
\end{figure}

In Fig.\ \ref{daWZfig} we present differential asymmetries for the cases
$V_l = W^+$, $Z^0$, and in Fig.\ \ref{daphfig} for $V_l=\gamma$.  
From Eqs.\ (\ref{J3WWeq})-(\ref{J3WZ2eq})
we see that the processes contributing to Figs.\ \ref{daWfig}a and 
\ref{daWZfig}a are
related by CP conjugation.  We therefore expect the asymmetries in
Fig.\ \ref{daWfig}a to be the same as asymmetries for $W^+$ production if the
produced {\sl antiquark} is observed at angle $\phi_3$, the relative
minus sign between the cross sections being compensated by the fact
that we assign the incoming quark the momentum $p_1$ in both cases.
However, Fig.\ \ref{daWZfig}a shows asymmetries 
with the quark observed at angle
$\phi_3$: these are opposite in sign and in general different in
magnitude from the corresponding charge conjugate processes in
Fig.\ \ref{daWfig}a. It is interesting that the asymmetries for the processes
$u\bar{u}\rightarrow u\bar{d}W^-$ and $d\bar{d}\rightarrow
d\bar{u}W^+$ are numerically almost equal, as are the asymmetries for
the CP conjugate pair of processes.  
The asymmetries in Fig.\ \ref{daWZfig}b for
$Z^0$ production are interesting in that they exhibit zeros at
intermediate values of $\phi_3$.  The figures also show that the
asymmetries when one or both of the vector bosons are $Z$s tend to be
smaller than when they are $W$s. 
Finally, Fig.\ \ref{daphfig} shows that the
asymmetries in the case of real photon production are almost an order
of magnitude larger than for production of massive vector boson pairs.
This difference is partly due to the fact that the threshold for real
photon production is lower than that for vector boson pair production.
Fig.\ \ref{daphfig}a is plotted for a subprocess energy of 110 GeV around which
the asymmetries tend to be largest.  Fig.\ \ref{daphfig}b shows that the
asymmetries at 250 GeV, which is a little above the weak boson pair
production threshold, are similar in magnitude to those in
Figs.\ \ref{daWfig} and \ref{daWZfig}.

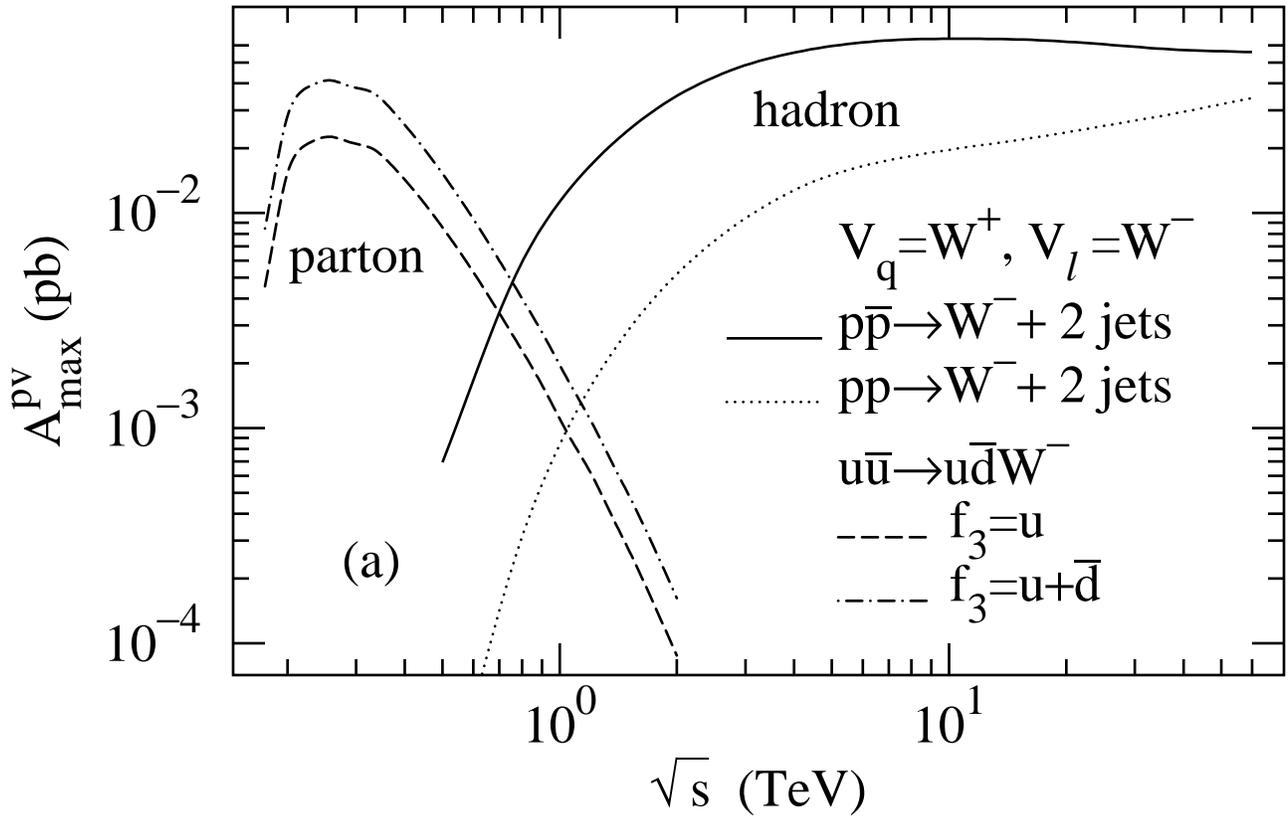
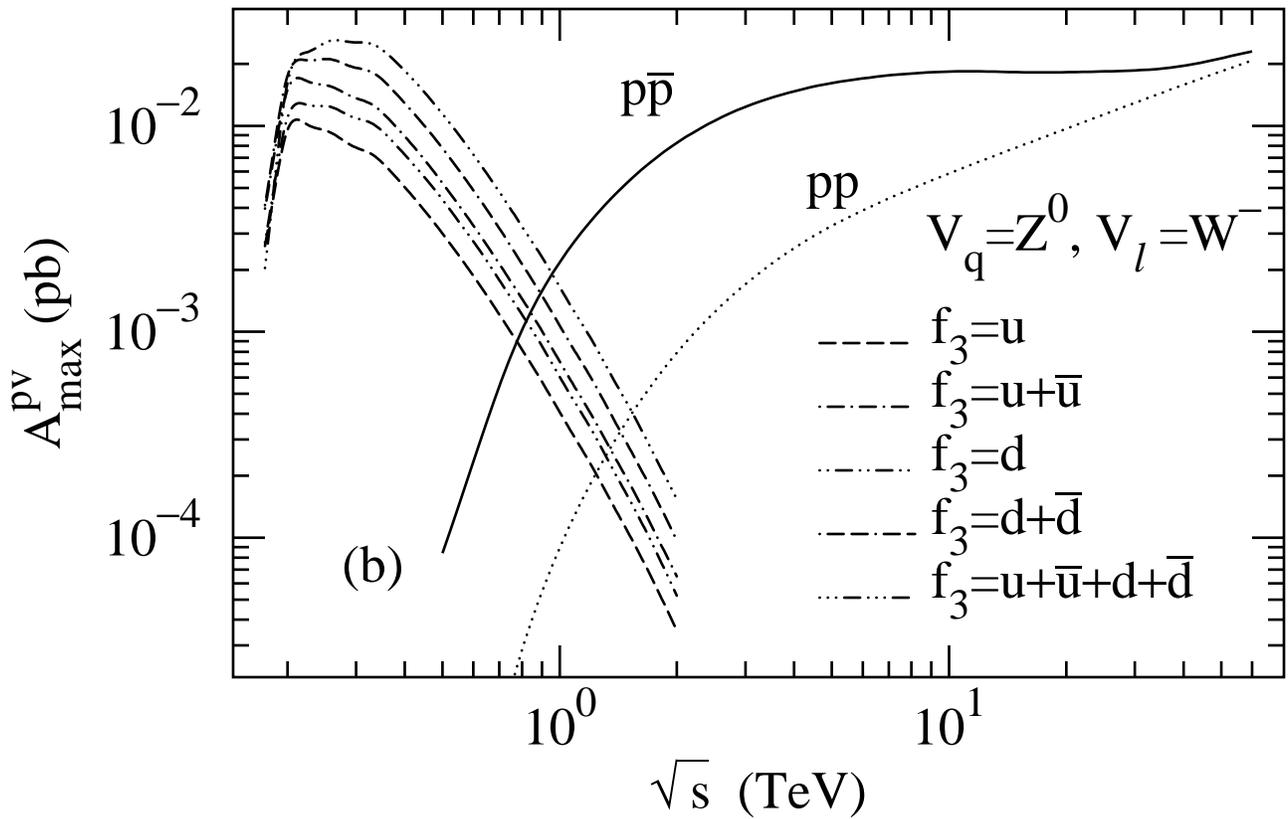
\begin{figure}
\caption{
Parton- and hadron-level maximum asymmetries as defined in 
Eq.\ (\protect\ref{Amax2eq}) in $W^-$ production for (a) $V_q=W^+$
and (b) $V_q=Z^0$. The parton level asymmetries shown illustrate
various assumptions concerning the distinguishability of the final
state quark jets.
The hadron-level asymmetries are ``flavor-blind''.
}
\label{AWfig}
\end{figure}

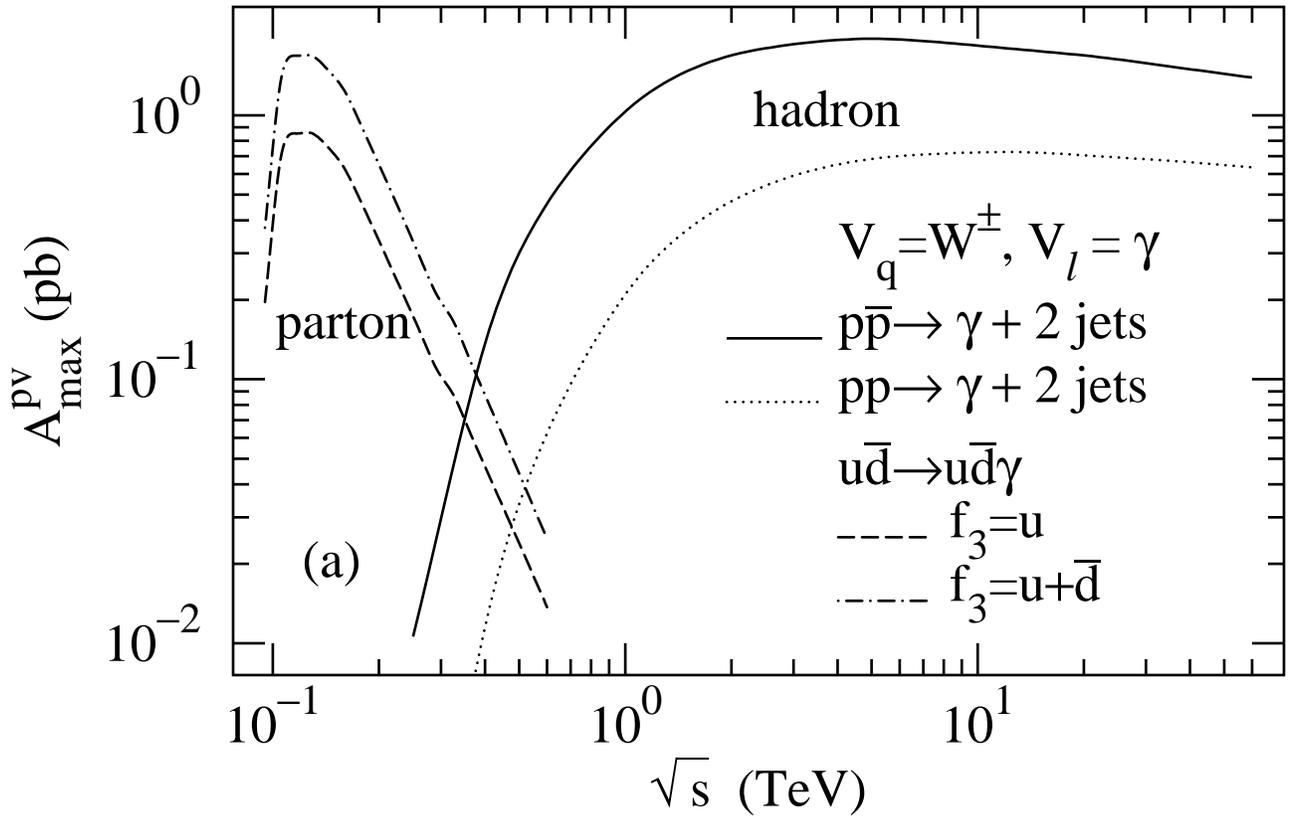
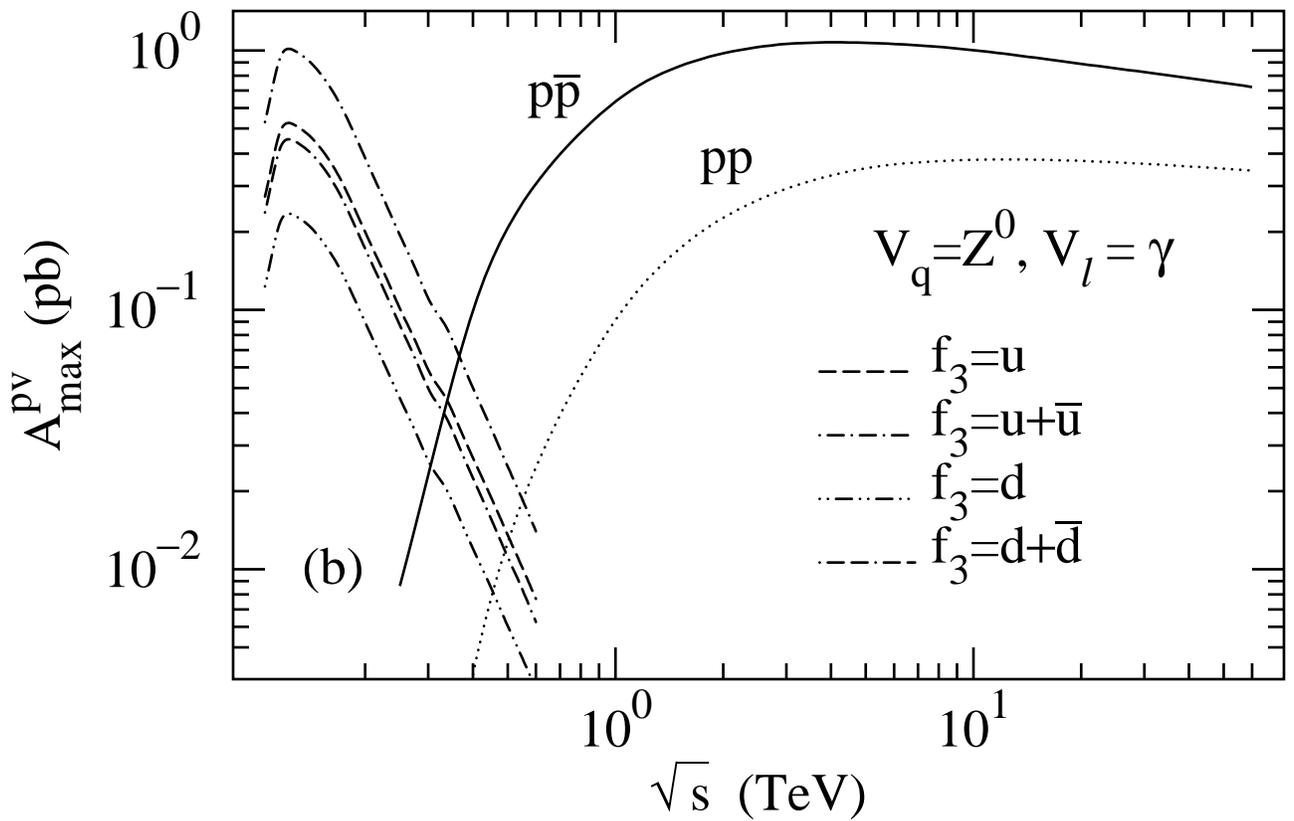
\begin{figure}
\caption{
Maximum asymmetries in real photon production. (a) $V_q=W^\pm$. At the parton
level, asymmetries for $u\bar d\rightarrow u\bar dZ^0$, i.e., for
$V_q=W^-$, are the same as for $d\bar u\rightarrow d\bar uZ^0$, and are
not shown. Both subprocesses are included at the hadron-level since
$W^\pm$ have the same mass. (b) $V_q=Z^0$.
}
\label{Aphfig}
\end{figure}

We next present results in Figs.\ \ref{AWfig}-\ref{Aphfig}
for the maximum observable
cumulative asymmetries $A^{\rm pv}_{\max} $ that were defined in
Eqs.\ (\ref{Amaxeq}) and (\ref{Amax2eq}).  
(The results for $pp$ and $p\bar{p}$ collisions,
which are presented for convenience in the same figures, will be
discussed in the following sub-section.)  
We remind the reader that the maximum asymmetry is an upper
bound on the cumulative observable asymmetry, a bound which can be
saturated only if one surrounds the interaction region with detectors
that are sufficiently small that the fully differential asymmetry (in
all kinematic variables and not just $\phi_3$) does not change sign
within the acceptance of each of them.  These figures show that the
asymmetries are largest at a few tens of GeV above the threshold for
production of the vector boson pair, and that they then fall quite
rapidly with energy.
We first discuss the two parton-level curves in Fig.\ \ref{daWfig}a.
The integrated value of the
differential asymmetry for the process $u\bar{u}\rightarrow
u\bar{d}W^-$ shown by the dashed curve in 
Fig.\ \ref{daWfig}a is $-0.0201$ pb, and the
corresponding value of $A^{\rm pv}_{\max} $ from the dashed curve in 
Fig.\ \ref{AWfig}a is $0.0225$ pb.
This comparison shows that oscillations in sign of the
asymmetry in variables other than the angle $\phi_3$ are negligible.
The dot-dashed curve in
Fig.\ \ref{AWfig}a shows the ``flavor-blind'' asymmetry for this subprocess.
We emphasize the fact that in forming this asymmetry, the fully-differential
cross sections for $u$ and for $\bar d$ to be observed at azimuth
$\phi_3$ are added {\em before} taking the absolute magnitude in
Eq.\ (\ref{Amax2eq}).
The corresponding parton-level cross sections for the process
$d\bar d\rightarrow u\bar dW^-$ are numerically very close to the
cross sections for $u\bar u\rightarrow u\bar dW^-$, and are therefore not
shown.
Fig.\ \ref{AWfig}b shows $A^{\rm
pv}_{\max} $ for the processes $d\bar{u}\rightarrow Z^0W^-$,
with $V_q=Z^0\rightarrow u\bar u$ or $d\bar d$.
The five parton-level curves give maximum asymmetries with various
assumptions concerning distinguishability of quark jets.
It is apparent that there is very little cancellation between asymmetries for
processes with quarks and antiquarks or quarks with different
weak-isospin quantum numbers; thus if any two of these processes
produce asymmetries of opposite sign, these must occur in different
regions of phase space. 
Maximum asymmetries in the production of a $Z^0$ are presented in
Fig.\ \ref{AZfig}, and for photon production in Fig.\ \ref{Aphfig}.
A comparison of Figs.\ \ref{AWfig}-\ref{Aphfig} shows that the
largest asymmetries occur in the case of a real photon and in
particular when $V_q=W^\pm$.

\subsubsection{Hadron-level asymmetries}

The parton-level asymmetries presented in the preceding section will
produce asymmetries at the hadron-level in the processes $p\bar{p}$ or
$pp\rightarrow V_l$ + 2 Jets according to Eq.\ (\ref{hadxceq}).  
In obtaining
numerical predictions for these asymmetries we shall assume that the
invariant mass of the 2-jet system can be reconstructed to within
approximately 5 GeV so that jets arising from the resonant decay of a
$W$ can be distinguished from jets arising from the decay of a $Z$.
We will take into consideration the contributions of five quark
flavors $u$, $d$, $s$, $c$, and $b$, using the MRSB structure function
parametrizations \cite{MRSB} to compute parton densities
$f_{a/h}(x,Q^2)$. We will assume that the flavors of parton jets
cannot be measured, and will sum differential asymmetries over all
possible incoming and outgoing light quark (and antiquark) flavors.

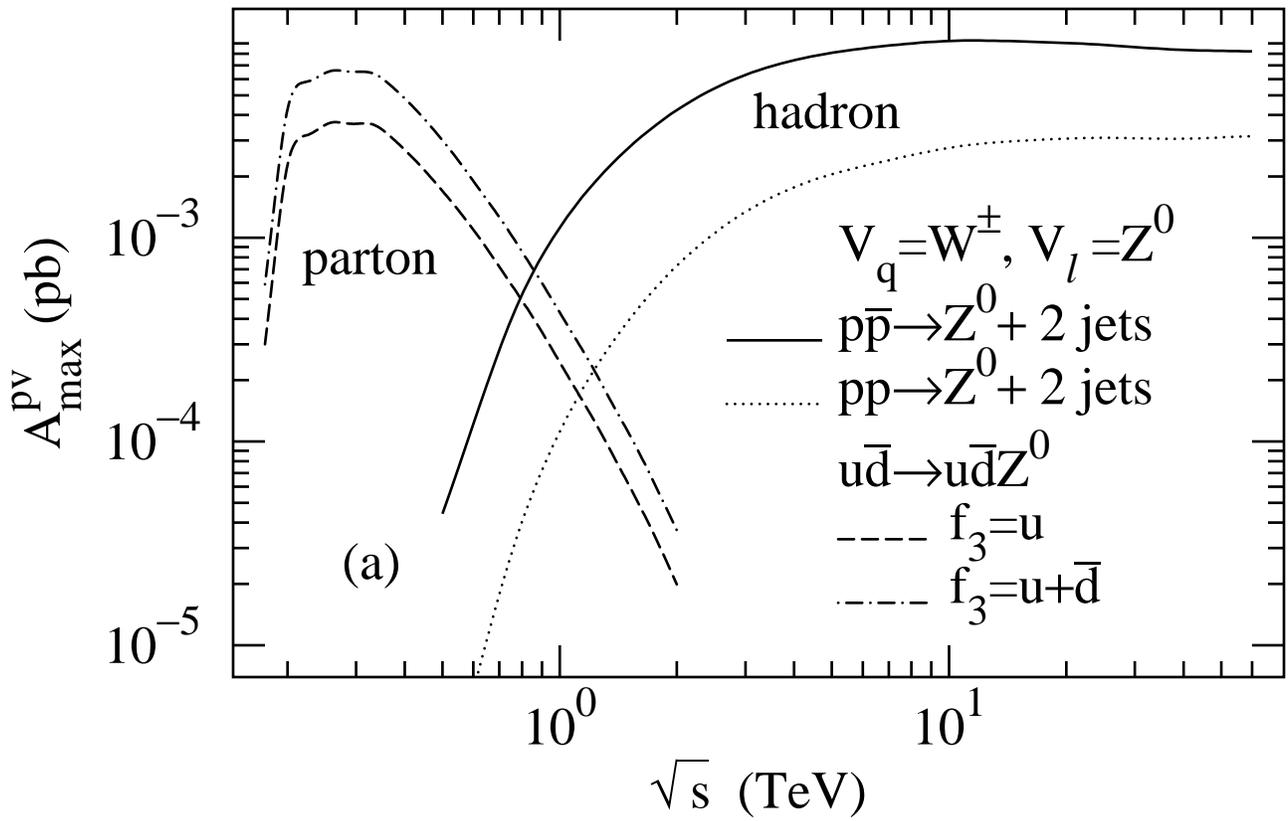
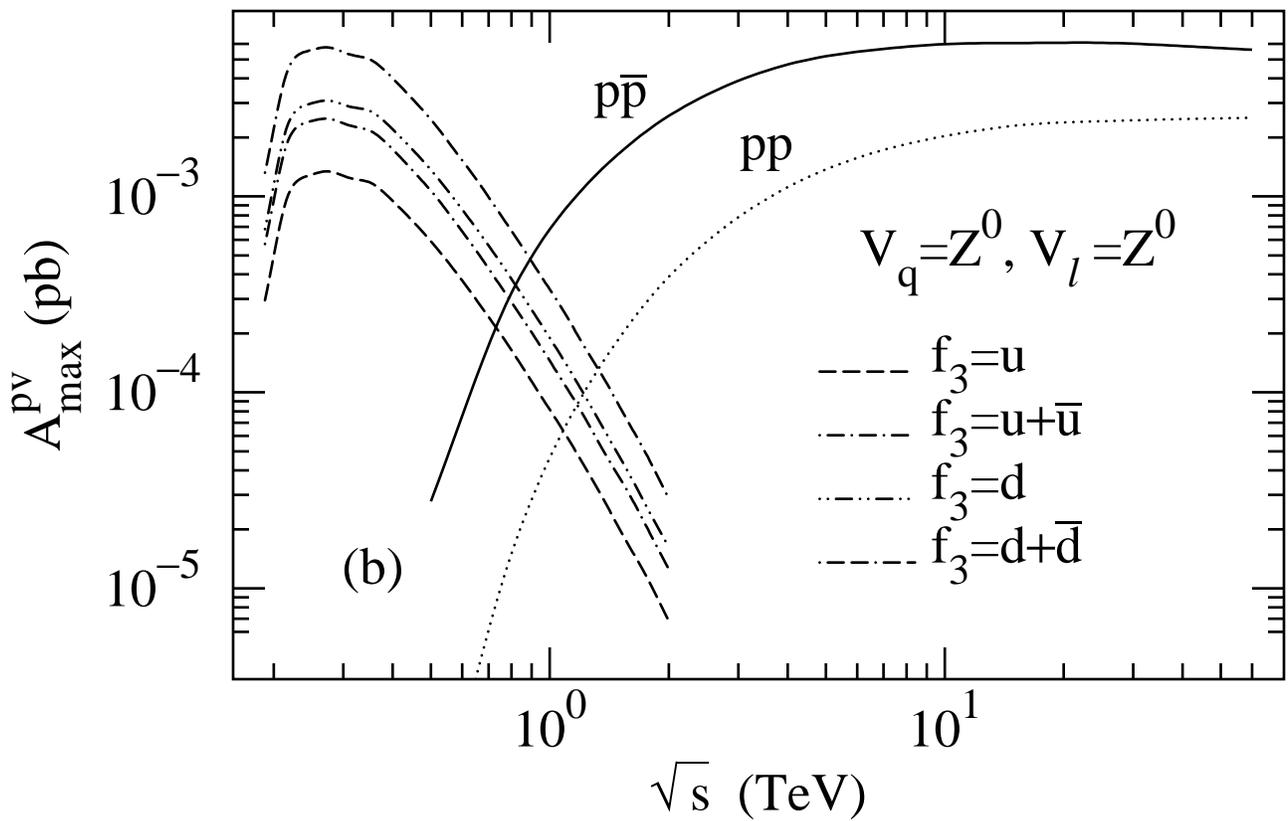
\begin{figure}
\caption{
Maximum asymmetries in $Z^0$ production. (a) $V_q=W^\pm$. At the parton
level, asymmetries for $u\bar d\rightarrow u\bar dZ^0$, i.e., for
$V_q=W^-$, are the same as for $d\bar u\rightarrow d\bar uZ^0$, and are
not shown. Both subprocesses are included at the hadron-level since
$W^\pm$ have the same mass. (b) $V_q=Z^0$.
}
\label{AZfig}
\end{figure}

Figs.\ \ref{AWfig}-\ref{Aphfig} show maximum 
observable cumulative asymmetries $A^{\rm
pv}_{\max}$ for $V_l=W$, $Z$ or $\gamma$.  The asymmetries begin to be
appreciable at values of $\sqrt{s}$ approximately 6 times larger than
the threshold for pair production of the corresponding pair of vector
bosons.  They are initially larger for $p\bar{p}$ collisions than for
$pp$ collisions because the contributing subprocesses involve
quark-antiquark annihilation.  As $\sqrt{s}$ increases, the densities
of sea quarks increase relative to those of valence quarks, and the
$pp$ and $p\bar{p}$ curves get closer to one another.  
A comparison of Figs.\ \ref{AWfig} and \ref{AZfig}
with Fig.\ \ref{Aphfig} would suggest that the prospects for observing
parity-violating asymmetries might be best in the case of real photon
production; however, the QCD background is also larger for real photon
production than it is for $W$ or $Z$ production, as we shall see in the
next sub-section.
It should be
emphasized that the cross sections in 
Figs.\ \ref{AWfig} and \ref{AZfig} for
production of a ``real'' $W$ or $Z$ have been multiplied by the
appropriate leptonic branching ratio (see Eq.\ (\ref{sigmahateq}))
for a single leptonic decay mode.
If more than one leptonic decay mode is observed, the signal will be
correspondingly enhanced.

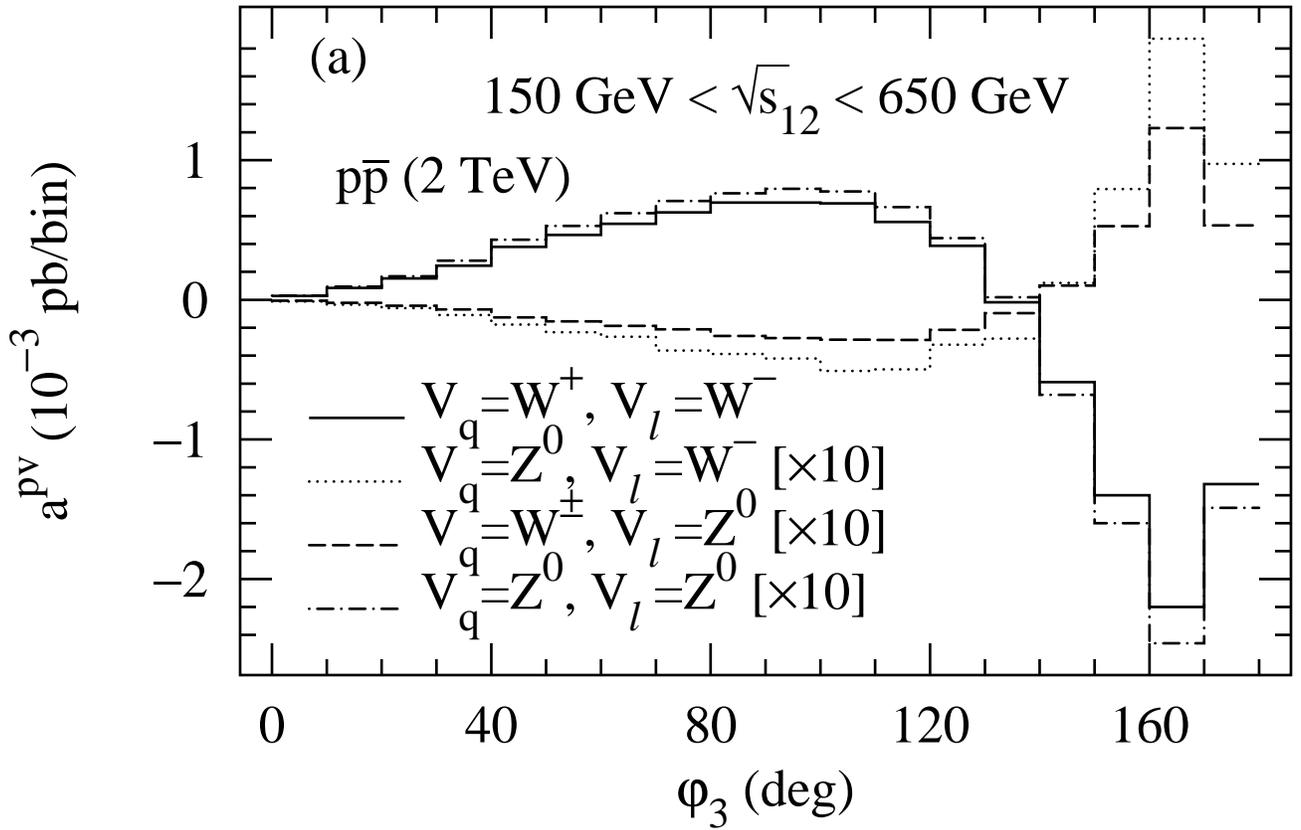
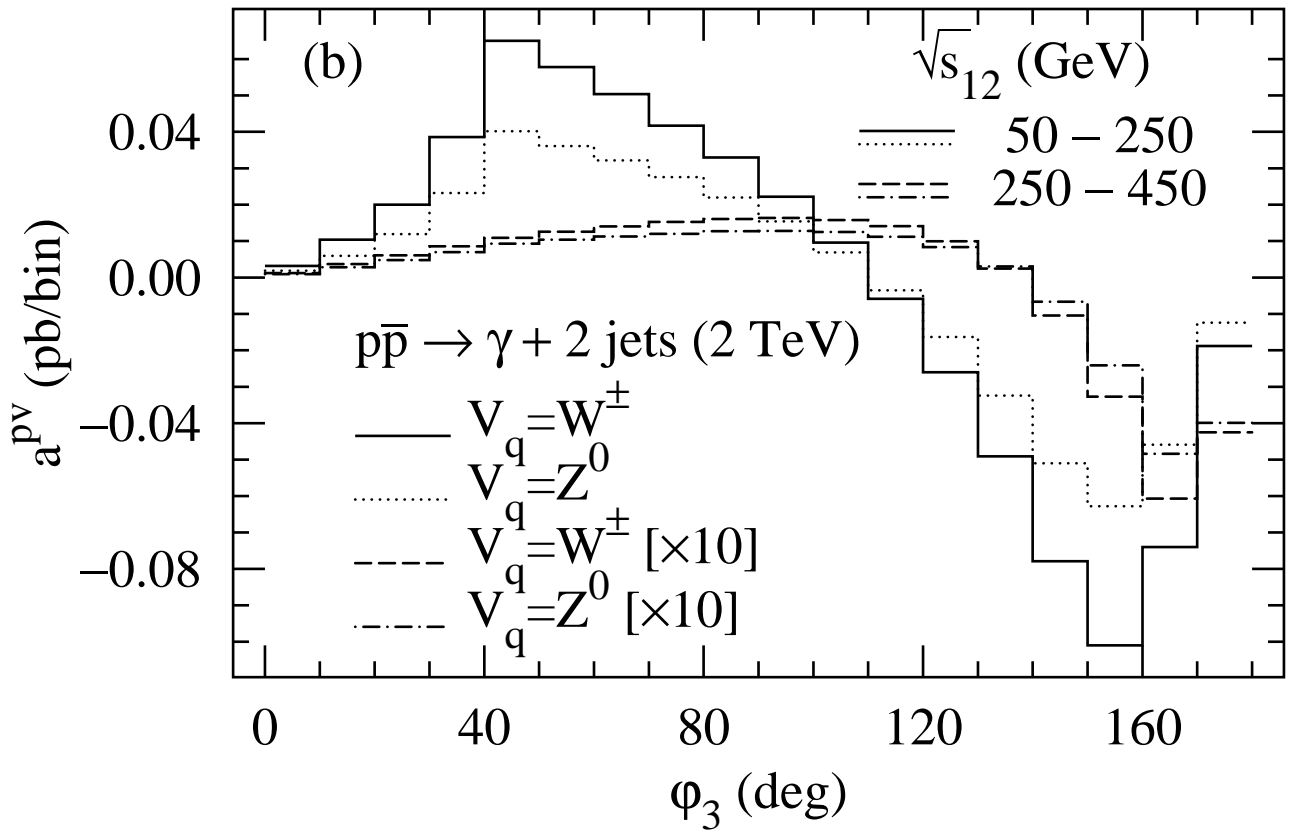
\begin{figure}
\caption{
Binned asymmetries in $p\bar p$ collisions at 2 TeV (a) in $W^-$ and $Z^0$
production for bins of width 500 GeV in the subprocess invariant mass
$\protect\sqrt{s_{12}}$, and (b) in real photon production for two
different choices of bins in $\protect\sqrt{s_{12}}$. 
The notation $[\times10]$ indicates that the histogram has been 
multiplied by a factor of 10 to show it on the same scale as the others.
Note that the algebraic sum of $a^{\rm pv}$ values in $\phi_3$ bins of any
each histogram is zero within numerical uncertainties.
}
\label{bin2fig}
\end{figure}

In Fig.\ \ref{bin2fig}a we present binned asymmetries as defined 
in Eq.\ (\ref{apveq}) for
$W^-$ and $Z$ production in $p\bar{p}$ collisions at 2 TeV.  We have
chosen bins of width $10^\circ$ in the angle $\phi_3$ and 500 GeV in
the subprocess invariant mass $\sqrt{s_{12}}$.  
Fig.\ \ref{bin2fig}b shows corresponding results for real photon production.
We note with reference to Fig.\ \ref{pvfig} that 
the proton beam has been chosen to define
the positive $z$ direction: thus, with a right handed coordinate
system, a positive asymmetry implies an excess of events in which the
jet assigned a momentum $p_3$ and azimuthal angle $\phi_3$ has a
positive $y$ component of momentum. Obviously, the existence of a
non-zero asymmetry depends on the fact that the proton and antiproton
beams differ from one another; specifically, they differ in their
valence quark content, and this produces a physical distinction
between the positive and negative $z$ directions, and thus the
handedness of the coordinate system is related to physical
observables.  If one sums the asymmetries represented by any of the
histograms in Fig.\ \ref{bin2fig} algebraically, one obtains a zero
net asymmetry.  This is due to the fact that we have assumed that the
jets are identified only by their momenta: momentum conservation
requires that if one of the jets has a positive $y$ component of
momentum, then the other has a negative $y$ component; if the bin
encompasses the whole of the hemisphere with $p_3^y \geq 0$, no
observed quantum number distinguishes between the two hemispheres, and
the asymmetry must vanish.  We note finally that the cumulative
asymmetries as defined by Eq.\ (\ref{Aeq}) for any of the histograms in these
figures are typically of order 20\% of the corresponding maximum
observable asymmetries in Figs.\ \ref{AWfig}-\ref{Aphfig} 
for $p\bar{p}$ collisions at 2 TeV. For example, the cumulative asymmetry
$A^{\rm pv}$ from the solid-line histogram in Fig.\ \ref{bin2fig}a is
0.010 pb, while $A^{\rm pv}_{\rm max}$ from the solid-line curve in
Fig.\ \ref{AWfig} at 2 TeV is 0.033 pb. There are evidently
considerably larger cancellations at the hadron level than at the parton
level in the bins we have chosen. 
These cancellations presumably occur in the sums over parton
types and the integrations over parton momentum fractions,
see Eq.\ (\ref{hadxceq}), required to compute the hadronic cross sections.
It is rather remarkable nevertheless that such a substantial signal of
the parton-level asymmetry survives at the hadron level.

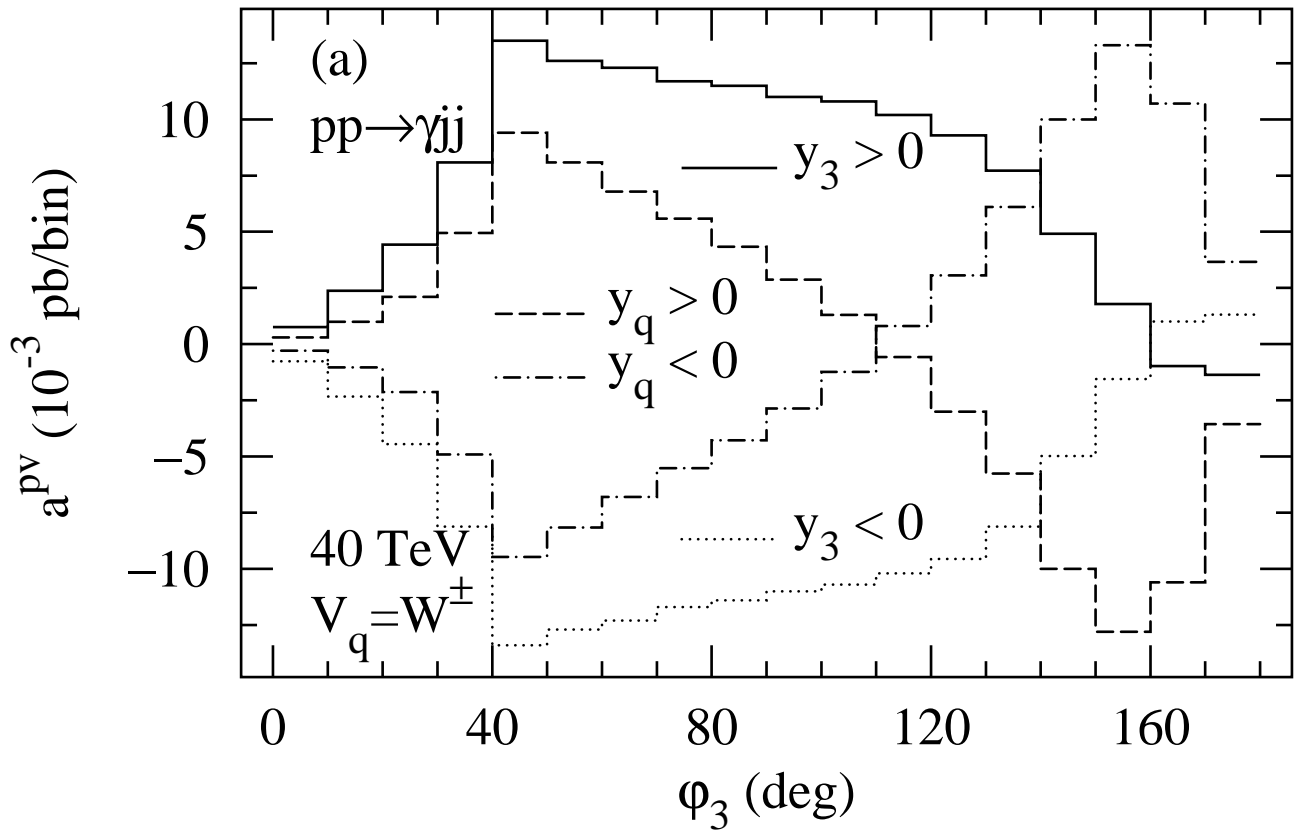
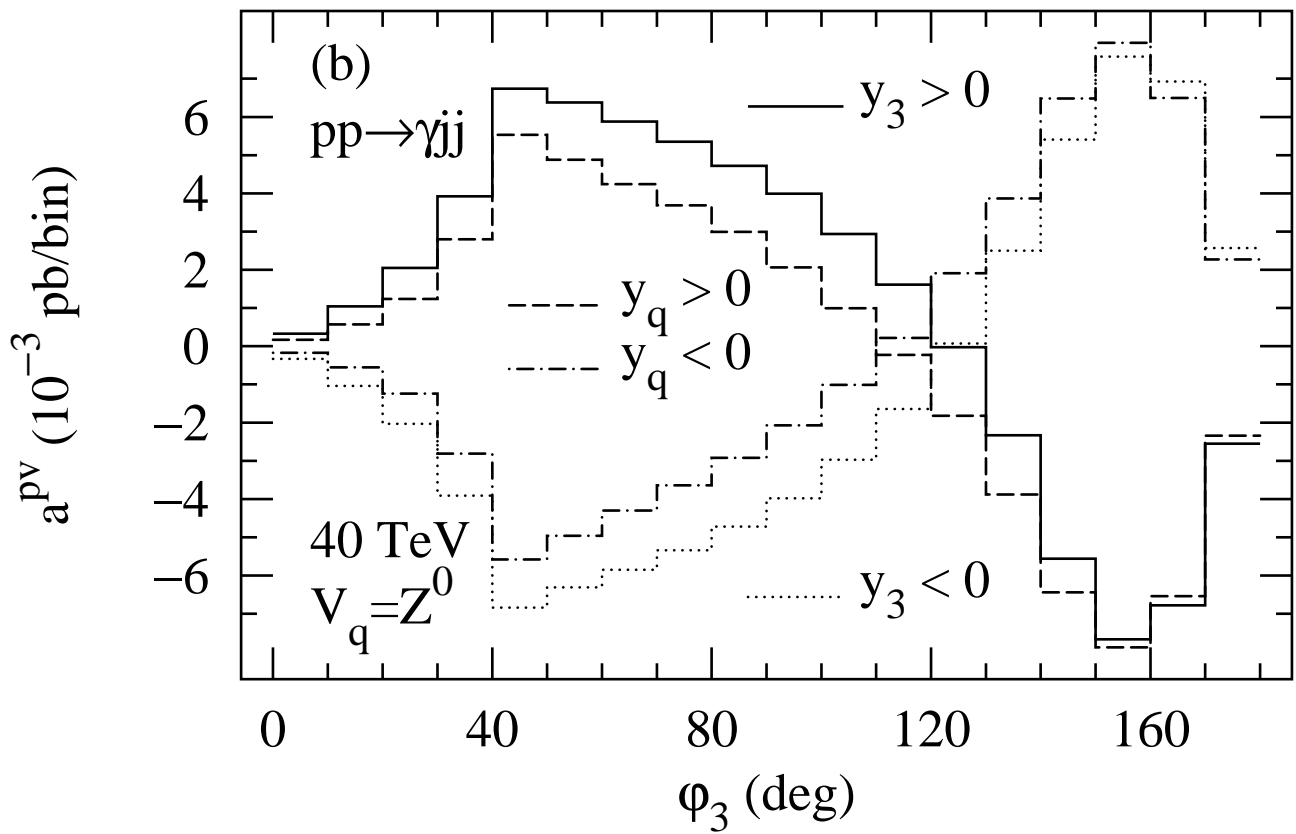
\begin{figure}
\caption{
Binned asymmetries for $pp\rightarrow\gamma$ + 2 jets at 40 TeV with
(a) $V_q=W^\pm$ and (b) $V_q=Z^0$. The subprocess energy is restricted
to the bin 80 GeV $<$ $\protect\sqrt{s_{12}}$ $<$ 280 GeV. The rapidities
$y_3$ and $y_q$ of the jet at azimuth $\phi_3$ and the photon are
further restricted as indicated to produce a non-vanishing asymmetry.
}
\label{bin40phfig}
\end{figure}

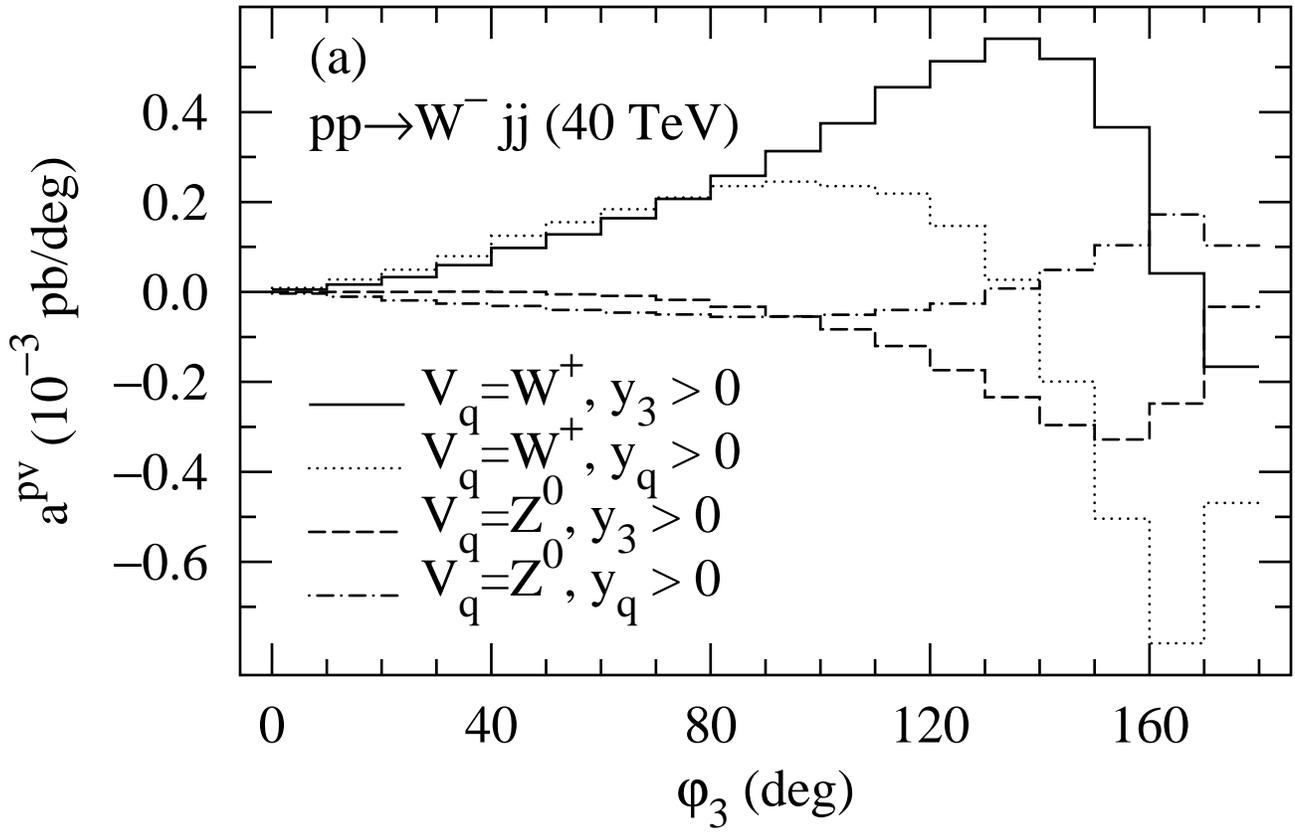
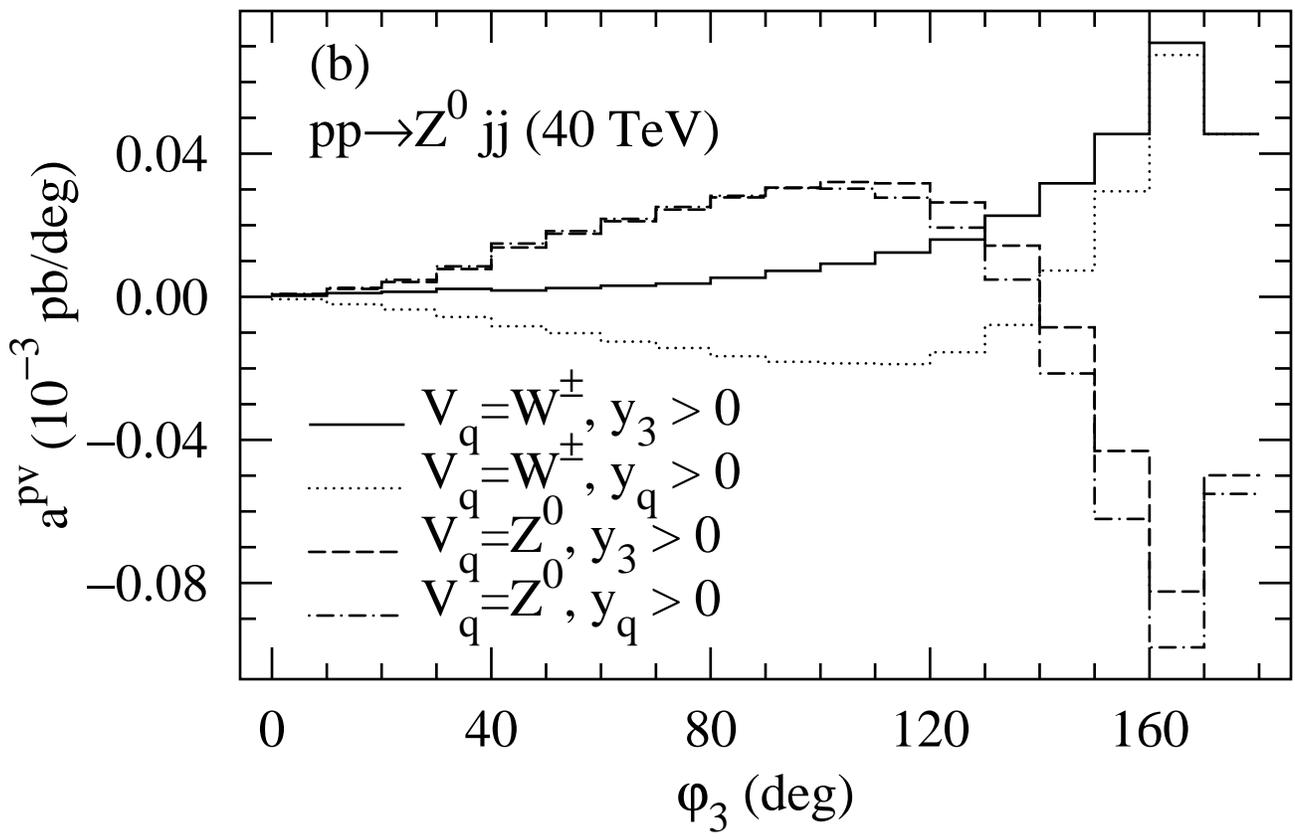
\begin{figure}
\caption{
Binned asymmetries at 40 TeV for (a) $pp\rightarrow W^-$ + 2 jets, and
(b) $pp\rightarrow Z^0$ + 2 jets. The subprocess energy is restricted
to the bin 150 GeV $<$ $\protect\sqrt{s_{12}}$ $<$ 350 GeV. Rapidities
$y_3$ and $y_q$ are restricted as shown to produce non-zero asymmetries.
}
\label{bin40WZfig}
\end{figure}

We next discuss asymmetries in $pp$ collisions.  It is apparent in this
case that the colliding beams will not provide a means for physically
distinguishing the positive and negative $z$ directions, since their
valence and sea quark contents are identical.  Thus if there were no
other observable that distinguishes the $\pm z$ directions, as was for
instance the case for the binned asymmetries in $pp$ collisions
presented in Fig.\ \ref{bin2fig}, one would observe no parity violation.
To observe a non-zero asymmetry, there must exist an observable that is
related to the handedness of the coordinate system.  The simplest way to
introduce such an observable is to demand that the vector boson have a
positive $z$ component of momentum in the laboratory (center-of-mass
frame of the colliding beams) in which the positive $z$ axis has been
chosen to coincide with one of the two proton beams chosen arbitrarily.
In other words, we choose bins for which the rapidity $y_q$ is $\geq0$. 
We could also choose bins in which $y_q\leq0$, or in general, bins that
are not symmetric in rapidity.  
Fig.\ \ref{bin40phfig} shows the binned asymmetries
that one obtains for $pp$ collisions at 40 TeV in the case of real
photon production with $y_q\geq0$ as well as for $y_q\leq0$.  It is
evident that the sum of the two histograms vanishes (within the
uncertainties of the Monte Carlo integration).  It is evidently also
possible to choose bins that are asymmetric in the rapidity $y_3$ of the
jet assigned the momentum $p_3$, and these asymmetries are also shown in
the figure.  In Fig.\ \ref{bin40WZfig} results are presented for $W^-$ and
$Z^0$ production in $pp$ collisions at 40 TeV.

\subsubsection{Parity-violating signal and QCD background}

The parity-violating asymmetries presented above will in practice 
be concealed in a substantial
parity-conserving background of electroweak boson + 2 jet
events \cite{background}.
This QCD background completely overwhelms the signal from electroweak
boson pair production in which one of the two bosons is observed in
its leptonic decay modes, and the other decays to two hadronic jets:
Stirling, Kleiss and Ellis \cite{stirling} found that the signal to
background ratio varies from $\sim1/30$ at 2 TeV to $\sim1/150$ at
40 TeV in $p\bar p$ collisions.
We find that similar signal to background ratios occur for the
chromo-electroweak interference contributions.
It should nevertheless be possible in principle to observe the interference
contributions by measuring parity-violating asymmetries to which the
QCD background does not contribute.
In practice, there will be statistical fluctuations in the background
which will tend to obscure such a signal: it is therefore essential to
estimate the signal to background ratio in order to decide whether
such asymmetries will be measurable.

To estimate the QCD background we use a set of compact analytic cross 
sections computed by R.K.\ Ellis and one of us (R.J.G.)
\cite{oregonProcs}.
We apply the same set of cuts given in Eqs.\ (\ref{cutseq}) 
to the QCD background events that we have used to compute the asymmetries.
In computing the latter, we have used the narrow width approximation
defined in Eq.\ (\ref{narrowWidthEq}) for the propagator of the
the electroweak boson $V_q$ which decays to two jets.
There is of course no such propagator in the amplitudes for the QCD
background, and the observed events are therefore expected to vary
quite slowly as a function of the invariant mass $\sqrt{s_{34}}$ of
the two jets with a small blip coming from the $V_q$ propagator in
the pair-production and interference contributions at
$\sqrt{s_{34}}=M_{V_q}^2$.
To extract this blip, events need only be sampled in
a small interval $(M_{V_q}-\Delta)^2 \leq s_{34} \leq (M_{V_q}+\Delta)^2$, 
with $\Delta \simeq \Gamma_{V_q}$.
In computing the background, we therefore apply an additional cut
$(M_{V_q}-\Gamma_{V_q})^2 \leq s_{34} \leq (M_{V_q}+\Gamma_{V_q})^2$
on the jet-jet invariant mass $\sqrt{s_{34}}$.

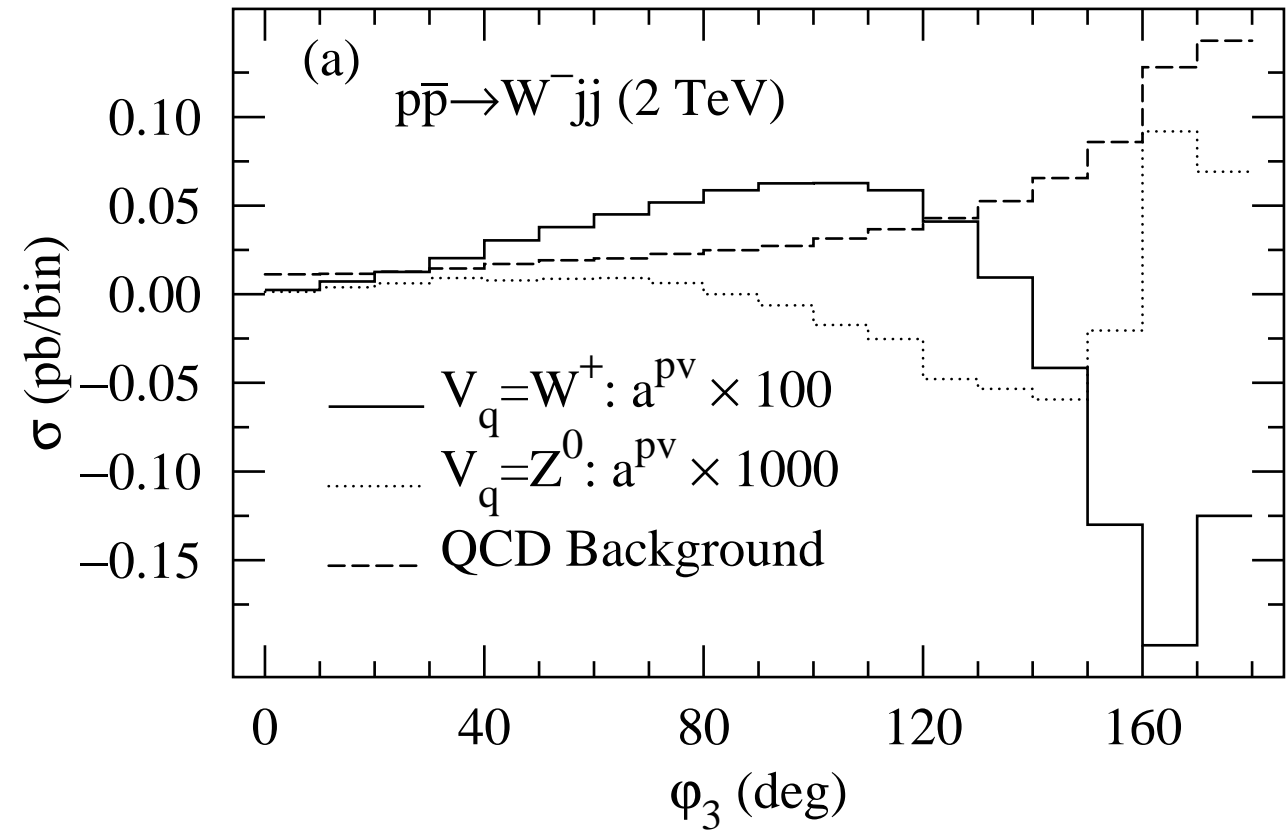
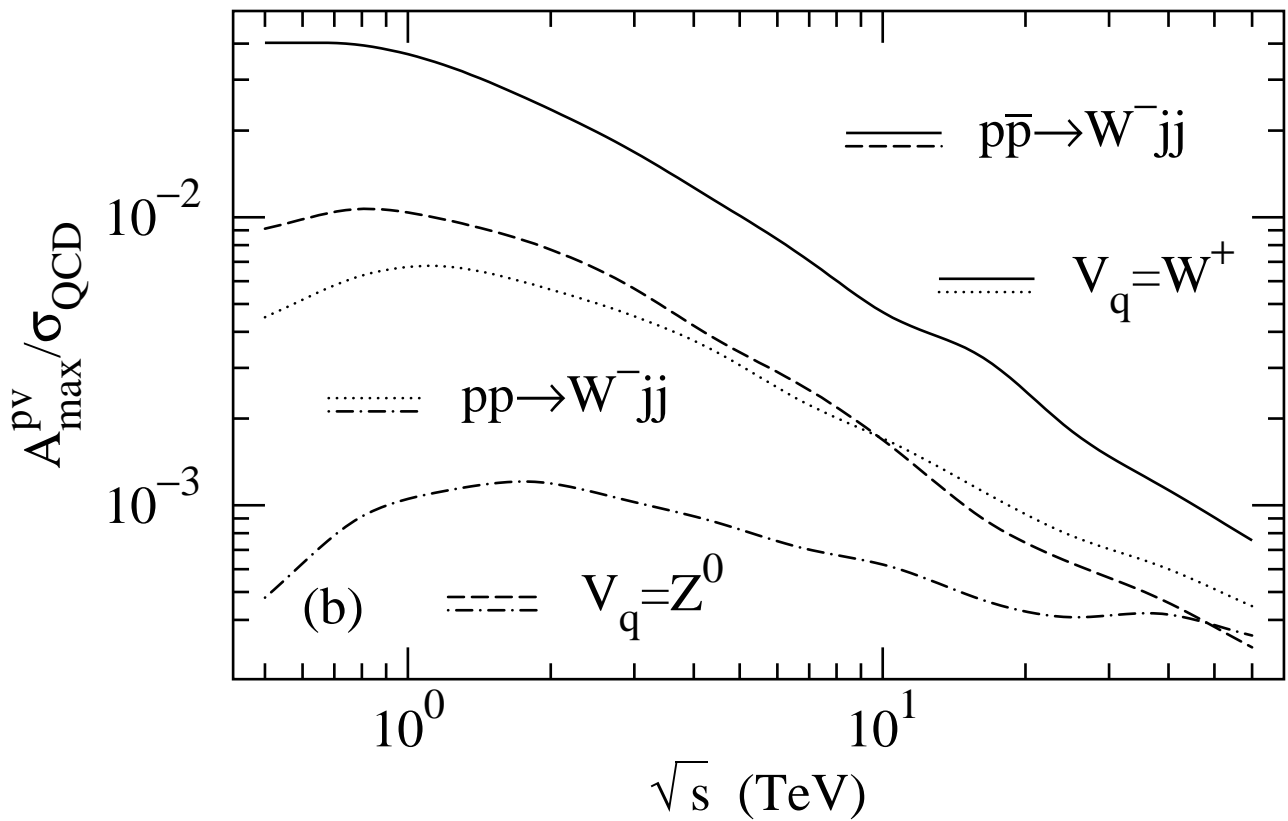
\begin{figure}
\caption{
Comparison of signal to background in $W^-$ + 2 jet production.
(a) Binned cross sections with 200 GeV $<$ $\protect\sqrt{s_{12}}$ 
$<$ 600 GeV. The parity-violating asymmetries have been multiplied
by the factors (100 and 1000) indicated to show them on the same scale
as the QCD background. Since the background histograms in the regions
$s_{34}\approx M_W^2$ and $s_{34}\approx M_Z^2$ are almost identical,
only one of them is shown.
(b) Variation of the signal to background
ratio in $pp$ and $p\bar p$ collisions as a function of the colliding
beam energy.
}
\label{Wbgfig}
\end{figure}

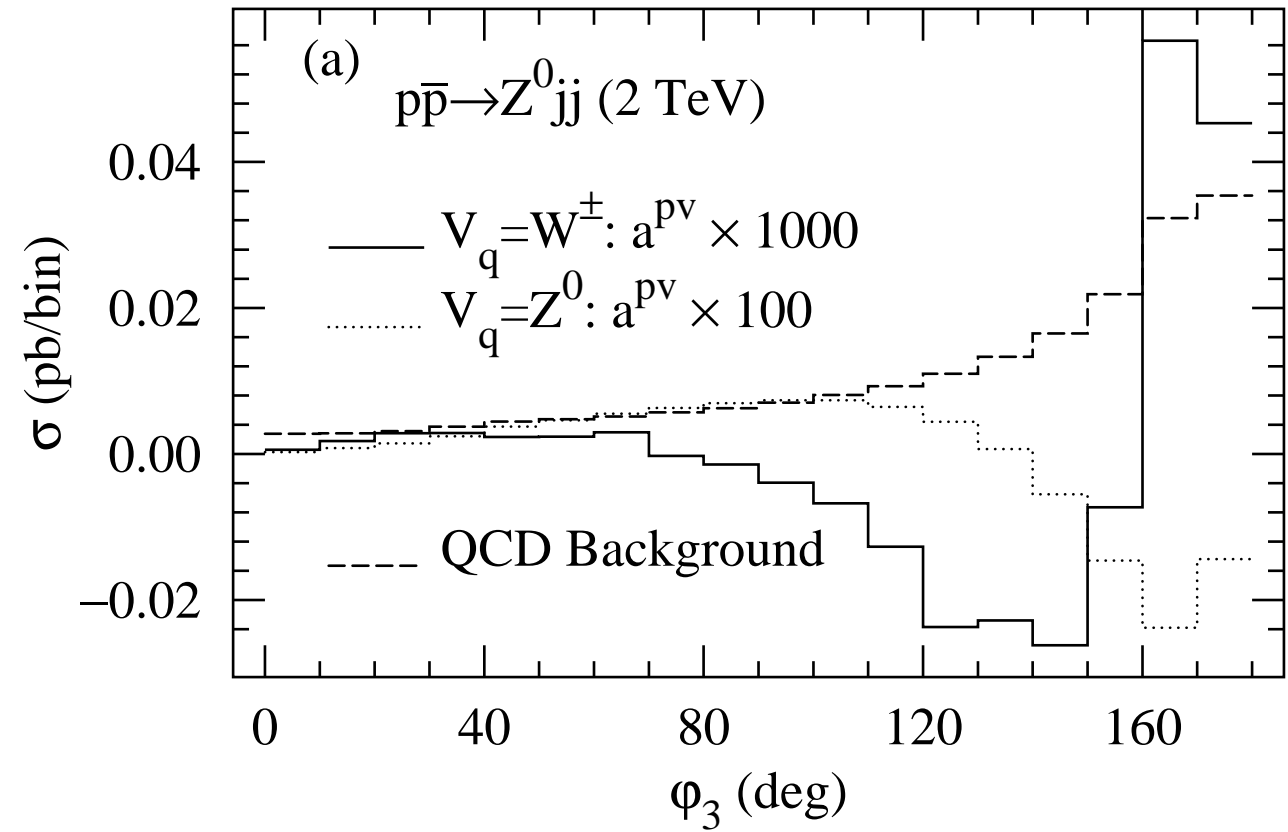
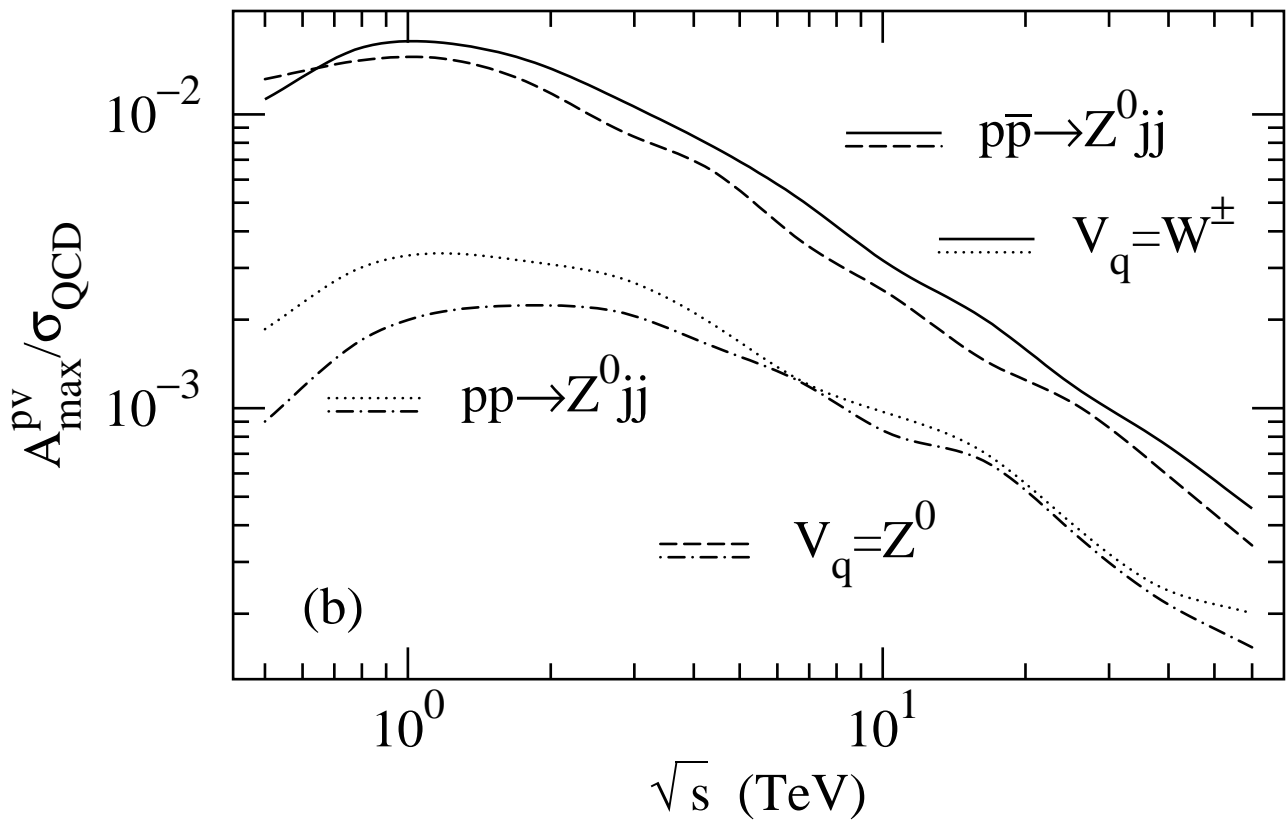
\begin{figure}
\caption{
Comparison of signal to background in $Z^0$ + 2 jet production.
(a) Binned cross sections with 200 GeV $<$ $\protect\sqrt{s_{12}}$ 
$<$ 600 GeV.  (b) Variation of the signal to background
ratio in $pp$ and $p\bar p$ collisions as a function of the colliding
beam energy.
}
\label{Zbgfig}
\end{figure}

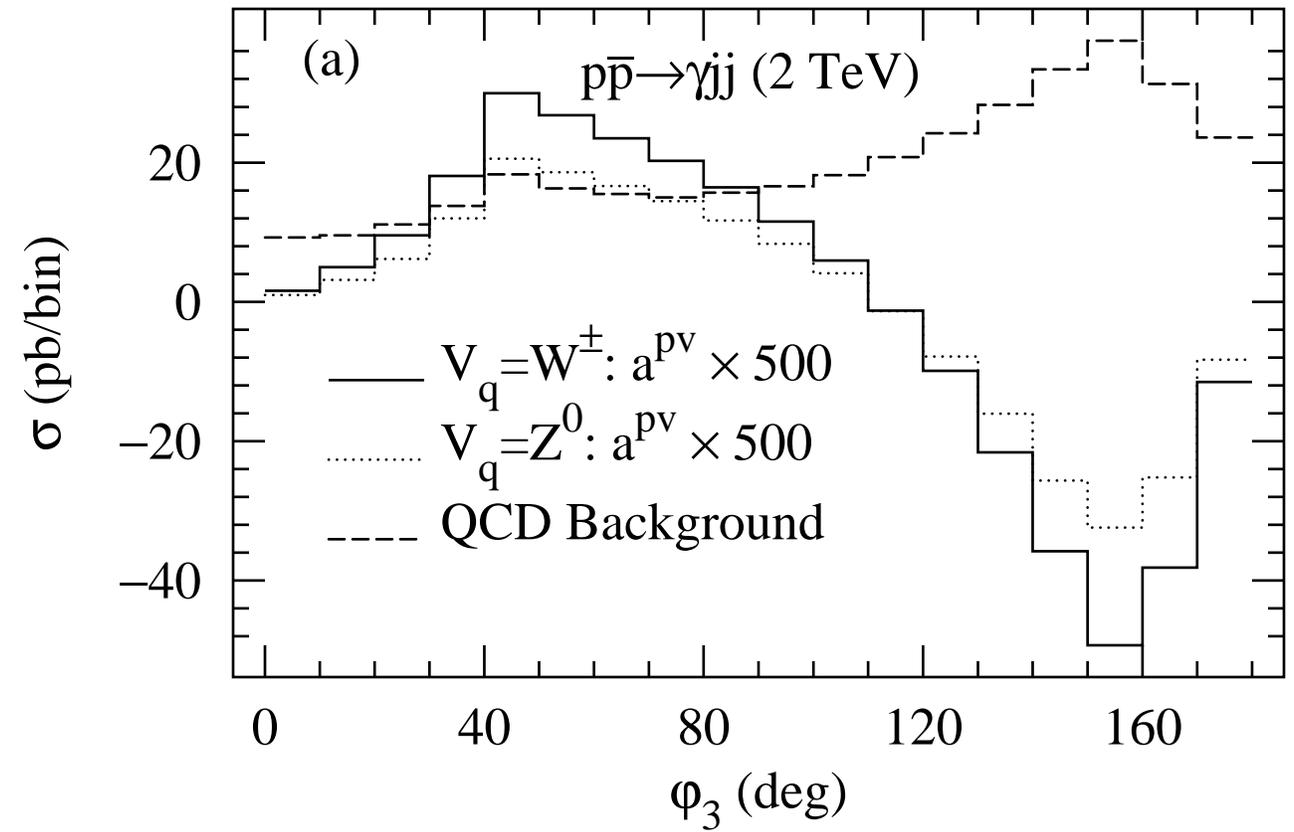
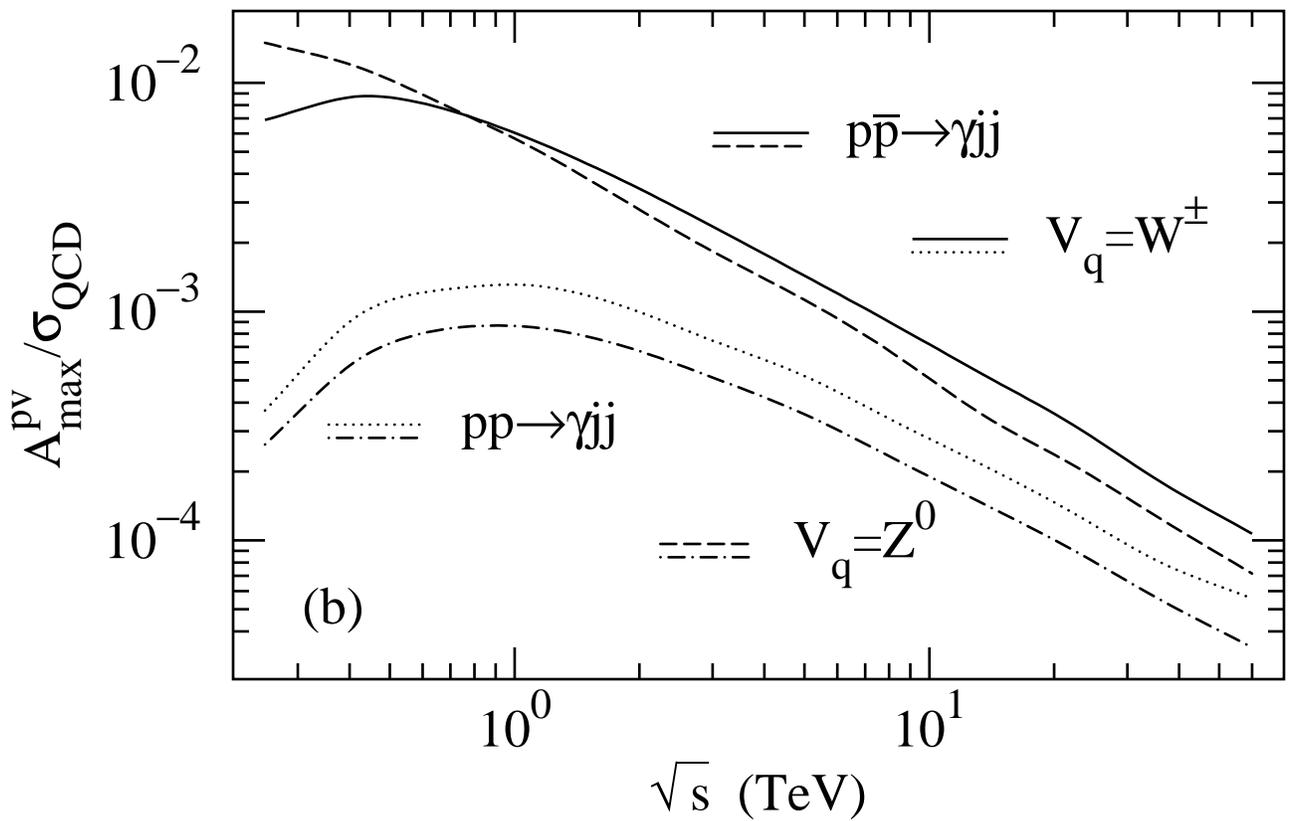
\begin{figure}
\caption{
Comparison of signal to background in real photon + 2 jet production.
(a) Binned cross sections with 100 GeV $<$ $\protect\sqrt{s_{12}}$ 
$<$ 500 GeV.  (b) Variation of the signal to background
ratio in $pp$ and $p\bar p$ collisions as a function of the colliding
beam energy.
}
\label{phbgfig}
\end{figure}

Figures \ref{Wbgfig}-\ref{phbgfig} show comparisons of the parity-violating
signal with the QCD background in electroweak boson + 2 jet production.
The signal to background ratios are rather small. They are generally
largest at $\sqrt s\sim 1$ TeV, and decrease with increasing $\sqrt s$
because the QCD background grows more rapidly than the asymmetries.
The magnitude and trend of the signal to background ratios are similar
to those observed in electroweak boson pair production 
\cite{stirling,gunion86}.
Some possible reasons why a larger resonant-peak effect is not observed
were discussed in the paragraphs following Eq.\ (\ref{weakpropeq}).
On a positive note, it is fairly remarkable that a signal comparable in
magnitude to the pair-production cross section does survive, in spite 
of the fact that we have imposed absolutely no requirements concerning
detection of the spins or flavors of the parton jets.

Let us next estimate the event rates required to observe these asymmetries
in $p\bar p$ collisions at $\sqrt s=2$ TeV, for example.
Suppose that one has chosen a bin as defined in Fig.\ \ref{pvfig}.
Let $n_{\rm bin}$ and $n_{\rm image\,bin}$ be the number of experimentally
observed events in this bin and its image, respectively.
Then,
\begin{equation}
a^{\rm pv}({\rm bin})={n_{\rm bin}-n_{\rm image\,bin}
	\over\int{\cal L}} \;,
\end{equation}
where $\int{\cal L}$ is the integrated luminosity used to obtain the sample
of events.
The difference $n_{\rm bin}-n_{\rm image\,bin}$ will be 
significant only if it is larger in absolute magnitude
than statistical fluctuations of
order $n_{\rm bin}^{1/2}$ in the number of events measured in the bin.
This means that an integrated luminosity
\begin{equation}
\int{\cal L} \gtrsim {\sigma_{\rm QCD}\over \left[a^{\rm pv}\right]^2}  \;,
\label{lumineq}
\end{equation}
is required to observe the parity-violating asymmetry in any particular
bin.

To estimate typical integrated luminosities required, consider
Fig.\ \ref{Wbgfig}a.
Choose a bin such that $140^\circ < \phi_3 < 180^\circ$.
In this bin, $a^{\rm pv}=0.0049$ pb from the solid curve with $V_q=W^+$,
and $\sigma_{\rm QCD}= 0.42$ pb from the dashed curve.
From Eq.\ (\ref{lumineq}), an integrated luminosity of 17 (nb)$^{-1}$
would be required to observe this asymmetry.
Approximately 13 (nb)$^{-1}$ would be required for a bin spanning
$0^\circ < \phi_3 < 140^\circ$ where the signal to background ratio
is smaller.
If it is {\em assumed} that $a^{\rm pv}$ changes sign around 
$\phi_3=140^\circ$, approximately 7.6 (nb)$^{-1}$ would be required to
obtain a statistically significant signal of parity violation from
the whole histogram.
It appears quite feasible to collect samples on the order of 5--25
(pb)$^{-1}$ in studies of $W$ + 2 jet events at the Fermilab
Tevatron \cite{cdfWjets}, 
and improvements in the beam luminosity might make it feasible
to collect samples of order 100 (pb)$^{-1}$.
However, it is apparent that very much higher integrated luminosites 
on the order of several (nb)$^{-1}$ would be
required to detect the parity-violating asymmetries studied in this paper.

Similarly large event rates are required to observe asymmetries in
$Z^0$ and real photon production.
While the predicted asymmetries 
in $\gamma$ production (see e.g., Fig.\ \ref{phbgfig}a)
are approximately an order of magnitude
larger in $p\bar p\rightarrow\gamma jj$ than they are in $W$ or
$Z$ production, the somewhat larger signal to background ratios
(see e.g., Fig.\ \ref{phbgfig}b), which enter
quadratically in Eq.\ (\ref{lumineq}), effectively cancel this 
this apparent advantage.

\subsection{Asymmetries in Lepton-pair + 2 Jet Production}

In the preceding section we presented asymmetries for final states
with three objects, an electroweak boson and two jets, produced at
large transverse momentum.  In this section we will present
asymmetries for final states with four objects, a lepton-pair and two
jets.  These asymmetries are interesting because leptons and jets are
detected directly experimentally, and the 4-particle final state
allows for a richer variety of parity-violating signatures.  However,
since we will demand that the lepton pair has an invariant mass equal
to that of a $W$ or $Z$ in order to obtain the largest asymmetries,
the dynamical origin of the asymmetries is really the same as in the
case of 3 particles.

\subsubsection{Kinematics}

Our choice of kinematic variables is a straightforward generalization
of the 3-particle case.  The colliding partons are assigned momenta
$p_1$ and $p_2$, and the jet partons $p_3$ and $p_4$.  With very
little loss of generality we may assume that leptons can be
distinguished from antileptons and that their momenta are measured
experimentally either by directly observing a charged lepton or by
inferring the momentum of a neutrino from missing momentum. (The case
$Z^0\rightarrow \nu\bar{\nu}$ presents possible ambiguities which we
will ignore).  Lepton and antilepton are assigned momenta $p_5$ and
$p_6$ respectively.  We then define a right-handed coordinate system
by choosing an $x$ axis such that $y_5^x>0$ and $y_5^y=0$.

If ${\cal J}_4$ is a contribution to a spin- and helicity-averaged
squared matrix element, the corresponding contribution to the
differential cross section at the parton level is
\begin{equation}
d\hat\sigma = {1\over2s_{12}}(2\pi)^4\delta^4(p_1+p_2-p_3-p_4-p_5-p_6)
  {\cal J}_4 {d^3p_3d^3p_4d^3p_5d^3p_6\over(2\pi)^{12} 16 
  p_3^0p_4^0p_5^0p_6^0}  \;.
\end{equation}
In analogy with Eq.\ (\ref{J3hateq}) we extract the delta functions
that arise from our use of the narrow width approximation for $V_l$
and $V_q$ from the cross section:
\begin{equation}
{\cal J}_4=\hat{\cal J}_4\delta(s_{34}-M_{V_q}^2)
  \delta(s_{56}-M_{V_l}^2)\;.
\label{J4hateq}
\end{equation}
On account of these delta functions and the invariance of the cross
section under rotations about the beam axis, the 4-particle phase
space is effectively 5-dimensional.  We will chose as independent
variables the rapidities $y_3$, $y_5$ and $y_6$ and the azimuthal
angles $\phi_3$ and $\phi_6$.  We first perform the $\vec p_4$
integration using 3-momentum conservation, and then perform the
$p_{3T}$, $p_{5T}$ and $p_{6T}$ integrations using energy conservation
and the delta functions from Eq.\ (\ref{J4hateq}).  This procedure
yields the following equations for $p_{5T}$ and $p_{6T}$:
\begin{equation}
p_{5T}\cosh(y_{12}-y_5)+p_{6T}\cosh(y_{12}-y_6) =
  {s_{12}+s_{56}-s_{34}\over2\sqrt{s_{12}}}\;,
\label{pTeq}
\end{equation}
\begin{equation}
p_{5T}p_{6T}={s_{56}\over2\left[\cosh(y_5-y_6)-\cos(\phi_6)
  \right]}\;,
\label{pT2eq}
\end{equation}
where $y_{12}$ was defined in connection with Eq\ (\ref{diffxceq}).
It is evident that these two equations will, in general, have two sets
of solutions for $p_{5T}$ and $p_{6T}$, both of which must be taken
into account in performing the non-trivial phase space integrations.
Finally, $p_{3T}$ is determined by the equation
\begin{equation}
p_{3T}={s_{34}\over 2\left[\sqrt{s_{12}}\cosh(y_{12}-y_3)
  -\sum\nolimits_{i=5}^6
  p_{iT}\left\{\cosh(y_i-y_3)-\cos(\phi_i-\phi_3)\right\}\right]}
  \;. 
\label{pT3eq}
\end{equation}
The final formula for the parton level cross section is
\begin{eqnarray}
\lefteqn{{d\hat\sigma\over dy_3dy_5dy_6d\phi_3d\phi_6}= } 
\quad \nonumber \\
&&  \sum_1^2 {(p_{3T}p_{5T}p_{6T})^2\hat{\cal J}_4\over2^{12}\pi^7
  s_{12}s_{34}s_{56}\sqrt{s_{12}}\bigl|p_{5T}\cosh(y_{12}-y_5)
  -p_{6T}\cosh(y_{12}-y_6)\bigr| }\;,
\end{eqnarray}
where the sum is over the two solutions of Eqs.\ (\ref{pTeq}) and
(\ref{pT2eq}).  The phase space boundaries are not easy to express in
analytic form, but they are straightforward to implement in a Monte
Carlo integration routine: given a set of values of the independent
variables, one solves for the transverse momenta in 
Eqs.\ (\ref{pTeq})- (\ref{pT3eq}) and accepts solutions that are real and
positive; the energies and momenta of the four final state particles
are then reconstructed using momentum conservation, and unphysical
solutions are rejected.  Hadron cross sections are computed using a
generalization of Eq.\ (\ref{hadxceq}).  A set of reflection-symmetric
phase space cuts, see Eq.\ (\ref{cutseq}), 
are imposed as in the case of $V$ + 2 jet production.

\subsubsection{Parton- and hadron-level asymmetries}

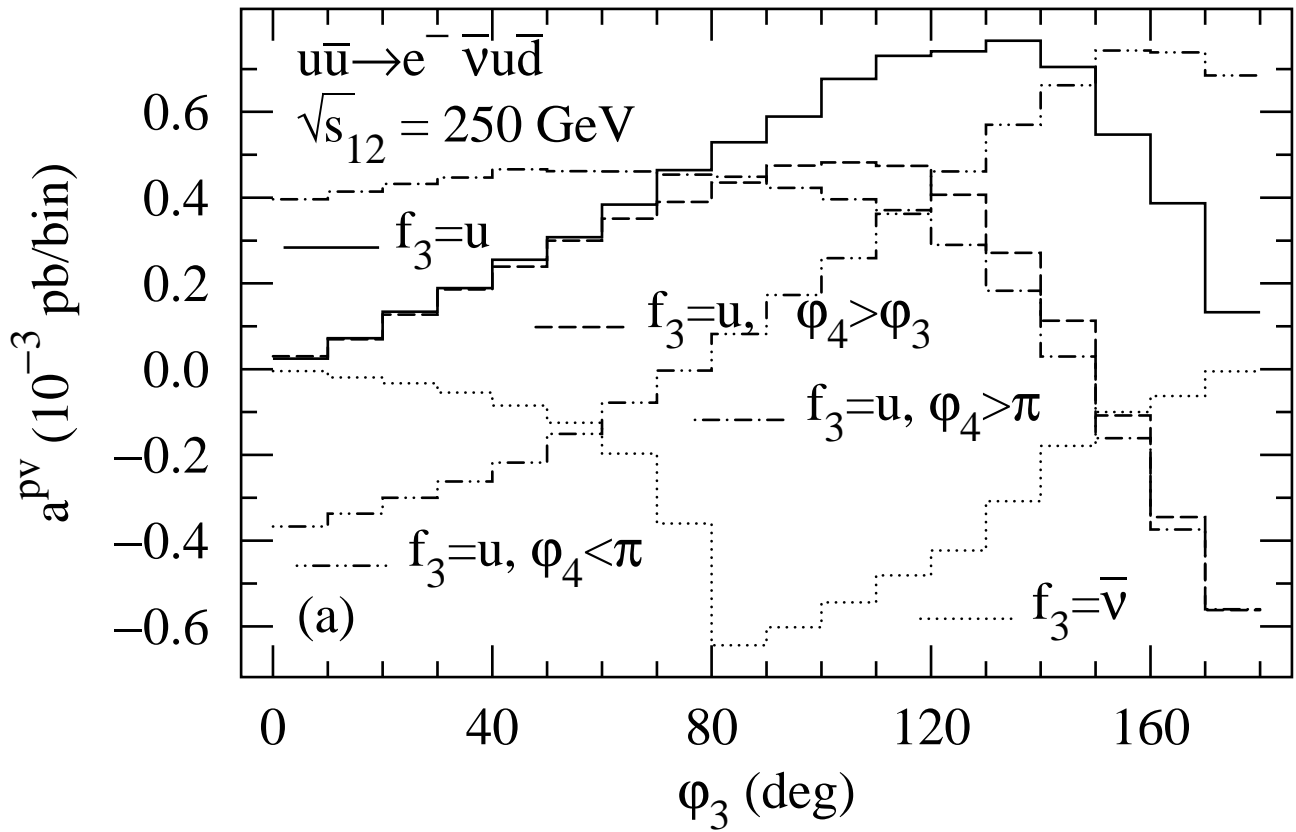
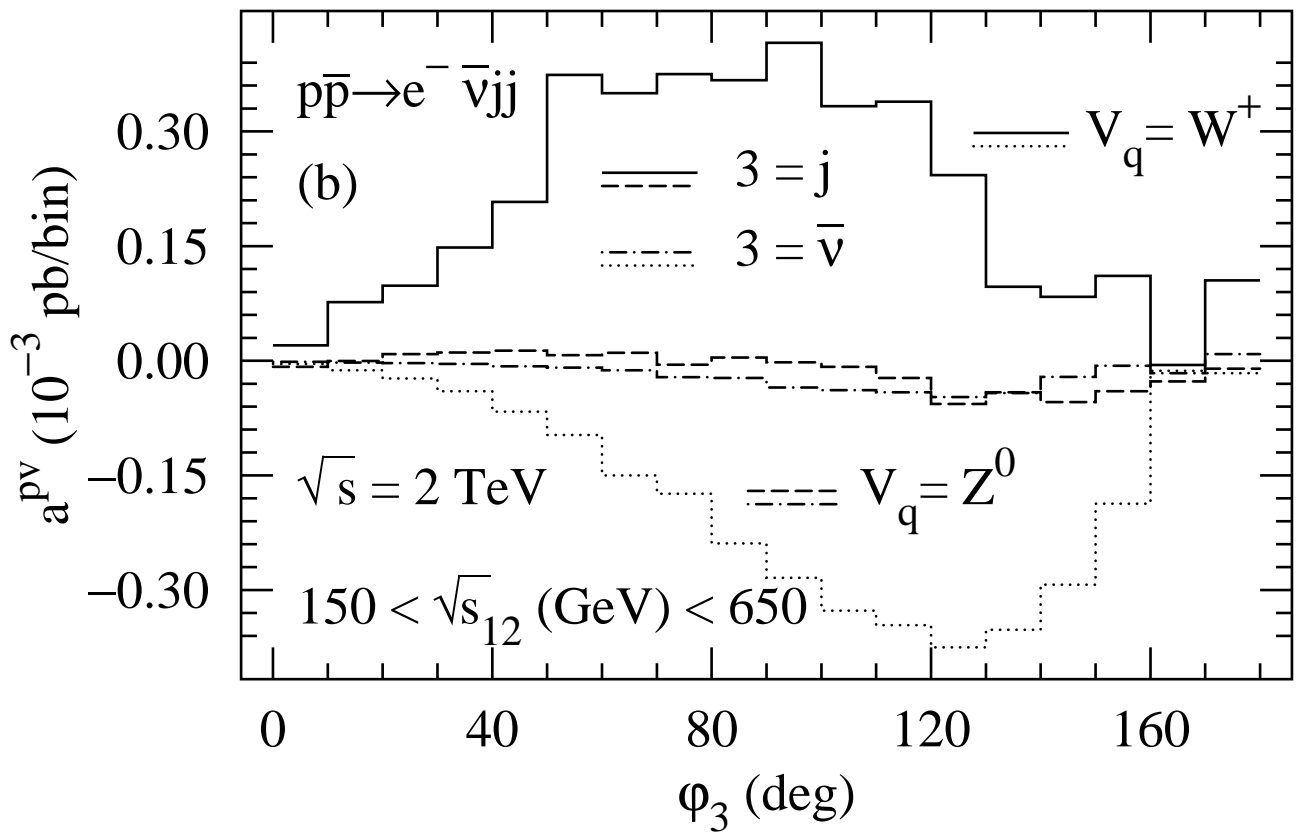
\begin{figure}
\caption{
Asymmetries from 4-particle final states. (a) At the parton level for the
process $u\bar u\rightarrow (W^-\rightarrow e^-\bar\nu)u\bar d$. The
$e^-$ momentum defines the direction $\phi=0$. The dotted histogram is
for observation of the $\bar\nu$ at azimuth $\phi_3$, and the other four
histograms for the $u$ quark at azimuth $\phi_3$ and with various cuts as
indicated on the azimuth $\phi_4$ of the $\bar d$. (b) At the hadron level
for the process $p\bar p\rightarrow (W^-\rightarrow e^-\bar\nu)$
+ 2 jets. For each of the two possibilities for $V_q= W^+,Z^0$, two
histograms are shown assuming either the $\bar\nu$ or one 
of the jets observed at azimuth $\phi_3$.
}
\label{W4fig}
\end{figure}

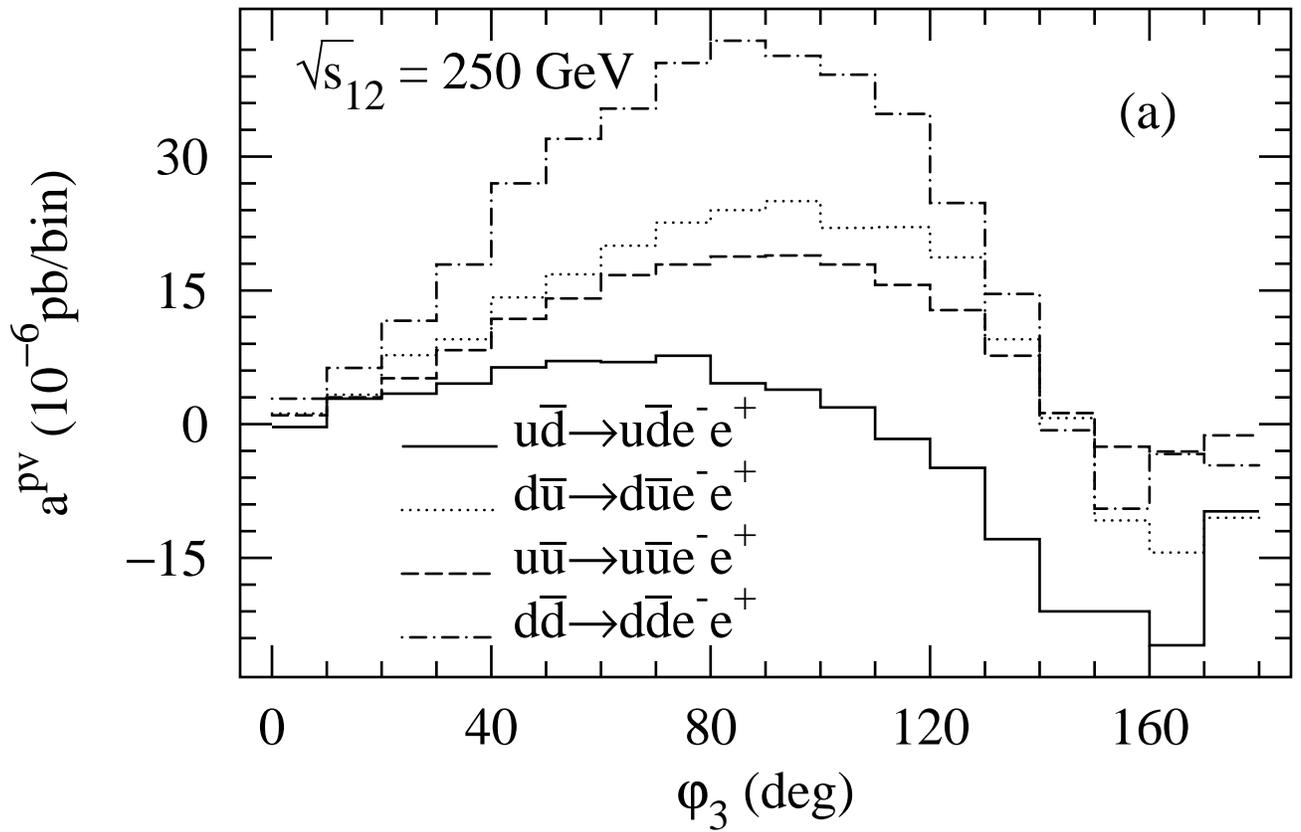
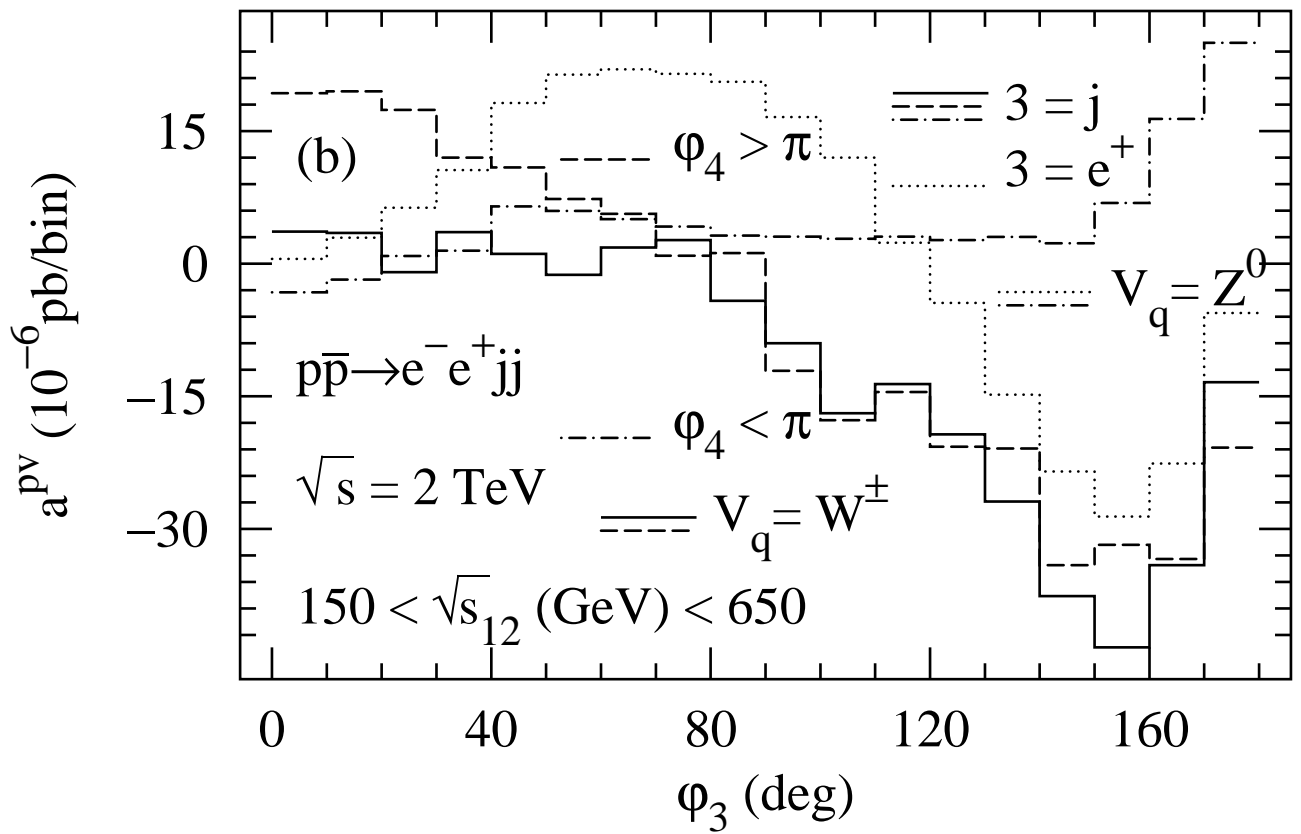
\begin{figure}
\caption{
Asymmetries from 4-particle final states. (a) Contributions from the four
different subprocess which produce a pair of $Z^0$s, one of which decays
to leptons and the other to quarks. The $e^-$ momentum defines azimuth
$\phi=0$, and the quark is observed at azimuth $\phi_3$. (b) Contributions
from the process $p\bar p\rightarrow (Z^0\rightarrow e^-e^+)$
+ 2 jets. 
Two histograms each for $V_q=W^\pm,Z^0$ are shown.
The dotted-line histogram assumes the $e^+$ is detected at azimuth
$\phi_3$. The other three histograms assume that one jet is
detected at azimuth $\phi_3$, and the azimuth $\phi_4$ of the other
jet is unrestricted (solid line), or restriced as indicated (dashed
and dot-dashed lines).
}
\label{Z4fig}
\end{figure}

In Figs.\ \ref{W4fig} and \ref{Z4fig}, we present parity-violating asymmetries
that involve 4-particle final states.
In general, the asymmetries tend to be a little smaller than those
presented in the preceding section for $V_l$ + 2 jet production.
This might be expected because the cuts in Eq.\ (\ref{cutseq}) applied
to each of the two decay leptons are more restrictive than the
same cuts applied to the single parent boson. 
In addition, the larger 4-particle phase space might be
expected to allow for more cancellations among the interference
contributions.

In Fig.\ \ref{W4fig}a, we present parton-level asymmetries for the
subprocess $u\bar u\rightarrow e^-\bar\nu u\bar d$, i.e., with
$V_l=W^-$ and $V_q=W^+$. The $e^-$ momentum is used to define the
$x{-}z$ plane. In the dotted-line histogram, the anti-neutrino is
detected in azimuthal bins of size $10^\circ$, and no restrictions, apart
from the standard cuts (\ref{cutseq}) are placed on the $u$ and $\bar d$
in the final state. In the other 4 histograms, the $u$ quark is
detected in the azimuthal bin, and no
restrictions, other than the standard cuts, are placed on the $\bar\nu$: 
in the solid-line histogram, no restriction, other than the standard
cuts, is placed on the $\bar d$; and in the remaining 3 histograms,
the azimuthal angle $\phi_4$ of the $\bar d$ is restricted as indicated.
The various histograms have characteristically different shapes.
It is evident that a much richer variety of asymmetries can be
defined in the case of 4-particle final states, when compared with
the corresponding 3-particle process---see the dashed curve in 
Fig.\ \ref{daWfig}a.
It should be possible to exploit this variety to reduce statistical
uncertainties in the signal-to-background ratio by rebinning an
experimental sample of events in different ways.
In Fig.\ \ref{W4fig}b, we present hadron-level asymmetries for the
process $p\bar p\rightarrow e^-\bar\nu$ + 2 jets---i.e., with
$V_l=W^-$ decaying to the lepton pair---at $\sqrt s=2$ TeV.
For each of the two possibilities, $V_q=W^+,Z^0$, two histograms are
presented assuming that a jet or the anti-neutrino, respectively, are
detected in the bin at azimuth $\phi_3$.
These histograms should be compared with the solid- and
dotted-line histograms presented in Fig.\ \ref{bin2fig}a.

In Fig.\ \ref{Z4fig}a, we present parton-level asymmetries for the
4 subprocesses in which $V_l=Z^0$ decays to an $e^+e^-$ pair.
In these histograms, the $e^-$ momentum is used to define the $x{-}z$ plane,
and the quark is detected in the azimuthal bin of size $10^\circ$ at
angle $\phi_3$, and no restrictions other than the standard cuts
are placed on the other particles.
In Fig.\ \ref{Z4fig}b, we present hadron-level asymmetries for the
process $p\bar p\rightarrow e^-e^+$ + 2 jets.
The histograms represent 4 different types of bins as defined in the
figure. It is interesting that the asymmetries for which $V_l=Z^0$
are somewhat smaller than those for $V_l=W^-$ presented in
Fig.\ \ref{W4fig}b.

\section{CONCLUSIONS}

In this paper we have presented predictions for parity-violating
asymmetries that arise from quantum mechanical interference effects
between the strong and electroweak components of the Standard Model.
The final states in which these asymmetries manifest themselves
involve an electroweak boson, either a real photon, or a $W$ or $Z$
observed say in its leptonic decay mode, and a pair of large $p_T$
jets with the invariant mass of the pair equal to the mass of a $W$ or
a $Z$.  These asymmetries might be observable in $pp$ and $p\bar{p}$
collisions above the threshold for production of pairs of electroweak
bosons, i.e., at center of mass energies in the TeV range and these
asymmetries are comparable in magnitude to the pair-production cross
sections. Unlike the pair production cross section which is parity
conserving, the interference contribution is parity-violating, and
this might make it easier to observe above a rather formidable QCD
background of electroweak boson + 2 jet events.
Actual observation of these effects will require somewhat higher
integrated luminosities than are currently available for example
at the Tevatron at Fermilab.
The results presented in this paper are therefore offered as examples of
the many interesting and subtle Standard Model predictions that can be
studied at a very-high luminosity TeV hadron collider;
they are also offered as examples of an unusual kind of
experimental signature involving multi-particle final states that
may have analogs in other high-energy collision processes.

\acknowledgments

This work has benefitted from very helpful remarks by and conversations with
M.~Chanowitz, R.K.~Ellis, S.D.~Ellis, I.~Hinchliffe and D.~Sivers, some
years ago, and with many other colleagues since.  One of us (R.J.G.) would
like to acknowledge the hospitality of D.~Sivers, C.~Zachos and the theory
group at Argonne National Laboratory where part of this work was done, as
well as the Argonne Division of Educational Programs for a Summer Faculty
Participation Award in 1986.  He would also like to thank W.A.~Bardeen,
R.K.~Ellis and the Theoretical Physics Department at Fermilab for their kind
hospitality in April and May of 1992.  This work was supported in part by
the National Science Foundation under grant numbers PHY83-10883,
PHY87-13231, and PHY92-12177.


\appendix

\section{Conventions for Massless Spinors}	

In this Appendix we define precisely the spinors in terms of which the
helicity amplitudes in this paper are presented.  We follow for the
most part the notation and conventions introduced in various of 
Refs.\ \cite{helamps}.  There are some subtle differences between some of
these conventions which we discuss for clarity and completeness.

\subsubsection{Massles fermions and anti-fermions with positive
energy}

The (four-component) spinor wave function associated with the
annihilation of a massless fermion with momentum $k_i$ and helicity
$\lambda=\pm{1\over2}$ is a solution of the two equations
\begin{eqnarray}
\not\!k_i u_\lambda(k_i)&=&0 \;, \nonumber \\
\gamma_5u_\lambda(k_i)&=&{\rm sign}(\lambda)\,u_\lambda(k_i)\;.
\label{Diraceq}
\end{eqnarray}
Here, $k^0>0$ and $k\cdot k=0$, and we use the $\gamma$ matrix
conventions of Bjorken and Drell \cite{bjdrell}.  
The solution of the two equations
(\ref{Diraceq}) is unique up to a multiplicative constant because the first
equation relates the two lower components $\chi$ of the spinor $u$ to
the upper components $\phi$, and the second then yields an eigenvalue
equation for the upper components:
\begin{equation}
\chi_\lambda={\vec\sigma\cdot\vec k_i\over k^0_i}\phi_\lambda
={\rm sign}(\lambda)\,\phi_\lambda\;.
\end{equation}
The multiplicative constant is determined up to an arbitrary phase
factor by normalizing the spinor so that
\begin{equation}
u_\lambda(k_i)\bar u_\lambda(k_i)=\omega_{{\rm sign}(\lambda)}
\not\!k_i\quad,\quad\omega_\pm\equiv{1\pm\gamma_5\over2}\;.
\label{normeq}
\end{equation}
The fermion spinors have thus been determined up to two arbitrary phase
factors, one for each helicity.

The four-component spinor wave function associated with the creation
of an anti-fermion with momentum $k_i$ and helicity $-\lambda$ is a
solution of the two equations
\begin{eqnarray}
\not\!k_i v_\lambda(k_i)&=&0\;, \nonumber \\
\gamma_5 v_\lambda(k_i)&=&-{\rm sign}(\lambda)\,v_\lambda(k_i)\;.
\label{Dirac2eq}
\end{eqnarray}
It is convenient and economical to determine the phases of the
antifermion spinors as follows:
\begin{equation}
u_\lambda(k_i)=v_{-\lambda}(k_i)\equiv |i,{\rm sign}(\lambda)
\rangle \;.
\end{equation}

\subsubsection{Charge conjugation}

We next fix the relative phase of spinors of opposite helicity using
charge conjugation, which is defined as follows:
\begin{equation}
u\rightarrow u^c=C\bar{u}^T\;,
\label{cceq}
\end{equation}
where the unitary matrix $C$ has the properties
\begin{equation}
C^{-1}\gamma_\mu C = -\gamma_\mu^T\quad,\quad C^{-1}=-C^*\;;
\label{cc2eq}
\end{equation}
the first of these properties 
ensures that charge conjugation reverses the signs of
$\not{\kern-3truept k_i}$ and $\gamma_5$ (i.e. interchanges 
Eqs.\ (\ref{Diraceq}) and (\ref{Dirac2eq})),
and the second ensures that $(u^c)^c=u$. Given a particular
choice of $C$, for example $C=i\gamma^2\gamma^0$, it follows that
\begin{equation}
u^c_\lambda(k_i) = e^{i\eta(k_i)}u_{-\lambda}(k_i)\;,
\label{cc3eq}
\end{equation}
where the phase constant $\eta(k_i)$ is independent of the
helicity $\lambda$.

Refs.\ \cite{helamps} propose to fix the phase $\eta(k_i)$ in slightly
different ways which, however, lead to the same amplitude expressions, as
we shall show.  Kleiss and Stirling relate spinors of opposite
helicity as follows: Choose a reference momentum $k_0$ with $k_0^0>0$
and $k_0\cdot k_0=0$ that is not parallel to any other external
momentum $k_i, i=1,2,3,...$, and an orthogonal space-like unit vector
$n_0$ with $n_0\cdot n_0=-1$ and $k_0\cdot n_0=0$, and define two
basic spinors
\begin{equation}
u_{+{1\over2}}(k_0)=\not\!n_0 u_{-{1\over2}}(k_0)\;.
\end{equation}
These two spinors are normalized as in Eq.\ (\ref{normeq}) but their
phase is not specified. Then, for any other massless spinor with
momentum $k_i, i=1,2,3,...$,
\begin{equation}
u_\lambda(k_i)={\not\!k_i\over\sqrt{2k_i\cdot
k_0}} u_{-\lambda}(k_0)\;.
\label{KSeq}
\end{equation}
This definition obviously determines the relative phase of spinors of
opposite helicity and hence the charge conjugation phase $\eta(k_i)$.
It is not difficult to show that in fact
\begin{equation}
e^{i\eta(k_i)}=-e^{i\eta(k_0)}\;,
\end{equation}
i.e., the charge conjugation phase depends only on the reference
momentum $k_0$ and not on the momentum $k_i$ of the fermion.  A simple
way of showing this is to relate the scalar product $\bar
u_{1\over2}(k_0)u_{1\over2}^c(k_i)$ to the product $\bar
u_{1\over2}(k_0)u_{-{1\over2}}(k_i)$ using Eqs.\ (\ref{cceq}),
(\ref{cc2eq}), and (\ref{cc3eq}).

Xu et al., and Gunion and Kunszt \cite{helamps}
fix the relative phase of spinors of
opposite helicity by the requirement
\begin{equation}
u_\lambda^c(k_i)=u_{-\lambda}(k_i)\;.
\label{XGKeq}
\end{equation}
This evidently corresponds to the special choice $\eta(k_0)=\pi$ in
the Kleiss-Stirling definition.

To summarize, Xu et al., and Gunion and Kunszt allow a single
arbitrary overall phase for each particle of momentum $k_i$, while
Kleiss and Stirling allow a choice of the reference vectors $k_0$ and
$n_0$ as well as a single arbitrary phase for say the spinor
$u_{1\over2}(k_0)$.  It may be important to take these differences
between these two definitions into account when evaluating helicity
amplitudes numerically since the different definitions, as well as
whatever choices are made in fixing the remaining unspecified overall
phases, will affect the sizes of various terms and thus the occurrence
of for example numerical errors due to roundoff.  However, both
defintions lead to the same analytical relations between spinor
products that we have used to simplify the analytical results
presented in this paper, and which we discuss next.

\subsubsection{Useful identities involving spinor products}

We next list, for reference, some identities that are extremely useful
in simplifying helicity-amplitude expressions.

Spinors can be written in compact bra-ket notation as follows:
\begin{eqnarray}
u_\lambda(k_i)& \equiv & |i,\sigma_i\rangle \equiv |i\sigma_i\rangle \;,
    \nonumber \\
\bar u_\lambda(k_i) & \equiv & \langle i\sigma_i| \;,
\label{braketeq}
\end{eqnarray}
where $\sigma_i={\rm sign}(\lambda_i)$, and we assume that $k^0>0$.

Let $\Gamma^{(n)}$ be a string consisting of a product of n Dirac
$\gamma$ matrices, and $\Gamma^{(n)}_R$ be the string with the
matrices written in reverse order.  The anti-commutation relation
$\{\gamma_\mu,\gamma_5\}=0$, together with 
Eqs.\ (\ref{Diraceq}), imply the helicity
selection rules
\begin{eqnarray}
\langle i\sigma|\Gamma^{(2n)}|j\sigma\rangle &=&0 \;, 
    \nonumber \\
  \langle i\sigma|\Gamma^{(2n+1)}|j,-\sigma\rangle &=& 0\;.
\label{selruleseq}
\end{eqnarray}
The phase choice implied in Eq.\ (\ref{XGKeq}), 
together with the the behavior of
$\gamma$ matrices under charge conjugation, Eq.\ (\ref{cc2eq}), imply the
following ``charge conjugation identity'':
\begin{equation}
\langle i\sigma_i|\Gamma^{(n)}|j\sigma_j\rangle = (-1)^{(n+1)}
  \langle j,-\sigma_j|\Gamma^{(n)}_R|i,-\sigma_i\rangle\;.
\label{linereveq}
\end{equation}
This identity is in fact also obeyed by spinors with phases chosen
according to the prescription of Kleiss and Stirling, 
Eq.\ (\ref{KSeq}), since
their charge conjugation phase is independent of momentum. (Kleiss and
Striling refer to Eq.\ (\ref{linereveq}) 
as the ``line reversal identity'', and prove
it without appealing directly to charge conjugation).

An extremely useful identity involving $\gamma$ matrices with their
indices contracted is the Chisholm identity:
\begin{equation}
\langle i\sigma|\gamma_\mu|j\sigma\rangle \gamma^\mu =
  2\left[\; |j\sigma\rangle\langle i\sigma| + 
  |i,-\sigma\rangle\langle j,-\sigma|\;\right]\;.
\label{chisholmeq}
\end{equation}
Proofs of this identity can be found in Refs.\ \cite{helamps}.

Another very useful identity is the ``cyclic identity'':
\begin{equation}
|i\pm\rangle\langle j{\mp}|k\pm\rangle +
  |j\pm\rangle\langle k{\mp}|i\pm\rangle +
  |k\pm\rangle\langle i{\mp}|j\pm\rangle = 0\;.
\label{cycliceq}
\end{equation}
A proof of this identity can be found in Xu et al.,
\cite{helamps}. We sketch here an
alternate simple proof which uses the Kleiss-Stirling spinor
definitions.  Using Eq.\ (\ref{KSeq}), we can write
\begin{eqnarray}
\lefteqn{|i-\rangle\langle {j+}| - |j-\rangle\langle {i+}|} \qquad
\nonumber \\
&& =  {1\over\sqrt{4k_i\cdot k_0 k_j\cdot k_0}} \bigl(\not\!k_i
  \not\!n_0\not\!k_0\not\!k_j - \not\!k_i\not\!n_0\not\!k_0\not\!k_j
  \bigr)\omega_- \nonumber \\ 
&& = a + b\gamma_5 + c^{\mu\nu}\sigma_{\mu\nu}
\label{cyceq}
\end{eqnarray}
The last line expresses the most general form of a product of an even
number of $\gamma$ matrices.  The coefficients can be found by forming
the appropriate traces.  It is easy to show that $a=-b= {1\over2}
\langle {j+}|i-\rangle$, and the fact that $c^{\mu\nu}=0$ is easily
proven using the charge conjugation identity Eq.\ (\ref{linereveq}).  
The cyclic identity (\ref{cycliceq}) 
with the lower signs follows when Eq.\ (\ref{cyceq}) is multiplied on the
right by $|{k-}\rangle$.  
The cyclic identity with the upper signs follows from
charge conjugation.  The cyclic identity is particularly useful in
simplifying helicity amplitudes that involve a non-abelian
three-vector-boson vertex.

\subsubsection{Negative energies and crossing}

If all external particles in a ``generic diagram'' are taken to be
outgoing, momentum conservation implies that some of these particles
must have negative energies.  It is evident, for example from the
square root factor in Eq.\ (\ref{KSeq}), that reversing the sign of the
energy is a discontinuous transformation and can introduce and
ambiguity in the phase of a spinor.  Gunion and Kunszt propose to fix
this ambiguity by requiring that a change in the 
sign of the momentum of a particle
in a helicity amplitude is equivalent to crossing that particle between
initial and final states.  It is easy to see that this might be
accomplished by defining
\begin{equation}
\matrix{
    |{-k},\sigma\rangle & = & i|k\sigma\rangle\;, \cr
    \langle-k,\sigma| & = & i\langle k\sigma| \;, \cr
}
\quad {\rm for}\; k^0 > 0\;.
\label{k0eq}
\end{equation}
An immediate consequence of this choice is that the spinor
normalization (\ref{normeq}) and the identities 
(\ref{selruleseq}-\ref{cycliceq}) preserve their forms
when the sign of any of the momenta $k_i$ involved is reversed.
Conversely, demanding that these identities be form invariant under
this sign change determines the choice of phases in (\ref{k0eq}) up to a
sign.  In this sense, the phase choice is implicit in the definitions
of Kleiss and Stirling.  The choice of phases in Eq.\ (\ref{k0eq}) requires
that we modify the definition of bra spinors, introduced 
in Eq.\ (\ref{braketeq}) for positive energy, as follows:
\begin{equation}
\langle k\sigma| \equiv {\rm sign}(k^0)\bar{u}_\lambda(k)
  = {\rm sign}(k^0)u^*_\lambda(k)^T\gamma_0
  \equiv u^+_\lambda(k)^T\gamma_0\;. 
\end{equation}
Following Ref.\ \cite{gunion86},
we have introduced a ``+ conjugation'' operation $(^+)$
which is the usual complex conjugation $(^*)$ operation, but followed
by a change of sign when it acts on a spinor with negative energy.
Note that complex conjugation is also to be replaced by ``+
conjugation'' in the definition of charge conjugation in 
Eq.\ (\ref{cceq}).
Most importantly, the process of ``squaring'' an amplitude ${\cal A}$
to obtain a probability (cross section) is done by multiplying the
amplitude by its ``+ conjugate'' rather than its complex conjugate.
The following elementary example illustrates that this leads to a
crossing-invariant result:
\begin{eqnarray}
\lefteqn{\langle k_1\sigma|k_2,-\sigma\rangle \langle
  k_1\sigma|k_2,-\sigma\rangle ^+} \qquad \nonumber \\
&& = {\rm sign}(k_1^0k_2^0)\;
  \left|\langle {\rm sign}(k_1^0)k_1,\sigma|{\rm sign}(k_1^0)k_2,
  -\sigma\rangle\right|^2 \nonumber \\
&& = 2k_1\cdot k_2\;.
\end{eqnarray}
In a practical computation, the following relation is useful in
forming a probability:
\begin{equation}
\langle i\sigma_i|\Gamma|j\sigma_j\rangle^+
  =\langle j\sigma_j|\Gamma_R|i\sigma_i\rangle\;.
\end{equation}
Here $\Gamma$ is a product of $\gamma$ matrices, and $\Gamma_R$ the
product with the matrices taken in reverse order.  This relation is
valid for positive and negative energies, and, when combined with the
completeness relation (\ref{normeq}), 
is also seen to produce crossing invariant probabilities.

\subsubsection{Photon wave functions}

To compute the helicity amplitudes for production of a real photon
plus 2 jets, we need to specify the wave function of a photon with
definite helicity $\lambda = \pm1$.  If the momentum $k$ of the
photon is given, its wave function is only determined up to a gauge
transformation and an arbitrary multiplicative constant.  We shall
follow Xu et al., \cite{helamps} 
to define the phase and gauge of the wave function in
terms of spinors associated with $k$ and with a momentum $p$ chosen
such that $k\cdot p \not=0$, $p\cdot p=0$ and $p^0>0$, as follows:
\begin{equation}
\epsilon_\lambda^\mu(k;p)=\lambda{\langle p,-\lambda|\gamma^\mu|
  k,-\lambda\rangle\over\sqrt2\langle p\lambda|k,-\lambda\rangle}
  \;. 
\label{poleq}
\end{equation}
If the spinors are normalized and their phase defined as in the
earlier in this Appendix, it is easy to verify that the
photon wave functions satisfy the following conditions:
\begin{eqnarray}
\epsilon_\lambda^\mu(k;p) &=& \epsilon_\lambda^*(k;p)\;, \nonumber\\
k\cdot\epsilon_\lambda(k;p) &=& 0\;, \nonumber\\
\epsilon_\lambda(k;p)\cdot\epsilon_{\lambda'}^*(k;p')
   &=& - \delta_{\lambda,\lambda'}\;, \nonumber\\
\epsilon_\lambda^\mu(k;p')&=&\epsilon_\lambda^\mu(k;p)
  +{\sqrt2\lambda\langle p',-\lambda|p\lambda\rangle
  \over\langle p',-\lambda|k\lambda\rangle\langle k,-\lambda|
  p\lambda\rangle}k^\mu\;.
\label{polpropseq}
\end{eqnarray}
In practice, the auxiliary momentum $p$ is chosen to simplify a given
helicity amplitude as much as possible.  
The last of Eqs.\ (\ref{polpropseq}) shows that any two
choices of $p$ yield polarization vectors that differ by a change of
gauge: thus the same auxiliary momentum must be used in all members of
a gauge invariant set of diagrams.

We note finally that $\epsilon_\lambda^\mu(k;p)$ is the wave function
of an outgoing photon, i.e., it is the polarization vector associated
with the creation operator in the photon field operator.  To verify
that $\lambda$ as defined in Eq.\ (\ref{poleq}) corresponds to the correct
circular polarization, let $n_1$ and $n_2$ be two linear polarization
vectors which satisfy
\begin{equation}
n_1^2=n_2^2=-1\;,\quad n_i\cdot k=0\;,\quad\vec n_1\times\vec n_2
 =\vec k\;.
\end{equation}
Then, the following phase relation between the transverse components
\begin{equation}
n_2\cdot \epsilon_\lambda(k;p) = i\lambda
 n_2\cdot \epsilon_\lambda(k;p)\;,
\end{equation}
shows that $\lambda=\pm1$ correspond to right and left circular
polarization respectively.\footnote{
This relation is easily verified by direct
computation e.g., by using a frame and gauge in which $k=k^0(1,0,0,1)$ and
$p=p^0(1,1,0,0)$.}  
The wave function of an incoming photon is
obtained either by crossing $k\rightarrow-k$ or by complex
conjugation:
\begin{equation}
\epsilon_\lambda(-k;p) = \epsilon^*_\lambda(k;p) \;.
\end{equation}
%


\section{Properties of Invariant Functions}	

In this Appendix, we collect and discuss several useful formulas
involving the various functions in terms of which the analytic results
of this paper have been presented.

\setcounter{subsubsection}{0}
\subsubsection{Properties of $F_{123456}$}

This function was defined in Eq.\ (\ref{Feq}).  
Using $\langle i|j\rangle = -\langle j|i\rangle$, see
Eq.\ (\ref{linereveq}), we see that
\begin{equation}
F_{123456}=-F_{321456}=-F_{163452}\;.
\label{Fprop1eq}
\end{equation}
Using the cyclic identity Eq.\ (\ref{cycliceq}), we see that
\begin{equation}
F_{123456}+F_{143652}+F_{163254} = 0\;.
\end{equation}
Finally, if the six momenta $k_i$ satisfy $\sum_{i=1}^6k_i = 0$,
it is easy to show using Eq.\ (\ref{normeq}) that
\begin{equation}
F_{123456} = -F^+_{216543}\;.
\label{Fprop3eq}
\end{equation}

\subsubsection{Triple-boson vertex contribution}

As an application of some of the identites listed in Appendix A we
show that the contribution of the diagram in Fig\ \ref{lljjdiags} with the
non-abelian triple-boson vertex can be expressed in terms of the $F$
functions.  This rather remarkable result is presented without a
derivation in \cite{gunion86}; since the derivation appears to be non-trivial,
we include it here for completeness.  It is not difficult to see that
the amplitude corresponding to this diagram, with all helicities
chosen to be negative for example, has the following Lorentz factor:
\begin{eqnarray}
&& \langle {1-}|\gamma^\lambda|2- \rangle 
  \langle {3-}|\gamma^\mu|4- \rangle 
  \langle {5-}|\gamma^\nu|6- \rangle 
\nonumber \\
&&\times\left[g_{\nu\lambda}(k_{12}-k_{56})_\mu
        +g_{\lambda\mu}(k_{34}-k_{12})_\nu
        +g_{\mu\nu}(k_{56}-k_{34})_\lambda\right] 
\nonumber \\
&& = 4 \langle {1-}|5+ \rangle \langle {6+}|2- \rangle 
  \left[ \langle {1-}|3+ \rangle \langle {4+}|1- \rangle +
  \langle {2-}|3+ \rangle \langle {4+}|2- \rangle \right]
\nonumber \\
&&-4 \langle {1-}|3+ \rangle \langle {4+}|2- \rangle 
  \left[ \langle {1-}|5+ \rangle \langle {6+}|1- \rangle +
  \langle {2-}|5+ \rangle \langle {6+}|2- \rangle \right]
\nonumber \\
&&-4 \langle {3-}|5+ \rangle \langle {6+}|4- \rangle 
  \left[ \langle {1-}|3+ \rangle \langle {3+}|2- \rangle +
  \langle {1-}|4+ \rangle \langle {4+}|2- \rangle \right]
\;.
\end{eqnarray}
To obtain the right hand side of this equation, 
we have used the Chisholm identity
(\ref{chisholmeq}) as well as 
Eqs.\ (\ref{Diraceq}-\ref{normeq}) and (\ref{linereveq}).  
While the right hand side
is not obviously expressible in terms of the $F$ functions, it is actually
equal to the difference $(F_{123456} - F_{125634})$, as can be seen
by examining the following expression:
\begin{eqnarray}
&&{\rm R.H.S.} - (F_{123456} - F_{125634}) \nonumber \\
&&= 4 \left\{
  \langle {3-}|5+ \rangle \langle {4+}|2- \rangle 
  \left[ \langle {1-}|4+ \rangle \langle {4+}|6- \rangle +
  \langle {1-}|5+ \rangle \langle {5+}|6- \rangle \right]
\right.
\nonumber \\
&&\quad -
  \langle {3-}|5+ \rangle \langle {1-}|3+ \rangle 
  \left[ \langle {6+}|4- \rangle \langle {3+}|2- \rangle +
  \langle {6+}|2- \rangle \langle {4+}|3- \rangle \right]
\nonumber \\
&&\quad \left. -
  \langle {4+}|2- \rangle \langle {6+}|2- \rangle 
  \left[ \langle {1-}|3+ \rangle \langle {2-}|5+ \rangle +
  \langle {1-}|5+ \rangle \langle {3-}|2+ \rangle \right] 
\right\} \;.
\end{eqnarray}
Using the cyclic identity (\ref{cycliceq}) on the terms in each of the 
second ane third
square brackets on the right hand side of the above equation, 
it is easy to see that the right hand side of the above
equation can be written
\begin{eqnarray}
&&  4
  \langle {3-}|5+ \rangle \langle {4+}|2- \rangle 
  \left[ \langle {1-}|2+ \rangle \langle {2+}|6- \rangle \right.
\nonumber \\
&&\quad \left. +
  \langle {1-}|3+ \rangle \langle {3+}|6- \rangle +
  \langle {1-}|4+ \rangle \langle {4+}|6- \rangle +
  \langle {1-}|5+ \rangle \langle {5+}|6- \rangle \right]
\;,
\end{eqnarray}
which can be seen to vanish by using Eq.\ (\ref{normeq}) and momentum
conservation.

\subsubsection{Properties of $(I,J,K)_{123456}$}

These functions were defined in Eqs.\ (\ref{Idefeq})-(\ref{Jdefeq}).  
Using the properties of the functions $F$, see 
Eqs.\ (\ref{Fprop1eq})-(\ref{Fprop3eq}) it is easy to
show that
\begin{equation}
I_{123456}=I^+_{123456}=I_{321456}=I_{163452}=I_{216543}\;.
\label{Ipropseq}
\end{equation}
Since $I$ is real, it does not contribute to the asymmetries presented
in this paper.  Nevertheless, we give here an expression for $I$ in
terms of invariants for completeness:
\begin{equation}
I_{123456} = 16s_{13}s_{26}\bigl[(s_{14}+s_{34})(s_{15}+s_{35})
   -s_{13}s_{45}\bigr]\;. 
\end{equation}

The function $J$ can likewise be shown to satisfy
\begin{equation}
J_{123456}=J^+_{125634}=J_{214365}\;.
\label{Jpropseq}
\end{equation}
Various equivalent expressions for $J$ in terms of invariants can be
derived by using Eqs.\ (\ref{Fprop1eq})-(\ref{Fprop3eq}).  
One such expression can be obtained by
writing out Eq.\ (\ref{Jdefeq}) using (\ref{Feq}), and rewriting
the various terms using the completeness relation (\ref{normeq}):
\begin{equation}
{1\over16}J_{123456} = -s_{13}s_{15}s_{24}s_{26} - s_{135}
  {\rm Tr}\left[{\gamma_R\not\!k_6\not\!k_2\not\!k_4\not\!k_3\not\!k_1
	\not\!k_5}\right] \;.
\end{equation}
The trace in this equation can be expressed in many equivalent forms,
one of which is
\begin{eqnarray}
&&{\rm Tr}\left[{\gamma_R\not\!k_6\not\!k_2\not\!k_4\not\!k_3\not\!k_1
\not\!k_5}\right] =
s_{15}T_{2436}-s_{25}T_{1346}-s_{35}T_{1426}+s_{45}T_{1326}
+s_{56}T_{1342} \nonumber \\
&&+i\bigl[s_{13}x_{2456}-s_{15}x_{2346}-s_{35}x_{1246}
       -s_{24}x_{1356}+s_{26}x_{1345}+s_{46}x_{1235}\bigr] \;,
\label{GKJeq}
\end{eqnarray}
where
\begin{equation}
  T_{1234} \equiv {\rm Tr}\left[{\not\!k_1\not\!k_2\not\!k_3\not\!k_4}\right]
  = s_{12}s_{34}-s_{13}s_{24}+s_{14}s_{23} 
  \;.
\end{equation}
The real part of $J$ in Eq.\ (\ref{GKJeq}) was given in 
Ref.\ \cite{gunion86}, which however
did not discuss the imaginary part.  A slightly more compact and
convenient expression for $J$ can be obtained by using 
Eq.\ (\ref{Fprop3eq}):
\begin{eqnarray}
&&  {1\over16}J_{1234556} = - F_{123456}F_{214365} \nonumber \\
&&  \quad = -s_{13}s_{24}{\rm Tr}\left[{\gamma_R\not\!k_3\not\!k_6
  \not\!k_4\not\!k_5}\right]
  +s_{13}(s_{26}+s_{46}){\rm Tr}\left[{\gamma_R\not\!k_3\not\!k_2
   \not\!k_4\not\!k_5}\right] \nonumber \\
&& \quad +(s_{15}+s_{35})s_{24}{\rm Tr}\left[{\gamma_R\not\!k_3
  \not\!k_6\not\!k_4\not\!k_1}\right]
   -(s_{15}+s_{35})(s_{26}+s_{46})
  {\rm Tr}\left[{\gamma_R\not\!k_3\not\!k_2\not\!k_4\not\!k_1}\right] \;. 
\end{eqnarray}
This yields the following expression for the imaginary part of $J$:
\begin{eqnarray}
&& {\rm Im}\,J_{123456} = 32\left[s_{13}s_{24}x_{3456} 
   +s_{13}(s_{26}+s_{46})x_{2345}\right. \nonumber \\
&& \quad\left. - (s_{15}+s_{35})s_{24}
  x_{1346}-(s_{15}+s_{35})(s_{26}+s_{46})x_{1234}\right] \;. 
\label{ImJeq}
\end{eqnarray}

The function $K$ satisfies 
\begin{equation}
 K_{123456} = K_{321456}\;.
\label{Kpropseq}
\end{equation}
Various equivalent expressions for $K$ can be derived, one of the
simplest of which is got by writing $F$ in terms of $F^+$:
\begin{eqnarray}
K_{123456}& = &F^+_{216543}F_{436521} \nonumber \\
& = & 16 s_{13}\left\{s_{246}{\rm Tr}\left[{\gamma_R\not\!k_6\not\!k_2
  \not\!k_5\not\!k_4}\right]-s_{26}s_{46}(s_{25}+s_{45}+s_{56})\right\} \;,
\end{eqnarray}
which yields 
\begin{equation}
 {\rm Im}\,K_{123456} = 32s_{13}s_{246}x_{2456}\;.
\label{ImKeq}
\end{equation}

The expressions for the imaginary parts of $X_{123456}$ and
$Y_{123456}$ given in Eqs.\ (\ref{ImXeq},\ref{ImYeq})
follow from the definitions in Eqs.\ (\ref{Xdefeq},\ref{Ydefeq}) and
from Eqs.\ (\ref{ImJeq}) and (\ref{ImKeq}).

\subsubsection{Properties of $(H,X^\gamma,Y^\gamma)_{12345}$}

These functions were introduced in Eqs.\ (\ref{Hdefeq}) and 
(\ref{Xpheq}-\ref{ImYpheq}).  Since
they involve five momenta which obey $\sum_{i=1}^5k_i=0$, they are
much simpler than the functions $I,J,K,X,Y$.  It is straightforward to
show that
\begin{eqnarray}
X^\gamma_{12345}& = & {8s_{13}^2{\rm Tr}\left[{\gamma_R
  \not\!k_1\not\!k_4\not\!k_3\not\!k_5}\right] \over s_{14}s_{15}s_{25}s_{35}
  }\;, \\
  Y^\gamma_{12345}&= & {8s_{13}^2{\rm Tr}\left[{\gamma_R
  \not\!k_2\not\!k_3\not\!k_4\not\!k_5}\right] \over s_{15}s_{23}s_{25}s_{45}
  }\;,
\end{eqnarray}
from which the imaginary parts can easily be extracted.


\newpage

\end{document}